\documentclass[11pt,a4paper]{article}
\pdfoutput=1
\usepackage{jheppub} 

\usepackage{amsmath}
\usepackage{subcaption}
\usepackage[T1]{fontenc}
\usepackage[utf8]{inputenc}
\usepackage{slashed}
\usepackage{placeins}

\DeclareUnicodeCharacter{3B3}{ }
\DeclareUnicodeCharacter{204E}{ }

\DeclareRobustCommand{\Eq}[1]{Eq.~\eqref{eq:#1}}
\DeclareRobustCommand{\eq}[1]{eq.~\eqref{eq:#1}}

\DeclareRobustCommand{\eqs}[2]{eqs.~\eqref{eq:#1} and \eqref{eq:#2}}
\DeclareRobustCommand{\eqss}[3]{eqs.~\eqref{eq:#1}, \eqref{eq:#2} and \eqref{eq:#3}}
\DeclareRobustCommand{\fig}[1]{Fig.~\ref{fig:#1}}

\DeclareRobustCommand{\app}[1]{appendix~\ref{app:#1}}
\DeclareRobustCommand{\sec}[1]{Sec.~\ref{sec:#1}}
\DeclareRobustCommand{\secs}[2]{Secs.~\ref{sec:#1} and \ref{sec:#2}}

\DeclareRobustCommand{\mycite}[1]{ref.~\cite{#1}}
\DeclareRobustCommand{\mycites}[1]{refs.~\cite{#1}}

\newcommand{\MS}{{\overline{\mathrm{MS}}}}
\newcommand{\df}{\mathrm{d}}
\newcommand{\img}{\mathrm{i}}
\newcommand{\eps}{\epsilon}
\newcommand{\cA}{\mathcal{A}}

\newcommand{\cI}{\mathcal{I}}

\newcommand{\cO}{\mathcal{O}}
\newcommand{\cP}{\mathcal{P}}

\newcommand{\cR}{{\tilde R}}
\newcommand{\nn}{\nonumber}
\newcommand{\bn}{{\bar{n}}}
\newcommand{\as}{\alpha_s}
\newcommand{\LQCD}{\Lambda_\mathrm{QCD}}

\newcommand{\tr}{\mathrm{tr}}
\newcommand{\Tr}{\mathrm{Tr}}
\newcommand{\fm}{\mathrm{fm}}
\newcommand{\GeV}{\mathrm{GeV}}

\newcommand{\pb}{{p{\cdot}b}}
\newcommand{\bt}{\vec b_T}

\newcommand{\qt}{\vec q_T}
\newcommand{\GammaC}{\Gamma_{\rm cusp}}
\newcommand{\ns}{{\rm ns}}
\newcommand{\MI}{\mathbf{\rm I}}
\newcommand{\RIpMOM}{RI$^\prime$/MOM}

\title{Renormalization and Matching for the Collins-Soper Kernel from Lattice QCD}

\author[a]{Markus A.~Ebert,}
\author[a]{Iain W.~Stewart,}
\author[a,b]{and Yong Zhao}

\affiliation[a]{Center for Theoretical Physics, Massachusetts Institute of Technology, Cambridge, Massachusetts 02139, USA}
\affiliation[b]{Physics Department, Brookhaven National Laboratory, Bldg. 510A, Upton, NY 11973, USA}%

\emailAdd{ebert@mit.edu}
\emailAdd{iains@mit.edu}
\emailAdd{yzhao@bnl.gov}

\abstract{
The Collins-Soper kernel, which governs the energy evolution of transverse-momentum dependent parton distribution functions (TMDPDFs), is required to accurately predict Drell-Yan like processes at small transverse momentum, and is a key ingredient for extracting TMDPDFs from experiment.
Earlier we proposed a method to calculate this kernel from ratios of the so-called quasi-TMDPDFs determined with lattice QCD, which are defined as  hadronic matrix elements of staple-shaped Euclidean Wilson line operators. Here we provide the one-loop renormalization of these operators in a regularization-independent momentum subtraction (RI$^\prime$/MOM) scheme, as well as the conversion factor from the RI$^\prime$/MOM-renormalized quasi-TMDPDF to the $\overline{\rm MS}$ scheme.
We also propose a procedure for calculating the Collins-Soper kernel directly from position space correlators, which simplifies the lattice determination.
}

\preprint{\vbox{
\hbox{MIT--CTP 5145}
}}

\arxivnumber{}

\begin{document}
\maketitle

\section{Introduction}

Transverse-momentum dependent parton distribution functions (TMDPDFs) describe the longitudinal and transverse momentum distribution of quarks and gluons in hadrons and nuclei, and thus are of vital interest to improving our understanding of hadronic and nuclear structure \cite{Boer:2011fh,Accardi:2012qut}.
They are also crucial to predicting transverse momentum distributions in the Drell-Yan process, a key observable both for the Tevatron~\cite{Affolder:1999jh,Abbott:1999yd,Abazov:2007ac,Abazov:2010kn} and the LHC~\cite{Aad:2011gj,Chatrchyan:2011wt, Aad:2014xaa, Khachatryan:2015oaa, Aad:2015auj,  Khachatryan:2016nbe},
as well as in semi-inclusive deep-inelastic scattering at low energies \cite{Ashman:1991cj,Derrick:1995xg,Adloff:1996dy,Aaron:2008ad,Airapetian:2012ki,Adolph:2013stb,Aghasyan:2017ctw}.

TMDPDFs measure the transverse momentum $q_T$ carried by the struck parton.
For perturbative $q_T \gg \LQCD$, they can be calculated in terms of collinear parton distribution functions, and the resulting matching formula is known to next-to-next-to-leading order (NNLO) \cite{Catani:2011kr,Catani:2012qa,Gehrmann:2014yya,Luebbert:2016itl,Echevarria:2015byo,Echevarria:2016scs,Luo:2019hmp,Luo:2019bmw}.
In contrast, for nonperturbative $q_T \lesssim \LQCD$, TMDPDFs become genuinely nonperturbative objects which so far have only been extracted from measurements by performing global fits to a variety of experimental data sets, see e.g.~\mycites{Landry:1999an,Landry:2002ix,Konychev:2005iy,DAlesio:2014mrz,Bacchetta:2017gcc,Scimemi:2017etj}. Since there are some issues associated to these extractions, in particular with reconciling low and high energy data, an independent determination from first principles is highly desirable.
This has motivated studies with lattice QCD that have been carried out in \mycites{Musch:2010ka,Musch:2011er,Engelhardt:2015xja,Yoon:2016dyh,Yoon:2017qzo}, primarily for ratios of moments in the longitudinal momentum fraction.

The TMDPDF $f_i(x,\bt,\mu,\zeta)$ for a parton of flavor $i$ depends on the longitudinal momentum fraction $x$ and the position space parameter $\bt$, which is Fourier-conjugate to $\qt$, as well as the renormalization scale $\mu$ and the Collins-Soper scale $\zeta$ \cite{Collins:1981va,Collins:1981uk}.
The latter encodes the energy dependence of the TMDPDF, {\it i.e.}\ the momentum of the hadron or equivalently the hard scale of the scattering process, and the associated evolution is governed by the Collins-Soper equation \cite{Collins:1981va,Collins:1981uk}
\begin{align}
 \zeta \frac{\df}{\df\zeta} f_i(x, \bt, \mu, \zeta) &
= \frac{1}{2} \gamma^i_\zeta(\mu,b_T) f_i(x, \bt, \mu, \zeta)
\,.\end{align}
The $\gamma^i_\zeta(\mu,b_T)$ that appears here is referred to as either the Collins-Soper (CS) kernel or rapidity anomalous dimension for the TMDPDF.  From consistency of the $\zeta$ and $\mu$ evolution equations, combined with information on the all order structure of the renormalization for Wilson line operators, it is known to have
the all-order structure
\begin{align} \label{eq:gamma_zeta}
 \gamma_\zeta^i(\mu,b_T) = -2 \int_{1/b_T}^\mu \frac{\df\mu'}{\mu'} \GammaC^i[\as(\mu')] + \gamma_\zeta^i[\as(1/b_T)]
\,,\end{align}
where $\GammaC^i(\as)$ is the cusp anomalous dimension and $\gamma_\zeta^i(\as)$ is the noncusp anomalous dimension,
both of which are known perturbatively in QCD at three loops, see~\mycites{Korchemsky:1987wg, Moch:2004pa, Vogt:2004mw} and~\mycites{Davies:1984hs,Davies:1984sp,deFlorian:2000pr,Becher:2010tm,Gehrmann:2014yya,Echevarria:2015byo,Luebbert:2016itl,Li:2016axz,Li:2016ctv,Vladimirov:2016dll}, respectively.%
\footnote{The four-loop cusp anomalous is also known numerically~\cite{Moch:2017uml, Moch:2018wjh}, and largely analytically~\cite{Grozin:2016ydd, Henn:2016men, Davies:2016jie, Lee:2016ixa, Grozin:2018vdn,  Lee:2019zop, Henn:2019rmi, Bruser:2019auj}.}
As should be evident from \eq{gamma_zeta}, the Collins-Soper kernel becomes genuinely nonperturbative when $b_T^{-1} \lesssim \LQCD$,
independent of the renormalization scale $\mu$.
Consequently, the scale evolution of TMDPDFs becomes nonperturbative itself, and relating TMDPDFs at different energies requires nonperturbative knowledge of $\gamma_\zeta^i(\mu,b_T)$, even if one chooses a perturbative $\mu\ge 1\,{\rm GeV}$. 

When extracting TMDPDFs from global fits, it is thus also necessary to fit $\gamma_\zeta^i$, which is typically achieved by splitting the kernel into a perturbative and nonperturbative piece,
\begin{align}\label{eq:gamma_zeta_split}
 \gamma_\zeta^i(\mu,b_T) &= -2 \int_{\mu_b}^\mu \frac{\df\mu'}{\mu'} \GammaC^i[\as(\mu')] + \gamma_\zeta^i(\mu, \mu_b)  + g^i(b_T,\mu_b)
\,,\end{align}
where $\mu_b \equiv \mu_b(\mu, b_T)$ is chosen to always be a perturbative scale, such that all nonperturbative physics is separated into the function $g^i(b_T,\mu_b)$.
A common parameterization of $g^i(b_T,\mu_b)$ is to assume a quadratic form, $g^i(b_T,\mu_b) = g^i_K b_T^2$ with constant $g_K^i$ \cite{Bacchetta:2017gcc,Scimemi:2017etj}, which has also been motivated by a renormalon analysis \cite{Scimemi:2016ffw}, but other forms have also been employed \cite{Bertone:2019nxa,Vladimirov:2019bfa}.
In the literature, there is a considerable discrepancy between \mycites{Scimemi:2017etj,Bertone:2019nxa} and \mycite{Bacchetta:2017gcc} on whether the nonperturbative part of $\gamma_\zeta^i$ is crucial to describe the measured data or not.
This is perhaps not surprising, as \mycites{Scimemi:2017etj,Bertone:2019nxa} are based on Drell-Yan data at relatively large $q_T$, where one expects nonperturbative effects to be suppressed, while they become much more important in the lower energy measurements included in \mycite{Bacchetta:2017gcc}.

The lack of precise knowledge of the nonperturbative part of $\gamma_\zeta^i(\mu,b_T)$ from global fits motivates an independent determination from lattice QCD.
Here, a key difficulty is that TMDPDFs are defined as lightcone correlation functions which depend on the Minkowski time, while first principles lattice QCD calculations are inherently restricted to the study of Euclidean time operators. 
Large-momentum effective theory (LaMET) was proposed to overcome this hurdle in a systematically improvable manner for collinear PDFs (and generalized PDFs)
by relating so-called quasi-PDFs, defined as equal-time correlators, through a perturbative matching to the physical PDF  \cite{Ji:2013dva,Ji:2014gla}.
For these collinear quasi-PDFs, significant progress has been made, in particular on their renormalization and matching onto PDFs~\cite{Xiong:2013bka,Ma:2014jla,Ma:2014jga,Ji:2015jwa,Ji:2015qla,Xiong:2015nua,Li:2016amo,Ishikawa:2016znu,Chen:2016fxx,Carlson:2017gpk,Briceno:2017cpo,Xiong:2017jtn,Constantinou:2017sej,Rossi:2017muf,Ji:2017rah,Ji:2017oey,Ishikawa:2017faj,Green:2017xeu,Wang:2017qyg,Chen:2017mie,Stewart:2017tvs,Wang:2017eel,Spanoudes:2018zya,Izubuchi:2018srq,Xu:2018mpf,Rossi:2018zkn,Zhang:2018diq,Li:2018tpe,Liu:2018tox}, and the study of power corrections to the matching relation~\cite{Chen:2016utp,Radyushkin:2017ffo,Braun:2018brg}, and first lattice calculations of the $x$-dependence of PDFs and distribution amplitudes have been carried out in~\mycites{Lin:2014zya,Alexandrou:2015rja,Chen:2016utp,Alexandrou:2016jqi,Zhang:2017bzy,Alexandrou:2017huk,Chen:2017mzz,Green:2017xeu,Chen:2017lnm,Chen:2017gck,Alexandrou:2018pbm,Chen:2018xof,Chen:2018fwa,Alexandrou:2018eet,Liu:2018uuj,Lin:2018qky,Fan:2018dxu,Liu:2018hxv,Cichy:2019ebf,Chai:2019rer}.
Recent lattice calculations at the physical pion mass have shown encouraging results for a precise determination of PDFs using the LaMET, including in particular those
of the European Twisted Mass Collaboration~\cite{Alexandrou:2018pbm,Alexandrou:2018eet}, and results reported by the Lattice Parton Physics Project Collaboration~\cite{Chen:2018xof,Lin:2018qky,Liu:2018hxv}.

The application of LaMET to obtain TMDPDFs from lattice has only been studied very recently~\cite{Ji:2014hxa,Ji:2018hvs,Ebert:2018gzl,Ebert:2019okf}.
A key difference to collinear PDFs is the necessity to combine a hadronic matrix element with a soft vacuum matrix element in order to obtain a well-defined (quasi) TMDPDF.
In \mycite{Ebert:2019okf} it was shown that this soft factor, which involves lightlike Wilson lines, can not be simply related to an equal-time correlation function computable on the lattice, and hence without a careful construction of the quasi-TMDPDF one can only generically expect to encounter a \emph{nonperturbative} relation between TMDPDFs and quasi-TMDPDFs, rather than a relation that is determined by a perturbatively calculable short distance coefficient.\footnote{A proposal for a potential quasi-TMDPDF definition that exibits a perturbative matching to the TMDPDF at one-loop was proposed in \mycite{Ebert:2019okf}, but it remains to be analyzed at higher orders.}
However, in certain ratios of TMDPDFs this soft factor, and physically related contributions in the TMD proton matrix elements, cancel out. Hence such ratios can be obtained from ratios of suitably defined quasi-TMDPDFs which can be obtained from lattice.
In particular, \mycite{Ebert:2018gzl} showed that the Collins-Soper kernel can be obtained from such a ratio using%
\footnote{Note that compared to \mycites{Ebert:2018gzl,Ebert:2019okf}, here we always drop superscripts ``TMD'' on $C_\ns^\mathrm{TMD}$ and $\tilde f_\ns^\mathrm{TMD}$.}
\begin{align} \label{eq:ratio_1}
 \gamma^q_\zeta(\mu, b_T) & = \frac{1}{\ln(P^z_1/P^z_2)}
  \ln \frac{C_\ns(\mu,x P_2^z)\, \tilde f_\ns(x, \bt, \mu, P^z_1)}
           {C_\ns(\mu,x P_1^z)\, \tilde f_\ns(x, \bt, \mu, P^z_2)}
\,,\end{align}
where $C_\ns$ is a perturbative matching coefficient given at one loop in \mycites{Ebert:2018gzl,Ebert:2019okf}, $\tilde f_\ns$ is the nonsinglet ($u{-}d$) quasi-TMDPDF, and $P^z_1 \ne P^z_2$ are two different proton momenta that are used for the corresponding quasi-TMDPDF calculations.

In order to obtain $\gamma_\zeta^i$ in the $\MS$ scheme, in \eq{ratio_1} the quasi-TMDPDF $\tilde f_\ns$ is assumed to be in the $\MS$ scheme as well.
Consequently, a critical step in this approach is the renormalization of the bare quasi-TMDPDF on the lattice and its subsequent scheme conversion into the $\MS$ scheme.
On the lattice, one employs the finite lattice spacing $a>0$ as UV regulator, and the renormalization should be performed in a scheme that is defined nonperturbatively to facilitate the removal of both linear and logarithmic divergences.
In contrast, the $\MS$ renormalization is defined by calculating in $d=4-2\eps$ dimensions and subtracting only poles in $1/\eps$.
Since the scheme conversion factor is defined as the difference of renormalized quantities, it is independent of the two underlying UV regulators.
In particular, this allows us to calculate it order-by-order in continuum perturbation theory in $d$ dimensions.

For the longitudinal quasi-PDFs, such nonperturbative renormalization, scheme conversions, and the associated matching to obtain the analog of $C_\ns$ have been studied and implemented in \mycites{Constantinou:2017sej,Alexandrou:2017huk,Chen:2017mzz,Stewart:2017tvs,Liu:2018uuj} with the regularization-independent momentum subtraction (RI/MOM) schemes~\cite{Martinelli:1994ty}.
Such a calculation has also been carried out in \mycite{Constantinou:2019vyb} for staple-shaped Wilson line operators at vanishing longitudinal separation, 
which is connected to the calculations needed for determining TMDPDFs.
In particular it corresponds to a special case of the quasi-TMDPDF operators studied here, which will involve staple-shaped Wilson lines but with an additional separation along the longitudinal direction.
In this paper, we determine the scheme conversion coefficient between the \RIpMOM~scheme\footnote{The \RIpMOM\ and RI/MOM correspond to two different schemes for the quark wave function renormalization, which we will discuss further in \sec{RIMOM}. We favor the \RIpMOM\ scheme here since it is the scheme most often adopted for this type of lattice calculation.} and $\MS$ for quasi-TMDPDFs, including the longitudinal separation,
and also calculate the corresponding one-loop matching coefficient $C_\ns$.

This paper is structured as follows.
In \sec{CS_Lattice} we briefly review the definition of (quasi) TMDPDFs and how the Collins-Soper kernel $\gamma_\zeta^q$ can be extracted from lattice \mycites{Ebert:2018gzl,Ebert:2019okf}.
In \sec{CS_position_space} we also propose a new improved method for obtaining $\gamma_\zeta^q$ which reduces systematic uncertainties in the lattice analysis by directly exploiting the quasi-TMPDF correlators in longitudinal position space.
We then proceed in \sec{RIMOM} to discuss the general structure of the \RIpMOM~renormalization and scheme conversion from the \RIpMOM~scheme to the $\MS$ scheme, before giving details on our one-loop calculation of the required renormalization and conversion factors in \sec{NLO}. The impact of these results are numerically illustrated in \sec{numerics}, before concluding in \sec{conclusion}. In \app{master_integrals} we collect formulae for the master integrals used in \sec{NLO}.

\section{Determination of the Collins-Soper kernel from lattice QCD}
\label{sec:CS_Lattice}

In this section we briefly review the definition of TMDPDFs and the construction of quasi-TMDPDFs computable on lattice,
as well as how the Collins-Soper kernel can be determined from these, and refer to \mycites{Ebert:2018gzl,Ebert:2019okf} for more details.
We also show how to determine the Collins-Soper kernel directly in position space, which is better suited to a lattice calculation than the method proposed in \mycites{Ebert:2018gzl} to obtain the kernel in momentum space.

\subsection{Definition of TMDPDFs}
\label{sec:def_TMDPDF}

We define the quark TMDPDF for a hadron moving close to the $n^\mu = (1,0,0,1)$ direction with momentum $P^\mu$ as
\begin{align} \label{eq:tmdpdf}
 f_q(x, \bt, \mu , \zeta) &
 = \lim_{\substack{\eps\to 0 \\ \tau\to 0}} Z_{\rm uv}^q(\mu,\zeta,\eps)\,
   \int\frac{\df b^+}{4\pi} e^{-\img \frac12 b^+ (x P^-)}
   B_q(b^+, \bt, \eps, \tau, x P^-) \Delta_S^q(b_T,\eps,\tau)
\,,\end{align}
where we use the lightcone coordinates $b^\pm = b^0 \mp b^z$ and $\bt$ are the transverse spatial coordinates
such that $\bt^2 = b_T^2 > 0$.
In \eq{tmdpdf}, the bare beam function $B_q$ is a hadronic matrix element encoding collinear radiation,
and the bare soft factor $\Delta_S^q$ is constructed from a soft vacuum matrix element, to be defined below.
The TMDPDF gives the probability to obtain a quark with lightcone momentum $p^- = x P^-$ and transverse momentum $\qt$,
which is Fourier-conjugate to the parameter $\bt$.
$Z_{\rm uv}^q$ is the UV renormalization constant, with $\eps$ being the UV regulator and $\mu$ the associated renormalization scale.
Beam and soft functions individually suffer from so-called rapidity divergences~\cite{Soper:1979fq,Collins:1981uk,Collins:1992tv,Collins:2008ht,Becher:2010tm,GarciaEchevarria:2011rb,Chiu:2011qc,Chiu:2012ir}, which are regulated by an additional regulator denoted as $\tau$, and these divergences give rise to the Collins-Soper scale $\zeta$.
However, the rapidity divergences cancel between beam and soft function as $\tau\to0$, giving rise to a well-defined TMDPDF.
For a detailed discussion of different rapidity regularization schemes, see e.g.~\mycite{Ebert:2019okf}.

The bare quark beam function is defined as
\begin{align} \label{eq:beam}
 B_q\bigl(b^+, \bt, \eps, \tau, x P^- \bigr) &=
 \Bigl<\! h(P) \Bigr|  \Bigl[ \bar q(b^\mu)
 W(b^\mu) \frac{\gamma^-}{2}
 W_{T}\bigl(-\infty\bn;\vec b_T,\vec 0_T\bigr)
 W^\dagger(0)  q(0) \Bigr]_\tau \Bigl| h(P) \!\Bigr>
\,,\end{align}
where $[...]_\tau$ denotes the rapidity regularization of the operator,
$h(P)$ denotes the hadron state of momentum $P^\mu$,
the quark fields are separated by $b^\mu = b^+ \bn^\mu/2 + b_T^\mu$ with $\bn^\mu=(1,0,0,-1)$,
and the Wilson lines are defined as%
\footnote{Note that we have changed the sign of the strong coupling $g$ compared to \mycites{Ebert:2018gzl,Ebert:2019okf} to agree with the convention that the covariant derivative is given by $D^\mu = \partial^\mu + \img g \cA^\mu$. This sign agrees with \mycite{Capitani:2002mp} which we use as a reference for Euclidean Feynman rules for our calculation.
}
\begin{align} \label{eq:Wilson_lines}
 W(x^\mu) &= P \exp\biggl[ \img g \int_{-\infty}^0 \df s\, \bn \cdot \cA(x^\mu + s \bn^\mu) \biggr]
\,, \nn\\
 W_{T}(x^\mu;\vec b_T,\vec 0_T) &=
   P \exp\left[ -\img g \int_{\vec 0_T}^{\vec b_T} \df \vec s_T \cdot \vec \cA_T(x^\mu + s_T^\mu) \right]
\,.\end{align}
\FloatBarrier

\begin{figure}[pt]
 \centering
 \begin{subfigure}{0.45\textwidth}
  \includegraphics[width=\textwidth]{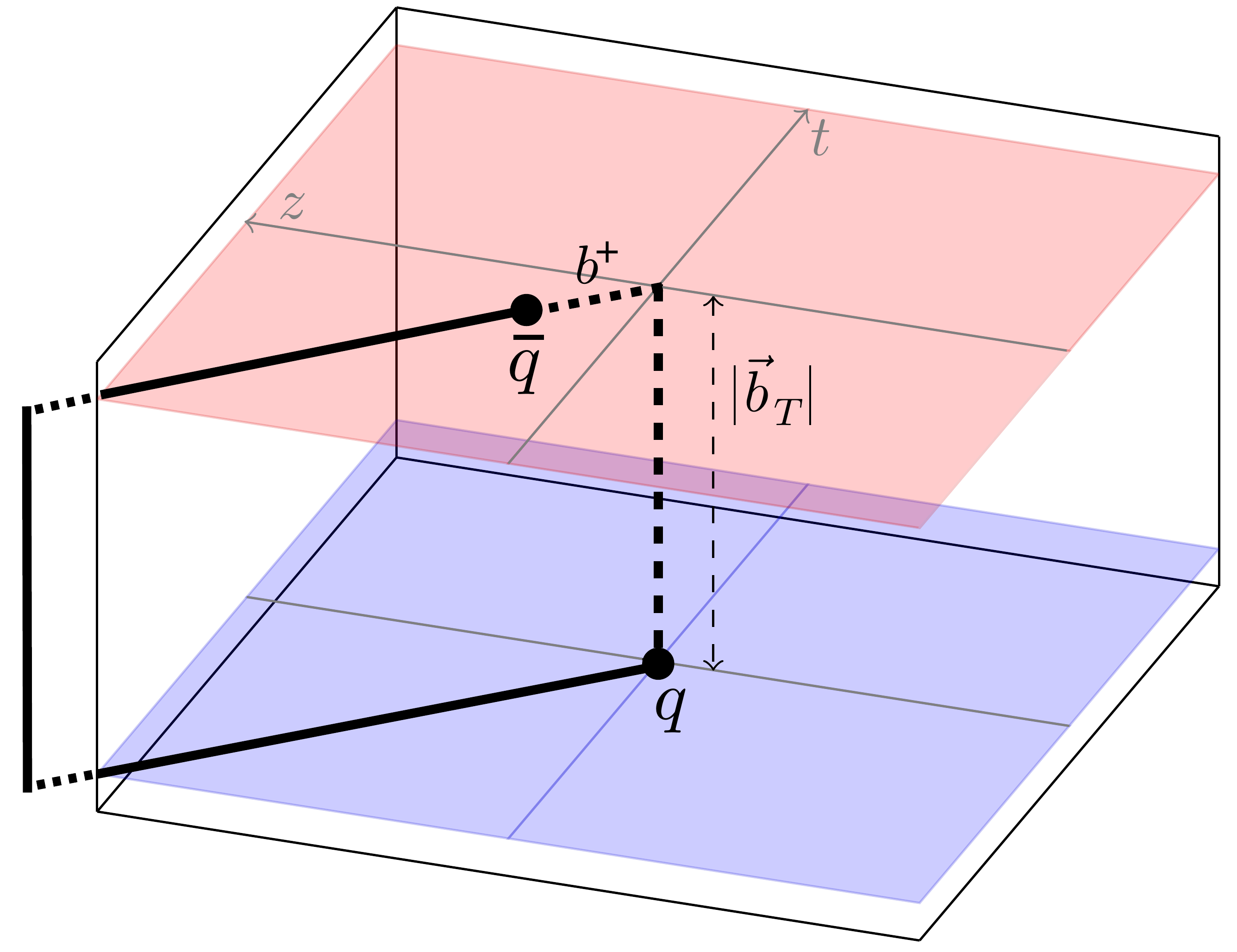}
  \caption{}
 \end{subfigure}
\hfill
 \begin{subfigure}{0.45\textwidth}
  \includegraphics[width=\textwidth]{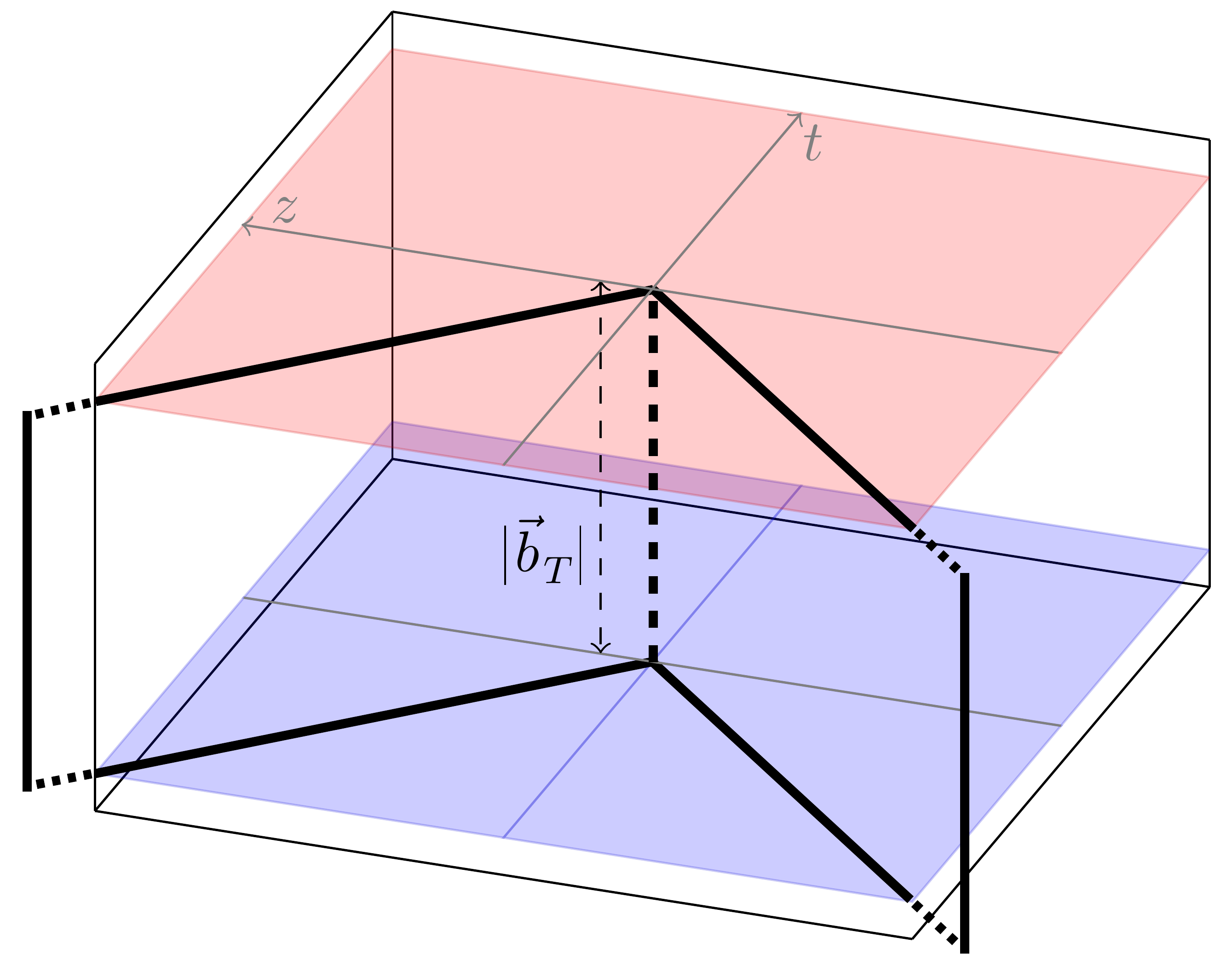}
  \caption{}
 \end{subfigure}
 \caption{Illustration of the Wilson line structure of the $n$-collinear beam function $B_{q}$ (a)
 and the soft function $S^q$ (b), as given in \eqs{beam}{soft}.
 The solid Wilson lines extend to infinity in the directions indicated. Adapted from \mycite{Li:2016axz}.}
 \label{fig:wilsonlines}
\end{figure}

The bare quark soft function is defined as
\begin{align} \label{eq:soft}
 S^q(b_T,\eps,\tau) &= \frac{1}{N_c} \bigl< 0 \bigr| {\rm Tr} \bigl[ S^\dagger_n(\bt) S_\bn(\bt)
   S_{T}(-\infty \bn;\vec b_T,\vec 0_T)
 \nn\\&\hspace{2cm}\times
 S^\dagger_\bn(\vec 0_T) S_n(\vec 0_T)
 S_{T}^\dagger\bigl(-\infty n;\vec b_T,\vec 0_T\bigr) \bigr]_\tau
 \bigl|0 \bigr>
\,,\end{align}
where as before $[...]_\tau$ denotes the rapidity regularization, and the Wilson lines are given by
\begin{align} \label{eq:soft_Wilson_lines}
 S_n(x^\mu) &= P \exp\biggl[ \img g \int_{-\infty}^0 \df s\, n \cdot \cA(x^\mu + s n^\mu) \biggr]
\,,\nn\\
 S_{T}(x^\mu;\vec b_T,\vec 0_T) &= W_{T}(x^\mu;\vec b_T,\vec 0_T) =
 P \exp\left[ -\img g \int_{\vec 0_T}^{\vec b_T} \df \vec s_T \cdot \vec \cA_T(x^\mu + s_T^\mu) \right]
\,.\end{align}
The Wilson line paths of both beam and soft function are illustrated in \fig{wilsonlines}.

Finally, the soft factor $\Delta_S^q$ entering \eq{tmdpdf} is defined as
\begin{align}
 \Delta_S^q(b_T,\eps,\tau) = \frac{\sqrt{S^q(b_T,\eps,\tau)}}{S_0^q(b_T,\eps,\tau)}
\,,\end{align}
where $S^q$ is the soft function defined in \eq{soft} and $S_0^q$ is a subtraction factor necessary to avoid double counting of soft physics in the beam and soft function.
Its definition depends on the employed rapidity regulator $\tau$, but as the notation indicates, it is typically closely related to $S^q$ itself.
For example, in the scheme of~\mycite{Chiu:2012ir} one has $S_0^q = 1$, while in the schemes of \mycite{GarciaEchevarria:2011rb,Li:2016axz} one has $S_0^q = S^q$.
For more details, see \mycite{Ebert:2019okf}.

\subsection{Definition of quasi-TMDPDFs}
\label{sec:def_quasi_TMDPDF}

The quasi-TMDPDF is defined analogous to \eq{tmdpdf}, but as an equal-time correlator rather than a lightcone correlation function, namely
\begin{align} \label{eq:qtmdpdf}
 \tilde f_q(x, \bt,\mu,P^z) =
 \lim_{\substack{a\to 0 \\ L\to \infty}}
 \int\! \frac{\df b^z}{2\pi} e^{\img b^z (x P^z)}
 &\tilde Z'_q(b^z,\mu,\tilde \mu) \tilde Z_{\rm uv}^q(b^z,\tilde \mu, a)
 \nn\\&\times
 \tilde B_q(b^z, \bt, a, P^z, L) \tilde\Delta_S^q(b_T, a, L)
\,.\end{align}
Here $\tilde B_q$ is the quasi-beam function, $\tilde \Delta_S^q$ includes the quasi-soft function together with subtractions, $\tilde Z_{\rm uv}^q$ carries out UV renormalization in a lattice-friendly scheme, where $\tilde \mu$ stands for any added scales introduced by this scheme choice, and $\tilde Z_q'$ converts the result perturbatively to the $\MS$ scheme with scale $\mu$.
Note that here, $L$ refers to the length of the Wilson lines in the definition of $\tilde B_q$ and $\tilde\Delta_S^q$ (see below), not the size of the lattice.
The quasi beam and soft functions will be constructed such that all Wilson line self energies proportional to $L/a$ and $b_T/a$, as well as divergences $\propto L/b_T$ which correspond to rapidity divergences in the lightlike case~\cite{Ji:2018hvs,Ebert:2019okf}, cancel between $\tilde B_q$ and $\tilde\Delta_S^q$. Therefore, the remaining UV renormalization $\tilde Z_{\rm uv}^q$ and the scheme conversion $\tilde Z'_q$ only depend on $b^z$, but not necessarily $b_T$ or $L$.%
\footnote{Depending on the lattice renormalization scheme, $\tilde\mu$ may induce dependence on other parameters, like $b_T$ and $L$.}
We keep implicit that finite lattice volume effects must be either removed or included as a systematic uncertainty.

The bare quasi-beam function is defined as
\begin{align} \label{eq:qbeam}
 \tilde B_q(b^z, \bt, a, P^z, L) =
 \Bigl<\! h(P) \Big| &\bar q(b^\mu) W_{\hat z}(b^\mu;L-b^z) \frac{\Gamma}{2}
 W_T(L \hat z; \bt, \vec{0}_T) W^\dagger_{\hat z}(0;L) q(0) \Big| h(P) \!\Bigr>
\,,\end{align}
where $b^\mu = (0, \bt, b^z)$, and the UV regulator is denoted as $a$, following the notation for the finite lattice spacing that acts as a UV regulator in lattice calculations.
Due to the finite lattice size, the longitudinal Wilson lines are truncated at a length $L$ less than the size of the lattice, which also regulates the analog of rapidity divergences \cite{Ji:2018hvs,Ebert:2019okf}.
Compared to \eq{beam}, we also replaced $\gamma^-$ by the Dirac structure $\Gamma$, which can be chosen as $\Gamma = \gamma^0$ or $\Gamma = \gamma^z$.
(Technically, one can also use a combination, for example $\gamma^0 + \gamma^z$.)
The Wilson lines of finite length $L$ are defined by
\begin{align} \label{eq:coll_Wilson_L}
 W_{\hat z}(x^\mu;L) &= P \exp\left[ -\img g \int_{L}^0 \df s \, \cA^z(x^\mu + s \hat z) \right]
\,,\end{align}
while the transverse gauge links are identical to those in \eq{Wilson_lines}.

\begin{figure}[pt]
 \centering
 \begin{subfigure}{0.45\textwidth}
  \includegraphics[width=\textwidth]{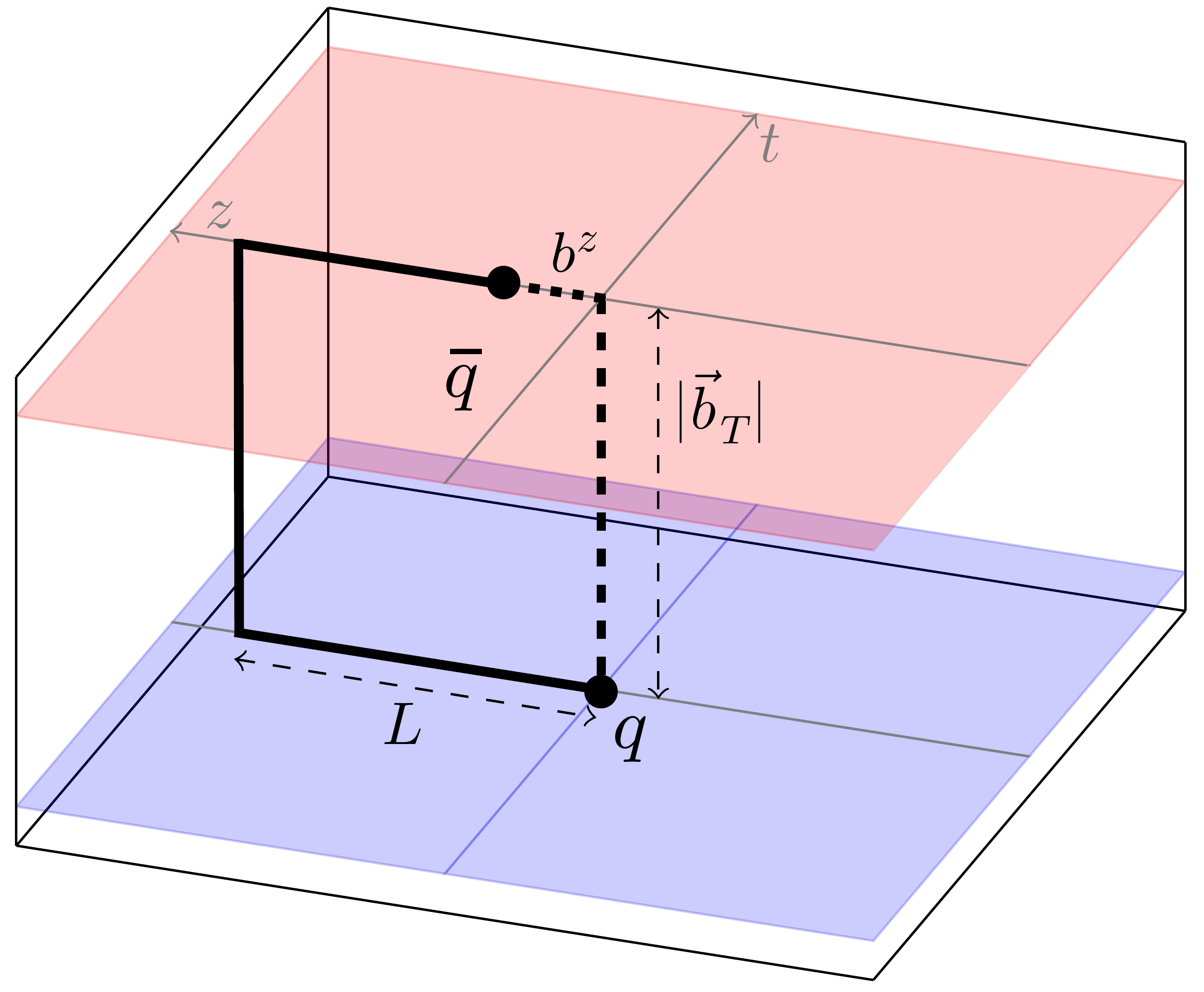}
  \caption{}
 \end{subfigure}
\hfill
 \begin{subfigure}{0.45\textwidth}
 \includegraphics[width=\textwidth]{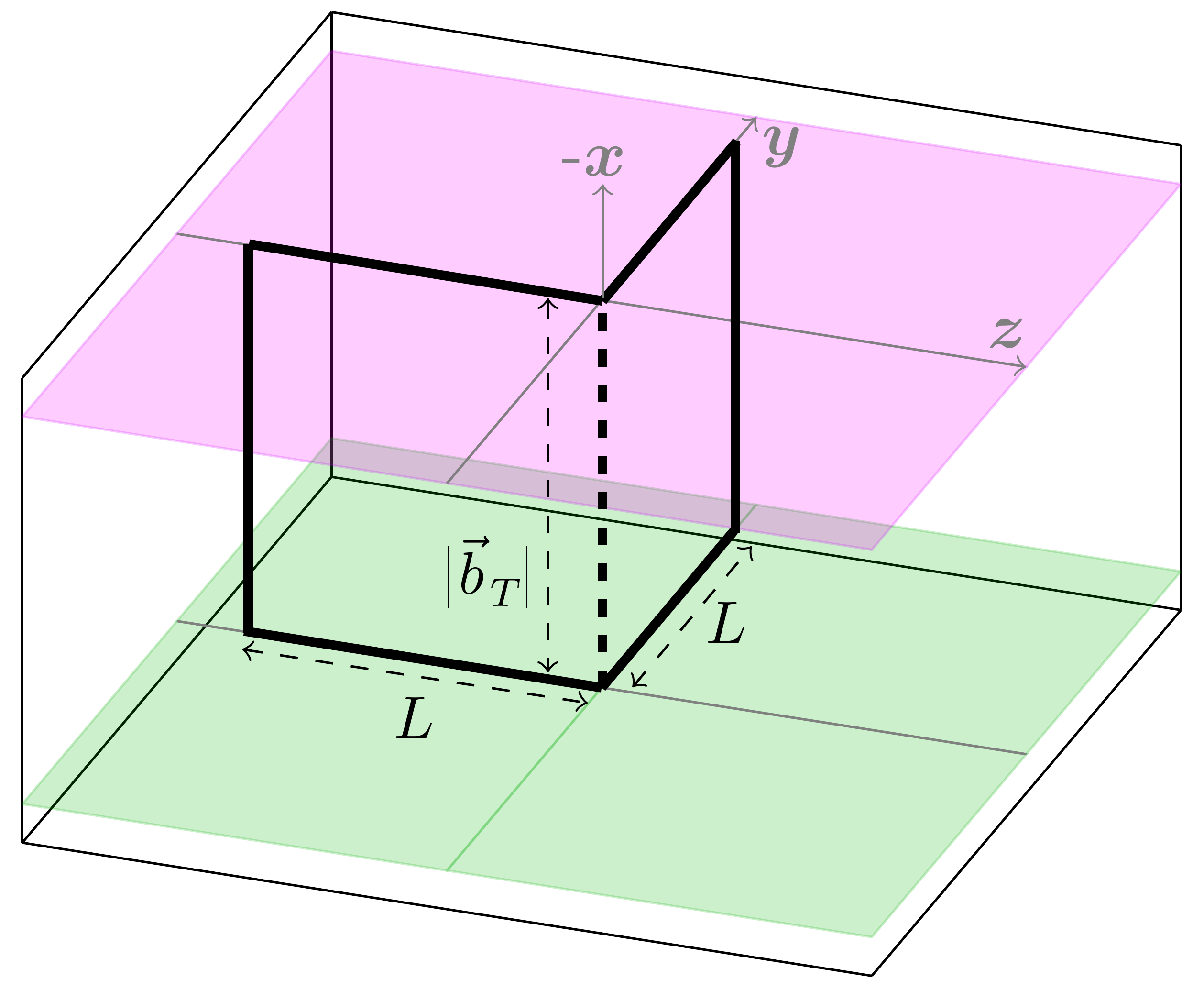}
  \caption{}
 \end{subfigure}
 \caption{Illustration of the Wilson line structure of the quasi beam function $\tilde B_{q}$ (a)
 and the bent quasi soft function $\tilde S^q_{\rm bent}$ (b), as given in \eqs{qbeam}{qsoft}.
 Note the different coordinate systems in the two figures: (a) is shown in $(z,t,x)$ space, while (b) is shown in $(z,y,x)$ space. In both cases, $\bt$ is aligned along the $x$ axis.}
 \label{fig:quasi_wilsonlines}
\end{figure}

For the quasi soft function, we use the bent soft function of \mycite{Ebert:2019okf}, defined as
\begin{align} \label{eq:qsoft}
 \tilde S^q_{\rm bent}(b_T, a, L) &
 = \frac{1}{N_c} \bigl< 0 \big|
  \Tr\bigl\{ S^\dagger_{\hat z}(\bt;L) S_{-\bn_\perp}\!(\bt;L)
   S_{T}(L\bn_\perp;\vec b_T,\vec 0_T)
   \nn\\&\hspace{2cm}\times
 S^\dagger_{-\bn_\perp}\!(\vec 0_T;L) S_{\hat z}(\vec 0_T;L)
 S_{T}^\dagger\bigl(-L \hat z;\vec b_T,\vec 0_T\bigr) \bigr\}
 \bigl|0 \bigr>
\,.\end{align}
Here, $\bn_\perp^\mu$ is the transverse unit vector orthogonal to $n_\perp^\mu = b_\perp^\mu / b_T$ and $\hat z$. Explicitly, if we parameterize $b_T^\mu = b_T (0,\cos\phi, \sin\phi,0)$, then $n_\perp^\mu = (0,\cos\phi,\sin\phi,0)$, and a valid choice for $\bn_\perp^\mu$ is $\bn_\perp^\mu = (0, -\sin\phi,\cos\phi,0)$.
The Wilson lines in \eq{qsoft} are defined as
\begin{align} \label{eq:soft_Wilson_L}
 S_{\hat z}(x^\mu; L) &= P \exp\biggl[ - \img g \int_{-L}^0 \df s \, \cA^z(x^\mu + s \hat z^\mu) \biggr]
\,,\nn\\
 S_{-\bn_\perp}\!(x^\mu;L) &= P \exp\biggl[ - \img g \int_{-L}^0 \df s\, \bn_\perp \cdot \cA(x^\mu - s \bn_\perp^\mu) \biggr]
\,,\nn\\
 S_{T}(x^\mu;\vec b_T,\vec 0_T) &= P \exp\biggl[  -\img g \int_{\vec 0_T}^{\vec b_T} \df \vec s_T \cdot \vec \cA_T(x^\mu + s_T^\mu) \biggr]
\,.\end{align}
The Wilson line paths of both quasi beam and quasi soft function are illustrated in \fig{quasi_wilsonlines} for the choice $\phi=0$.

The quasi-soft factor is obtained from the bent soft function as
\begin{align} \label{eq:qDeltaS}
 \tilde\Delta_S^q(b_T,a,L)
 = \frac{\sqrt{\tilde S^q_{\rm bent}(b_T, a, L)}}{\tilde S_0^q(b_T, a, L)}
 = \frac{1}{\sqrt{\tilde S^q_{\rm bent}(b_T, a, L)}}
\,,\end{align}
where $\tilde S_0^q = \tilde S^q_{\rm bent}$ is the subtraction factor which avoids double counting between quasi beam and soft functions. 
The overall length of the Wilson lines appearing in $\tilde \Delta_S^q$ must be chosen to ensure the cancellation of Wilson line self energies in \eq{qtmdpdf}~\cite{Ebert:2019okf}, whereas implementing this with the specific choice $\tilde S_0^q = \tilde S^q_{\rm bent}$ corresponds to a particular scheme.

Note that for the construction of the quasi-TMDPDF, different definitions of the quasi soft function could be employed as well.
This yields different definitions of the quasi-TMDPDF, which will affect the (possibly nonperturbative) kernel relating quasi-TMDPDFs and TMDPDFs, see \mycite{Ebert:2019okf} for a more detailed discussion.
With the bent soft function in \eq{soft}, this relation was shown to be short distance dominated and hence perturbative at one loop, which motivates its use here.
Importantly, for the determination of the Collins-Soper kernel the soft factor always cancels, such that this precise definition does not matter.

The spacelike Wilson lines of $\tilde B_q$ as given in \eq{qbeam} and those  of $\tilde S_q$ as given in \eq{qsoft} give rise to self energies that yield power law divergences proportional to $e^{\delta m \, L_{\rm tot}}$.
Here, $\delta m$ is a mass correction that absorbs divergences as $a\to0$,
and the total lengths of the Wilson line structures are given by
$L_{\rm tot}^B = L + |L-b^z| + b_T$ for $\tilde B_q$ and $L_{\rm tot}^S = 4L + 2b_T$ for $\tilde S_q$, respectively.
After combining the quasi beam function with the square root of the quasi soft function,
the Wilson line self-energies yield the overall power-law divergence
\begin{align} \label{eq:deltam}
 e^{\delta m \, \bigl(L_{\rm tot}^{\rm B} - \frac12 L_{\rm tot}^{\rm S}\bigr)} = e^{-\delta m \, b^z}
\,,\qquad (b^z < L)
\,,\end{align}
which has to be absorbed by $\tilde Z^q_{\rm uv}(b^z,\tilde\mu,a)$.
To cancel this divergence on the lattice, the nonperturbative UV renormalization has to be applied before the Fourier transform, as shown in \eq{qtmdpdf}, while in the lightlike case it is independent of $b^z$ and can be pulled out, see \eq{tmdpdf}.
This distinction is important, implying in the ratio of TMDPDFs the UV renormalization factor $Z_{\rm uv}^q$ cancel out,
whereas this is not possible for ratios of quasi-TMDPDFs.

\subsection{Determination of the Collins-Soper kernel in momentum space}
\label{sec:CS_momentum_space}

In this section, we briefly review the method proposed in \mycite{Ebert:2018gzl} for calculating the Collins-Soper kernel from lattice QCD.
As discussed in \mycite{Ebert:2018gzl}, and in more detail in \mycite{Ebert:2019okf}, in general there is a mismatch between the infrared structure of the quasi-beam function and beam function due to the fact that the latter requires a dedicated rapidity regulator, whereas the former has rapidity divergences regulated by the finite length $L$. This spoils the simplest boost picture from LaMET (even when supplemented by short distance corrections), for relating these proton matrix elements.  Nevertheless, when combined with the quasi-soft and soft functions, these divergences and the $L$ dependence cancel, enabling the possibility of a matching equation between the quasi-TMDPDF and TMDPDF. However, even after these cancellations there can still be a mismatch in the remaining infrared structure of the quasi-soft and soft functions, leaving a relation of the form
\begin{align} \label{eq:relation_qTMD_1}
 \tilde f_{\ns}(x, \bt, \mu, P^z) &
 = C_{\ns}\bigl(\mu, x P^z\bigr) g^S_{q}(b_T, \mu)
   \exp\biggl[ \frac12  \gamma_\zeta^q(\mu, b_T) \ln\frac{(2 x P^z)^2}{\zeta} \biggr]
   f_{\ns}(x, \bt, \mu, \zeta)
 \nn\\&\quad
 + \cO\biggl(\frac{b_T}{L}, \frac{1}{b_T P^z}, \frac{1}{P^z L}\biggr)
\,.\end{align}
Here, $C_{\ns}$ is a perturbative kernel for the nonsinglet $\ns{=}u{-}d$ channel,
$g_q^S$ is a nonperturbative contribution which reflects the mismatch in soft physics,
and $\gamma_\zeta^q$ is the standard Collins-Soper kernel, which allows one to relate the TMDPDF
at the scale $\zeta$ to the quasi-TMDPDF at proton momentum $P^z$.
We assume the hierarchy of scales that $1/P^z \ll b_T \ll L$, such that
corrections to this matching relation are suppressed for large $L$ and $P^z$,
as indicated, and will be suppressed in the following.
In \mycite{Ebert:2019okf} it was shown that the bent soft function yields $g^S_{q}(b_T,\mu) = 1 + {\cal O}(\alpha_s^2)$. To demonstrate that \eq{relation_qTMD_1} is a true matching equation requires an all-order proof that $g^S_{q}(b_T,\mu) = 1$, which has not been demonstrated. However the lack of this proof does not impact the determination of the anomalous dimension $\gamma_\zeta^q(\mu,b_T)$, to which we now turn.

Evaluating \eq{relation_qTMD_1} at two different proton
momenta $P^z_1 \ne P^z_2$ but the same $\zeta$,
and taking the ratio of the results yields
\begin{align} \label{eq:relation_qTMD_2}
 \frac{\tilde f_{\ns}(x, \bt, \mu, P^z_1)}{\tilde f_{\ns}(x, \bt, \mu, P^z_2)}
 = \frac{C_{\ns}\bigl(\mu, x P^z_1\bigr)}{C_{\ns}\bigl(\mu, x P^z_2\bigr)}
   \exp\biggl[ \gamma_\zeta^q(\mu, b_T) \ln\frac{P^z_1}{P^z_2} \biggr]
\,.\end{align}
Here $g_q^S$ and $f_{\ns}$ have dropped out.
In \mycite{Ebert:2018gzl}, this was solved for $\gamma_\zeta^q$ as
\begin{align} \label{eq:gamma_zeta_1}
 \gamma^q_\zeta(\mu, b_T) &= \frac{1}{\ln(P^z_1/P^z_2)}
  \ln \frac{C_\ns(\mu,x P_2^z)\, \tilde f_\ns(x, \bt, \mu, P^z_1)}
           {C_\ns(\mu,x P_1^z)\, \tilde f_\ns(x, \bt, \mu, P^z_2)}
\,.\end{align}

On the lattice one obtains the quasi-TMDPDF by Fourier transforming a position-space correlation function to momentum space, as given in \eq{qtmdpdf}.
Inserting \eq{qtmdpdf} into \eq{gamma_zeta_1}, one then obtains \cite{Ebert:2018gzl}
\begin{align} \label{eq:gamma_zeta_2}
 \gamma^q_\zeta(\mu, b_T) &= \frac{1}{\ln(P^z_1/P^z_2)}
 \\\nn&\times
 \ln \frac{C_\ns(\mu,x P_2^z)\, \int\! \df b^z\, e^{ib^z xP_1^z}\, \tilde Z_q^\prime(b^z, \mu, \tilde\mu)\,
 \tilde Z_{\rm uv}^{q}(b^z,\tilde \mu, a) \, \tilde B_\ns(b^z, \bt, a, P_1^z, L)}
 {C_\ns(\mu,x P_1^z)\, \int\! \df b^z\, e^{ib^z xP_2^z}\, \tilde Z_q^\prime(b^z, \mu, \tilde\mu)\,
 \tilde Z_{\rm uv}^{q}(b^z,\tilde \mu, a) \, \tilde B_\ns(b^z, \bt, a, P_2^z, L)}
\,.\end{align}
Note that here we have canceled the quasi soft factor $\tilde\Delta_S^q(b_T,a,L)$ in the ratio, as it is independent of $b^z$.
The advantage of doing so is that one needs to calculate one less nonperturbative function from lattice QCD.
The price to pay is that $\tilde B_\ns$ still contains Wilson line self energies $\propto L/a, b_T/a$ and divergences $\propto L/b_T$, which now only cancel in the ratio rather than in the numerator and denominator, respectively.
To achieve the separate cancellation, we can simultaneously insert a $b^z$-independent factor $\cR_B$ in both the numerator and denominator to separately cancel these leftover divergences,
\begin{align} \label{eq:gamma_zeta_2new}
 &\gamma^q_\zeta(\mu, b_T) = \frac{1}{\ln(P^z_1/P^z_2)}
 \\\nn&\times
 \ln \frac{C_\ns(\mu,x P_2^z)\, \int\! \df b^z\, e^{ib^z xP_1^z}\, \tilde Z_q^\prime(b^z, \mu, \tilde\mu)\,
 \tilde Z_{\rm uv}^q(b^z,\tilde \mu, a) \, \cR_B(b_T,\tilde\mu,a,L) \,  \tilde B_\ns(b^z, \bt, a, P_1^z, L)}
 {C_\ns(\mu,x P_1^z)\, \int\! \df b^z\, e^{ib^z xP_2^z}\, \tilde Z_q^\prime(b^z, \mu, \tilde\mu)\,
 \tilde Z_{\rm uv}^q(b^z,\tilde \mu, a) \, \cR_B(b_T,\tilde\mu,a,L) \, \tilde B_\ns(b^z, \bt, a, P_2^z, L)}
\,.\end{align}
This factor has to be constructed such that it exactly removes all divergences that would normally be canceled by $\tilde\Delta_S^q(b_T,a,L)$, \emph{i.e.}\ all power-law divergences not yet absorbed by $\tilde Z_{\rm uv}^q(b^z,\tilde\mu,a)$.
One trivial choice for this factor is thus to use the soft factor that was canceled before, $\cR_B =\tilde\Delta_S^q$, while another simple choice would be $\cR_B = [\tilde B_\ns(0, \bt, a, P_R^z, L)]^{-1}$, \emph{i.e.}\ the quasi beam function at vanishing separation $b^z=0$ and some reference momentum $P_R^z$.
In \sec{RIMOM}, we will construct a more refined expression by using the nonperturbative \RIpMOM\ renormalization factor in a similar fashion, and the final definition for the combination $\tilde Z_{\rm uv}^q \cR_B$ will be given in \eq{ZqDeltaZB}.

As stressed in \mycite{Ebert:2018gzl}, \eqs{gamma_zeta_2}{gamma_zeta_2new} are formally independent of $x$, $P_1^z$ and $P_2^z$, up to power corrections as indicated in \eq{relation_qTMD_1}, such that one can use any residual dependence of the lattice results
on these parameters to assess systematic uncertainties.

The one-loop result for the matching coefficient $C_\ns$ that enters \eq{gamma_zeta_2}, 
when $\tilde f_{\rm ns}$, $\tilde Z_q^\prime$ and $\gamma_\zeta^q$ are in the $\MS$ scheme,
has been calculated in \mycites{Ebert:2018gzl,Ebert:2019okf} and is given by
\begin{align} \label{eq:C_nlo}
 C_{\ns}\bigl(\mu, x P^z\bigr) &
 = 1 + \frac{\as C_F}{4\pi} \biggl[ - \ln^2\frac{(2 x P^z)^2}{\mu^2}
     + 2 \ln\frac{(2 x P^z)^2}{\mu^2} -4 + \frac{\pi^2}{6} \biggr] + \cO(\as^2)
\,.\end{align}
This short distance coefficient can also be extracted from the results of \mycite{Ji:2018hvs}.
Note that $C_{\ns}$ is an even function of its second argument, $C_{\ns}(\mu,-xP^z) = C_{\ns}(\mu,xP^z)$.

\subsection{Determination of the Collins-Soper kernel in position space}
\label{sec:CS_position_space}

A potential drawback of using \eq{gamma_zeta_2} and \eq{gamma_zeta_2new} is that one has to Fourier transform
the position-space correlator $\tilde B_{\ns}(b^z,\vec b_T,a,P^z,L)$.
This can be a limiting factor, as only a finite number of $b^z$ values are available from lattice, which thus does not fully determine the quasi beam function (often referred to as an inverse problem).
We hence propose in this section a related but modified formula which enables the matching to be performed directly in position space, thus providing an alternate method to carry out the calculation and test systematic uncertainties.

To derive this relation, we need the Fourier transforms of the quasi TMDPDF,
\begin{align} \label{eq:FT_fTMD}
 \tilde f_\ns(x, \bt, \mu, P^z) &= \int\frac{\df b^z}{2\pi} e^{\img (x P^z) b^z} \tilde f_\ns(b^z, \bt, \mu, P^z)
\,.\end{align}
Here, $\tilde f_\ns(x, \bt, \mu, P^z)$ is the quasi-TMDPDF defined previously in momentum space,
which is now expressed in terms of its Fourier-transform $\tilde f_\ns(b^z, \bt, \mu, P^z)$ in position space
on the right hand side of \eq{FT_fTMD}. The advantage of working with the latter is its direct connection
to the quasi-beam function $\tilde B(b^z,\bt,a,P^z,L)$, which is the object actually calculated on the lattice.
Note that for simplicity we distinguish the quasi-TMDPDF in position and momentum space only by their arguments,
as it is always clear from context which one we refer to.
It will be convenient to work with the Fourier transform of the \emph{inverse} of the kernel $C_\ns$, defined through
\begin{align} \label{eq:FT_C}
 \bar C_\ns(\mu, b^z P^z, P^z) &\equiv \int\!\df x \, e^{-\img x (b^z P^z)} \, \bigl[ C_\ns(\mu, x P^z)\bigr]^{-1}
\,,\nn\\
 \bigl[ C_\ns(\mu, x P^z)\bigr]^{-1} &= \int\frac{\df (b^z P^z)}{2\pi} \, e^{\img x (b^z P^z)} \, \bar C_\ns(\mu, b^z P^z, P^z)
\,.
\end{align}
Plugging \eqs{FT_fTMD}{FT_C} back into \eq{relation_qTMD_2}, we get
\begin{align} \label{eq:relation_qTMD_3} 
&
 P^z_1 \int \df b^z_1 \, \df{b^z_1}' \, e^{\img x P^z_1 (b^z_1 + {b^z_1}')}
 \bar C_\ns(\mu, {b^z_1}' P^z_1, P^z_1) \tilde f_\ns(b^z_1, \bt, \mu, P^z_1)
 \\\nn
=~&
 P^z_2 \int \df b^z_2 \, \df{b^z_2}' \, e^{\img x P^z_2 (b^z_2 + {b^z_2}')}
 \bar C_\ns(\mu, {b^z_2}' P^z_2, P^z_2) \tilde f_\ns(b^z_2, \bt, \mu, P^z_2)
 \ \exp\biggl[ \gamma_\zeta^q(\mu, b_T) \ln\frac{P^z_1}{P^z_2} \biggr]
\,.\end{align}
Next, we Fourier transform both sides from momentum fraction $x$ to a dimensionless position $y$, by multiplying by $e^{-\img x y}$ and integrating over $x$, obtaining
\begin{align} \label{eq:relation_qTMD_4} 
 &
 \int \df b^z_1 \, \bar C_\ns(\mu, y - b^z_1 P^z_1, P^z_1) \tilde f_\ns(b^z_1, \bt, \mu, P^z_1)
 \\\nn
 =~&
 \int \df b^z_2 \, \bar C_\ns(\mu, y - b^z_2 P^z_2, P^z_2) \tilde f_\ns(b^z_2, \bt, \mu, P^z_2)
 \ \exp\biggl[ \gamma_\zeta^q(\mu, b_T) \ln\frac{P^z_1}{P^z_2} \biggr]
\,.\end{align}
This can trivially be solved for $\gamma_\zeta^q$ as
\begin{align} \label{eq:gamma_zeta_3}
 \gamma_\zeta^q(\mu, b_T) = \frac{1}{\ln(P^z_1/P^z_2)}
 \ln\frac{\int\!\df b^z \, \bar C_\ns(\mu, y - b^z P^z_1, P^z_1) \tilde f_\ns(b^z, \bt, \mu, P^z_1)}
         {\int\!\df b^z \, \bar C_\ns(\mu, y - b^z P^z_2, P^z_2) \tilde f_\ns(b^z, \bt, \mu, P^z_2)}
\,.\end{align}
As expected, Fourier transforming the product in \eq{relation_qTMD_2} yields a convolution in position space.
In \eq{gamma_zeta_3}, $\tilde f_\ns$ is the renormalized nonsinglet quasi-TMDPDF in position-space as calculated on lattice.

Using the expression \eq{qtmdpdf} for $\tilde f_\ns$, we obtain the final expression
\begin{align} \label{eq:gamma_zeta_4}
 &\gamma_\zeta^q(\mu, b_T) = \frac{1}{\ln(P^z_1/P^z_2)}
 \\\nn
&\times
 \ln\frac{\int\!\df b^z \, \bar C_\ns(\mu, y - b^z P^z_1,P^z_1)\, \tilde Z'_q(b^z,\mu,\tilde \mu)\,
 	\tilde Z_{\rm uv}^q(b^z, \tilde \mu, a) \, \cR_B(b_T,\tilde\mu,a,L) \, \tilde B_{\rm ns}(b^z, \bt, a, P^z_1, L)}
    {\int\!\df b^z \, \bar C_\ns(\mu, y - b^z P^z_2, P^z_2)\, \tilde Z'_q(b^z,\mu,\tilde \mu)\,
    \tilde Z_{\rm uv}^q(b^z, \tilde \mu, a) \, \cR_B(b_T,\tilde\mu,a,L) \,  \tilde B_{\rm ns}(b^z, \bt, a, P^z_2, L)}
\,,\end{align}
where we suppress the explicit limits $L\to\infty$ and $a\to0$ for simplicity.
As in \eq{gamma_zeta_2new}, we have inserted a factor $\cR_B$ that cancels all divergences in $L/a, b_T/a$ and $L/b_T$ separately in the numerator and denominator, which otherwise would only cancel in the ratio.
Again, formally the dependence of the right hand side of \eq{gamma_zeta_4} on $y$, $P^z_1$ and $P^z_2$ cancels up to power corrections, such that one can use any residual dependence of the lattice results on these parameters to assess systematic uncertainties.
To use the improved formula in \eq{gamma_zeta_4} one only needs the position-space proton matrix element $\tilde B_{\rm ns}$ (directly obtained on the lattice),
its renormalization factor $\tilde Z_{\rm uv}^q$ (also obtained on the lattice) combined with
the factor $\cR_B$ (discussed in \sec{RIMOM}),
the $\MS$-conversion factor $\tilde Z'_q$
(which we calculate in \secs{RIMOM}{NLO} of this paper),
and the Fourier-transformed matching kernel $\bar C_\ns$ (which we obtain below).

In both \eqs{gamma_zeta_2new}{gamma_zeta_4} the dominant contributions to the integrals come from the small $b^z$ region. 
In the convolution in \eq{gamma_zeta_4} the kernel $\bar C_\ns(y - b^z P^z)$ given below is peaked around $b^z P^z\sim y$, while contributions from the region $|b^zP^z-y|\gg1$ are suppressed by this kernel.
In comparison, the Fourier transform in \eq{gamma_zeta_2new} is dominated by $xP^z b^z\sim 1$ and becomes less sensitive to $b^z\gg 1/(xP^z)$ due to suppression by the phase factor $\exp(ixP^zb^z)$. In practice, we can implement both methods on the lattice and compare their systematic uncertainties.

Note that we have chosen the definition \eq{FT_C} of the position-space kernel $\bar C_\ns$ to be determined by the transform of the inverse of $C_\ns$ in order to make \eq{gamma_zeta_4} particularly simple, with a numerator depending only on the momentum $P^z_1$ and the denominator only on $P^z_2$.
For comparison and completeness, we present in \app{CS_position_space} the corresponding derivation when using a position-space kernel $\bar C_\ns^\prime$ that is defined by the transform of $C_\ns$ itself, in which case numerator and denominator would both depend on $P^z_1$ and $P^z_2$.

The Fourier transform $\bar C_\ns$ can be further simplified by employing that in the physical limit $L, P^z \to\infty$, $\tilde f_q$ has limited support $x \in [0,1]$ for quarks and $x \in [-1,0]$ for antiquarks.
Hence, we can make different choices for the integration range in \eq{FT_C} which lead to formally equivalent results when the resulting coefficients $\bar C_{\rm ns}$ are employed in \eq{gamma_zeta_4}.
To exploit this freedom we consider the two natural choices, defining
\begin{align} \label{eq:FT_C_2}
 \bar C_\ns^{[0,1]}(\mu, b^z P^z, P^z) &= \int_0^1\!\df x \, e^{-\img x (b^z P^z)} \, \bigl[ C_\ns(\mu, x P^z)\bigr]^{-1}
\,,\\ \label{eq:FT_C_2alt}
 \bar C_\ns^{[-1,1]}(\mu, b^z P^z, P^z) &= \int_{-1}^1\!\df x \, e^{-\img x (b^z P^z)} \, \bigl[ C_\ns(\mu, x P^z)\bigr]^{-1}
\,,\end{align}
where the superscript in $\bar C_\ns^D$ denotes the integration domain $D$.

Physically, $\bar C_\ns^{[0,1]}$ as defined in \eq{FT_C_2} corresponds to the kernel for a quark quasi-TMDPDF,
while $\bar C_\ns^{[-1,0]} = \bigl(\bar C_\ns^{[0,1]}\bigr)^*$ would correspond to an antiquark.
\Eq{FT_C_2alt} thus corresponds to the sum of quark and antiquark contributions.
Since in \eq{gamma_zeta_4} we only employ the nonsinglet channel $\ns{=}u{-}d$, the antiquark contribution must cancel, and one can equivalently employ the unrestricted $x$ integration in \eq{FT_C}, or one of the restricted versions in \eq{FT_C_2} or \eq{FT_C_2alt}, for the kernel entering \eq{gamma_zeta_4}.
In practice, there will be a remnant contribution from antiquarks since one does not work in the physical limit $L, P^z \to\infty$. Hence one can employ the difference between \eqss{FT_C}{FT_C_2}{FT_C_2alt} as a further handle to probe systematic uncertainties from working at finite momentum.
Note that since $C_\ns$ depends logarithmically on $x P^z / \mu$, its Fourier transform according to \eq{FT_C} with unconstrained integration range will involve plus distributions which are complicated to implement numerically, so here we will refrain from advocating for using the unrestricted integration, and hence only present the simpler results obtained using \eqs{FT_C_2}{FT_C_2alt}.

\paragraph{Matching kernel in position space.}
Next we explicitly calculate the Fourier transform of $C_\ns$ to position space as defined in \eqs{FT_C_2}{FT_C_2alt}.
$C_\ns$ was calculated at next-to-leading order (NLO) in the $\MS$ scheme in \mycites{Ebert:2018gzl,Ebert:2019okf} and is given in \eq{C_nlo}.
Perturbatively inverting it gives the one-loop result
\begin{align} \label{eq:C_nlo_inv}
 \bigl[C_{\ns}\bigl(\mu, x P^z\bigr)\bigr]^{-1} &
 = 1 + \frac{\as C_F}{4\pi} \biggl[ \ln^2\frac{(2 x P^z)^2}{\mu^2}
     - 2 \ln\frac{(2 x P^z)^2}{\mu^2} + 4 - \frac{\pi^2}{6} \biggr] + \cO(\as^2)
\,.\end{align}
Fourier transform according to \eqs{FT_C_2}{FT_C_2alt}, we obtain
\begin{align} \label{eq:FT_Cnv_nlo}
 \bar C_\ns^D(\mu, y, P^z) &
 = f_0^D(y) + \frac{\as C_F}{4\pi} \Bigl[f_2^D(y) + (2 L_z - 2) f_1^D(y) + \Bigl( L_z^2  - 2 L_z + 4 - \frac{\pi^2}{6} \Bigr)  f_0^D(y) \Bigl]
 \nn\\&\quad
 + \cO(\as^2)
\,,\end{align}
where we abbreviated $L_z = \ln[(2 P^z / \mu)^2]$, and as before the superscript $D$ on the three required functions $f_i^D(y)$ denotes the integration range of the Fourier transform.

For the case of integrating over $D = [0,1]$ the auxiliary integrals are
\begin{align}
 f_n^{[0,1]}(y) &= \int_0^1 \df x \, e^{-\img x y} \ln^n(x^2)
 \nn\\&
 = (-2)^n n! ~ {}_{n+1}F_{n+1}\bigl(\{1,\cdots,1\}, \{2, \cdots, 2\}, -\img y\bigr)
\,.\end{align}
Here, $_n F _n$ is a hypergeometric function.
The results for $n=0$ and $n=1$ can be expressed using standard functions,
\begin{align}
 f_0^{[0,1]}(y) &= \frac{1 - e^{-\img y}}{\img y}
\,,\qquad
 f_1^{[0,1]}(y) = \frac{2 \img}{y} \bigl[ \Gamma(0, \img y) + \ln(\img y) + \gamma_E \bigr]
\,.\end{align}
On the other hand, for the case of  integrating over $D=[-1,1]$ the auxiliary integrals are
\begin{align}
f_n^{[-1,1]}(y) &= \int_{-1}^1 \df x \, e^{-\img x y} \ln^n(x^2)= 2 \, {\rm Re} \bigl[ f_n^{[0,1]}(y) \bigr]
\,.\end{align}
For $n=0$ and $n=1$, we obtain the simple results
\begin{align}
 f_0^{[-1,1]}(y) &= 2 \frac{\sin y}{y}
\,,\qquad
 f_1^{[-1,1]}(y) = -4 \frac{\mathrm{Si}(y)}{y}
\,,\end{align}
where ${\rm Si}(y)=\int_0^y dt\, \sin(t)/t$ is the sine integral function.

Note that the above functions  behave as $f_n^D(y) \sim 1/y$ for large $y$ and oscillate,
and hence the dominant contribution to the convolutions in \eqs{gamma_zeta_3}{gamma_zeta_4}
is given by $y - b^z P^z \sim 1$. This naturally limits the impact of the quasi beam function
from large $b^z$.

\section{\texorpdfstring{\RIpMOM}{RI'/MOM}~renormalization and matching}
\label{sec:RIMOM}

The determination of $\gamma_\zeta^q$ using either \eq{gamma_zeta_2new} or \eqref{eq:gamma_zeta_4} requires calculating the quasi-beam function $\tilde B_q$ from lattice, a renormalization of UV divergences with $\tilde Z_{\rm uv}^q$, a definition of $\cR_B$ to cancel remaining power-law divergences (one choice would be $\tilde\Delta_S^q$), and finally $\tilde Z_q^\prime$ to convert to the $\MS$ scheme.
Here, we specify in detail a preferred choice for how to construct these nonperturbative renormalization factors in the \RIpMOM\ scheme, and how the conversion factor $\tilde Z_q^\prime$ can be calculated perturbatively.
$\tilde Z_q^\prime$  is then calculated at one loop in \sec{NLO}.

Note that for $b^z{=}0$, the corresponding $\MS$ conversion kernel for the quasi beam function $\tilde B_q$ has been calculated in \mycite{Constantinou:2019vyb}, which is sufficient for the lattice studies of the $x$-moments of TMDPDFs carried out in \mycites{Musch:2010ka,Musch:2011er,Engelhardt:2015xja,Yoon:2016dyh,Yoon:2017qzo}, but does not suffice for the determination of the Collins-Soper kernel which requires the calculation for nonvanishing $b^z$.

To renormalize the staple-shaped Wilson line operators entering the quasi beam function on the lattice, we need to prove their renormalizability first. Under lattice regularization, Lorentz symmetry group is broken into the hypercubic group, so it is more involved to employ standard field theory techniques to make this proof. Nevertheless, it has been proven that lattice gauge theory is renormalizable to all orders of perturbation theory within the functional formalism~\cite{Reisz:1988kk}, which also stands for the case with a background gauge field~\cite{Luscher:1995vs}. Therefore, the counterterms to the lattice action are only those allowed by gauge and hypercubic symmetries. This proof is also applicable to composite operators, which is the basis for their nonperturbative renormalization on the lattice.
Therefore, we expect the renormalization of staple-shaped Wilson line operators to be similar in both continuum and lattice perturbation theories, except that in the latter there can be novel counterterms allowed by lattice symmetries.
To begin with, we argue that in continuum theory the staple-shaped quark Wilson line operator can be renormalized multiplicatively in position space as
\begin{align} \label{eq:ren}
 \cO_0^\Gamma(b^\mu,\Lambda,L) &
 \equiv\bar \psi_0(b^\mu) W_{\hat z}\frac{\Gamma}{2} W_T W^\dagger_{\hat z} \psi_0(0)
 \nn\\&
 = Z_{q,\rm wf} \, e^{\delta m (L + |L - b^z| + b_T)} \Bigl(\bar \psi(b^\mu) W_{\hat z}\frac{\Gamma}{2} W_T W^\dagger_{\hat z} \psi(0)\Bigr)_R
 \nn\\&
 \equiv \tilde Z_B \Bigl(\bar \psi(b^\mu) W_{\hat z}\frac{\Gamma}{2} W_T W^\dagger_{\hat z} \psi(0)\Bigr)_R
\,,\end{align}
where $\Lambda$ is a generic UV regulator that respects Lorentz invariance and gauge invariance.
Here, $W_{\hat{z}}$ and $W_T$ are Wilson lines as defined in \eqs{Wilson_lines}{coll_Wilson_L},
and for brevity we suppressed their explicit arguments, which are given in \eq{qbeam}.
In the first line in \eq{ren}, we work with bare quark fields $\psi_0$ and Wilson lines
built of bare gluon fields and bare couplings, while in the second line we work with renormalized fields and couplings, indicated by the subscript $R$.
$Z_{q,\rm wf}$ includes all the logarithmic UV divergences originating from the wave function renormalization
and the quark-Wilson-line vertices. The exponential absorbs all linear power divergences
from the self-energies of the spacelike Wilson lines, where $L + |L - b^z| + b_T$
is the total length of the staple. 

\Eq{ren} resembles the multiplicative renormalization for the straight Wilson line operators,
for which the renormalization in the \RIpMOM~scheme has been studied in~\mycite{Constantinou:2017sej,Stewart:2017tvs,Chen:2017mzz,Alexandrou:2017huk,Liu:2018uuj}.
For the staple-shaped operators discussed here, the multiplicative renormalization has also been used in \mycite{Constantinou:2019vyb}, which carried out the \RIpMOM~renormalization for the special case $b^z=0$, \emph{i.e.} vanishing longitudinal separation of the staple. The proof of \eq{ren} is analogous to that for the straight Wilson line operators, where one employs the auxiliary field formalism~\cite{Dorn:1986dt,Ji:2017oey,Green:2017xeu,Zhang:2018diq}. This auxiliary field formalism is also commonly used to derive Wilson line operators in the Soft Collinear Effective Theory, see the original work in Refs.~\cite{Bauer:2001yt,Bauer:2002nz}.
For the TMD, by using three independent auxiliary ``heavy quark'' fields for each edge of the
staple-shaped Wilson line, the nonlocal quark Wilson line operator can be reduced to the
product of four composite operators in the effective theory that includes these auxiliary fields.
BRST invariance implies that this effective theory is renormalizable through multiplicative counterterms
to all orders in perturbation theory~\cite{Ji:2017oey}. It follows that in continuum QCD,
the staple-shaped quark Wilson line operator can indeed be renormalized multiplicatively as shown in \eq{ren}.

On the discretized lattice where $\Lambda\to 1/a$, we can also use the auxiliary field theory to replicate the proof, and hypercubic symmetry does not allow the operator to have UV-divergent mixings with other operators with the same or lower dimensions. Though mixing with higher-dimensional operators is allowed, it is power suppressed and not relevant when one takes the continuum limit $a\to0$. However, as pointed out in~\mycites{Constantinou:2019vyb,Yoon:2017qzo}, due to the breaking of chiral symmetries, there will be mixing with other operators on a discretized lattice, and thus the renormalization on lattice requires an independent study~\cite{Yoon:2017qzo,Shanahan:2019zcq}. After lattice renormalization of the quasi beam function, its continuum limit can be taken and the result is independent of the UV regulator, which allows us to calculate the scheme conversion factors in continuum perturbation theory with dimensional regularization. 
In this work, we will discuss how to renormalize the quasi beam function in the \RIpMOM~scheme on the lattice, and then focus on its conversion to the $\MS$ scheme in continuum perturbation theory. Since one-loop lattice perturbation theory~\cite{Constantinou:2019vyb} suggests that the mixing due to chiral-symmetry breaking is zero for certain choices of $\Gamma$, while for the other choices the mixings can be reduced by tuning the parameters of lattice action, we do not consider this effect in our calculation by assuming that either a proper choice of $\Gamma$ is made or the mixings are sufficiently small with fine-tuned lattice parameters.

To implement the \RIpMOM~scheme for the quasi beam function, one first computes the amputated Green's function of the operator given in \eq{ren},
\begin{align} \label{eq:Lambda}
 \Lambda^{\Gamma}_0(b, a, p, L) \equiv
 \left[S^{-1}_0(p,a)\right]^\dagger \sum_{x,y}e^{ip\cdot (x-y)}
 \big\langle 0 \big|T\bigl[\psi_0(x,a) {\cal O}_0^\Gamma(b^\mu,a,L) \bar{\psi}_0(y,a)\bigr]\big|0\big\rangle S^{-1}_0(p,a)
\,,\end{align}
which is also referred to as the vertex function. Here and below $b$ indicates dependence on $b^z$ and $b_T$.
In \eq{Lambda}, $S_0(p,a)$ is the bare quark propagator that can be calculated nonperturbatively on the lattice.
$\Lambda^{\Gamma}_0(b, a, p, L)$ is a linear combination of Dirac matrices that are allowed by the symmetries of space-time and the operator itself.
For off-shell quarks, there will also be finite mixing with equation-of-motion operators that vanish in the on-shell limit.
Furthermore, the off-shell matrix element is not gauge invariant, and thus one has to fix a particular gauge choice as part of the renormalization scheme, which in lattice QCD is typically chosen as the Landau gauge.

In practice, one needs to choose a projection operator $\cP$ to define the off-shell matrix element of the quasi-beam function from the amputated Green's function,
\begin{align} \label{eq:q}
 \tilde{q}^{\Gamma\cal, P}_0(b, a, p, L) = {\rm tr}\bigl[\Lambda^\Gamma_0(b, a, p, L) \cP\bigr]\,.
\end{align}
The choice of $\cP$ is not unique~\cite{Stewart:2017tvs,Liu:2018uuj}, but it must have overlap with $\Gamma$ to project out all the UV divergences as $a\to0$. In \mycite{Constantinou:2017sej,Constantinou:2019vyb}, the choice is ${\cP}=\Gamma$, while in \mycites{Stewart:2017tvs,Liu:2018uuj} both the choice ${\cP}=\slashed p$, and a choice for $\cP$ that effectively projects out the coefficient of $\Gamma$ in the covariant decomposition of $\Lambda^\Gamma_0$, were considered. In principle, the dependence on the projection $\cP$ will be canceled by the scheme conversion factor, since the $\MS$ renormalization constant is unique. But in practice, since the conversion factor is computed at fixed orders in perturbation theory, there can still be remnant $\cP$ dependence at higher orders, which is part of the systematic uncertainty.

In the \RIpMOM~scheme, the renormalization constant $\tilde Z_B^{\Gamma,\cP}$ of the
bare operator $\cO_0^\Gamma$ defined in \eq{ren} is determined by requiring that at
a specific momentum $p_R^\mu$, the projection defined in \eq{q} reduces to its value
at tree level in perturbation theory.
Here, we actually need to define the \RIpMOM\ condition for the quasi-TMDPDF, which also includes the soft factor. It reads
\begin{align}\label{eq:rimomTMDPDF}
 &\tilde Z^{\Gamma,\cP}_q(b^z, b_T^R, p_R, a) \, Z_{\rm wf}(p_R,a) \,
 \lim_{L\to\infty}  \tilde q^{\Gamma,\cP}_0(b, a, p, L) \, \tilde\Delta_S^q(b_T,a,L) \bigg|_{\stackrel{p^\mu = p_R^\mu}{b_T=b_T^R}}
\nn\\&
 = \tilde{q}^{\Gamma,\cP}_{\rm tree}(b^z,b_T^R,p_R)
\,.\end{align}
Here, $q^{\Gamma,\cP}_{\rm tree}$ is the value of \eq{q} at tree-level in perturbation theory,
which is nonzero only for particular choices of $\Gamma$ and $\cP$,
and each such pair $(\Gamma,\cP)$ define a particular $\tilde{Z}_q^{\Gamma,\cP}$.
The tree level soft factor is given by $\tilde \Delta_{S\,\rm tree}^q=1$ and hence not explicitly given in \eq{rimomTMDPDF}.
Here the choice for the scales $p_R$ and $b_T^R$ are part of the definition of the \RIpMOM\ scheme.

In \eq{rimomTMDPDF}, the wave function renormalization factor $Z_{\rm wf}$ arises to compensate
for the renormalization of the bare quark fields in \eq{Lambda}.
It is determined independently with the following condition on the quark propagator,
\begin{align}\label{eq:wf}
 & \bigl[Z_{\rm wf}(p,a)\bigr]^{-1} S_0(p,a)\big|_{p^2=p_R^2} = S_{\rm tree}(p_R,a)
\nn\\\Rightarrow~&
 \bigl[Z_{\rm wf}(p_R,a)\bigr]^{-1} = \frac14 {\rm tr}\bigl[S^{-1}_0(p,a)S_{\rm tree}(p,a)\bigr]_{p^2=p_R^2}
\,,\end{align}
where the $1/4$ arises from the trace over Dirac indices.%
\footnote{In the literature, one often includes a trace over color indices, in which case the prefactor $1/4$ in \eq{wf} is replaced by $1/12$. For simplicity, we keep this normalized color trace implicit.}
The use of \eq{wf} in \eq{rimomTMDPDF} defines the RI$^\prime$ scheme, while in the closely related RI scheme $Z_{\rm wf}$ is defined by imposing vector current conservation using Ward identities~\cite{Martinelli:1994ty}.

From \eqs{qtmdpdf}{rimomTMDPDF}, it follows that
\begin{align} \label{eq:ZqRIMOM}
 \tilde Z_{\rm uv}^q(b^z,\tilde\mu,a) &\equiv \tilde Z^{\Gamma,\cP}_q(b^z, b_T^R, p_R, a)
 = \lim_{L\to\infty} \frac{\tilde Z^{\Gamma,\cP}_B(b^z, b_T^R, p_R, a, L)}{\tilde\Delta_S^q(b_T^R,a,L)}
\,,\end{align}
where we have split $\tilde Z_q^{\Gamma,\cP}$ into a piece $\tilde Z^{\Gamma,\cP}_B$ arising from the \RIpMOM\ prescription applied to the quasi beam function only, and the quasi soft factor $\tilde\Delta_S^q(b_T^R,a,L)$. The $\tilde Z^{\Gamma,\cP}_B$ is given by the \RIpMOM\ condition
\begin{align} \label{eq:ZBRIMOM}
 \tilde Z^{\Gamma,\cP}_B(b^z, b_T^R, p_R, a, L) &\equiv \bigl[Z_{\rm wf}(p_R,a)\bigr]^{-1} \frac{\tilde{q}^{\Gamma,\cP}_{\rm tree}(b^z,b_T^R,p_R)}{\tilde q^{\Gamma,\cP}_0(b^z,b_T^R, a, p, L)}
\,.\end{align}
From \eq{ZqRIMOM} we can also identify the \RIpMOM~renormalization scale $\tilde\mu=(b_T^R,p_R)$, which contains a choice for both the momentum $p_R^\mu$ and the transverse separation $b_T^R$ to be used when defining the renormalization constant.
This is unusual for a \RIpMOM~scheme, where one would normally only specify $p_R^\mu$, but not $b_T^R$.
The reason to also specify $b_T = b_T^R$ here is that $b_T$ itself can become a nonperturbative scale, and hence must not enter the perturbative scheme conversion factor $\tilde Z'_q(b^z,\mu,\tilde\mu)$.
In contrast, $b_T^R$ can always be chosen to be a perturbative scale, similar to $p_R^\mu$, thus ensuring that this scheme conversion factor to $\MS$ remains perturbatively calculable.

Using $\tilde{Z}^{\Gamma,\cP}_q$, the bare quasi-TMDPDF can be renormalized in position space as
\begin{align} \label{eq:f_RIMOM}
 \tilde f_q^{\rm OM}(b^z, b_T, P^z, b_T^R, p_R, L)
 = \lim_{a\to 0} \, \tilde{Z}^{\Gamma,\cP}_q(b^z,b_T^R,p_R,a) \, \tilde B_q(b^z, \bt, a, P^z, L) \tilde \Delta_S^q(b_T,a,L)
\,.\end{align}
The \RIpMOM-renormalized quasi-TMDPDF obtained from \eq{f_RIMOM} is independent of the UV regulator,
and therefore can be matched perturbatively onto the $\MS$~renormalized quasi-TMDPDF, which is given by
\begin{align} \label{eq:f_MS}
 \tilde f_q^\MS(b^z, b_T, P^z, \mu, L)
 = \lim_{\eps\to 0} \, \tilde Z_q^\MS(\mu,\eps) \, \tilde B_q(b^z, \bt, \eps, P^z, L) \tilde \Delta_S^q(b_T,\eps,L)
\,.\end{align}
$\tilde Z_q^\MS$ is calculated in the continuum theory with dimensional regularization using $d=4-2\eps$ dimensions and subtracts poles in $\eps$ only.
Comparing \eqs{f_RIMOM}{f_MS}, we can read off the relation between the \RIpMOM~and $\MS$ schemes,
\begin{align} \label{eq:Zf_RIMOM_MS}
 \tilde f_q^\MS(b^z, b_T, P^z, \mu, L) &= Z_q^{\prime \Gamma,\cP}(b^z,\mu,b_T^R,p_R) \, \tilde f_q^{\rm OM}(b^z, b_T, P^z, b_T^R, p_R, L)
\,,\nn\\
 \tilde Z_q^\prime(b^z, \mu, \tilde\mu) &\equiv \tilde Z_q^{\prime \Gamma,\cP}(b^z,\mu,b_T^R,p_R)
 = \lim_{\eps\to0} \frac{\tilde Z_q^\MS(\mu,\eps)}{\tilde Z^{\Gamma,\cP}_q(b^z, b_T^R,p_R, \eps)}
\,.\end{align}

Note that in \eq{f_RIMOM}, all divergences in $L/a$, $b_T/a$ and $L/b_T$ cancel among $\tilde B_q$ and $\tilde\Delta_S^q$,
rather than being absorbed by $\tilde Z_q^{\Gamma,\cP}$. However, for the determination of the Collins-Soper kernel $\gamma_\zeta^q$ as suggested in \eqs{gamma_zeta_2new}{gamma_zeta_4}, it was advantageous to cancel out the soft factor in the ratio so it does not have to be calculated on the lattice. In this case these power law divergences also only cancel in the
ratio. Such power law divergences can be problematic since it is generally unwise to attempt to extract a signal only after canceling out large contributions.
This can be remedied by constructing the factor $\cR_B$ to precisely cancel these divergences.
A convenient choice in the \RIpMOM\ scheme is
\begin{align} \label{eq:Delta_ZB}
 \cR_B(b_T,b_T^R,p_R,a,L) = \tilde\Delta_S^q(b_T^R,a,L) \frac{\tilde Z^{\Gamma,\cP}_B(0, b_T, p_R, a, L)}{\tilde Z^{\Gamma,\cP}_B(0, b_T^R, p_R, a, L)}
\,.\end{align}
Hence the combination that enters \eqs{gamma_zeta_2new}{gamma_zeta_4} is given by
\begin{align} \label{eq:ZqDeltaZB}
 \tilde Z_{\rm uv}^q(b^z,\tilde \mu, a) \, \cR_B(b_T,\tilde\mu,a,L)
 &\equiv \tilde Z^{\Gamma,\cP}_q(b^z, b_T^R, p_R, a) \cR_B(b_T,b_T^R,p_R,a,L)
\nn\\&
 = \frac{\tilde Z^{\Gamma,\cP}_B(b^z, b_T^R, p_R, a, L)}{\tilde Z^{\Gamma,\cP}_B(0, b_T^R, p_R, a, L)}
   \tilde Z^{\Gamma,\cP}_B(0, b_T, p_R, a, L)
\,.\end{align}
In this expression, the quasi soft factor $\tilde\Delta_S^q$ has canceled between $\tilde Z^{\Gamma,\cP}_q$ and $\cR_B$,
as desired so that it does not need to be calculated on the lattice. The factor $\tilde Z^{\Gamma,\cP}_B(0, b_T, p_R, a, L)$
in \eq{ZqDeltaZB} cancels all divergences in $L/a$, $b_T/a$ and $L/b_T$ that are present
in $\tilde B_q(b^z,b_T,a,P^z,L)$, while the remaining fraction in \eq{ZqDeltaZB} removes
all leftover UV divergences, in particular those proportional to $b^z/a$. Thus 
this result fulfills all requirements of \eqs{gamma_zeta_2new}{gamma_zeta_4}. 

As a result, both the numerator and denominator in \eqs{gamma_zeta_2new}{gamma_zeta_4} have well defined continuum ($a\to0$) and $L\to\infty$ limits before one calculates their ratio. Nevertheless, if one loosens the requirement for convergence by taking the continuum ($a\to0$) and $L\to\infty$ limits after calculating the ratio, then it is important to note that the combination $\tilde Z^{\Gamma,\cP}_B(0, b_T, p_R, a)/ \tilde Z^{\Gamma,\cP}_B(0, b_T^R, p_R, a)$ will cancel out in the ratios in \eqs{gamma_zeta_2new}{gamma_zeta_4}. If this is done, then the UV and $L/b_T$ divergences it accounts for will still cancel out between the numerator and denominator in those limits.

\section{One-loop results}
\label{sec:NLO}

In this section, we provide details on the one-loop calculation of the quasi-beam function in an off-shell state,
as well as the resulting \RIpMOM~renormalization factor $\tilde Z_{\rm uv}^q$ and the \RIpMOM~to $\MS$ conversion factor $\tilde Z'_q$ for the quasi-TMDPDF.
Throughout this section, we work in Euclidean space with $d=4-2\eps$ dimensions, as our aim is to calculate $\tilde Z_q^\prime$ with a lattice friendly definition of the Lorentz indices.

\subsection{Quasi-beam function with an off-shell regulator}
\label{sec:qbeam_nlo}

For completeness, we first give the Feynman rules in Euclidean space,
following the notation of \mycite{Capitani:2002mp}.
For a covariant gauge with gauge parameter $\xi$, the gluon propagator reads
\begin{align} \label{eq:prop}
 \raisebox{-1ex}{\includegraphics[width=3cm]{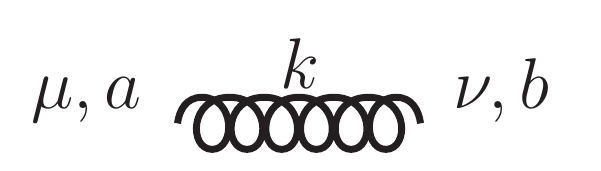}}
 = \frac{\delta^{ab}}{k_E^2} \biggl[ \delta^{\mu\nu} - (1-\xi) \frac{k_E^\mu k_E^\nu}{k_E^2} \biggr]
\,,\end{align}
The Feynman rules for quark propagators and the QCD vertex are given by
\begin{align} \label{eq:prop2}
 \raisebox{-1ex}{\includegraphics[width=3cm]{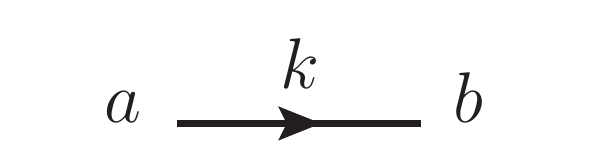}}
 = -\img \delta^{ab} \frac{\slashed{k}_{\!E}}{k_{\!E}^2}
\quad,\qquad
 \raisebox{-4ex}{\includegraphics[width=3cm]{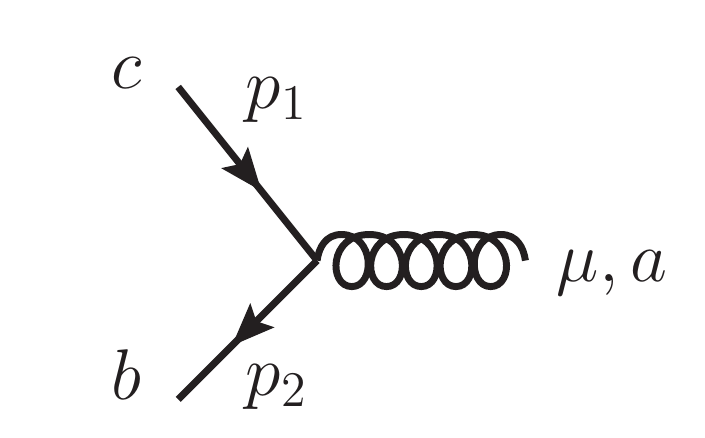}}
 = -\img g \, t^a \gamma_E^\mu
\,,\end{align}
where the sign of the strong coupling constant $g$ is such that the covariant derivative
is given by $D_\mu = \partial_\mu + \img g A_\mu$.
In \eqs{prop}{prop2}, $k_E$ is a Euclidean momentum such that $k_E^2 = \sum_i (k^i)^2$.
The $\gamma_E^\mu$ in \eq{prop2} are Dirac matrices in Euclidean space,
which are related to the Dirac matrices $\gamma_M^\mu$ in Minkowski space by
$\gamma_E^0 = \gamma_M^0$ and $\gamma_E^i = \img \gamma_M^i$, and obey $\gamma_E^\mu = \gamma_{E\,\mu}$.
The Euclidean $\gamma_E^5$ is defined as $\gamma_E^5 \equiv \gamma_E^0 \gamma_E^1 \gamma_E^2 \gamma_E^3$.
In the remainder of this section, we suppress the explicit subscript ``E'',
as we will always work in Euclidean space.

We consider the matrix element of the quasi TMD beam function operator in \eq{ren}
with an off-shell quark state $|q_s(p)\rangle$ of Euclidean momentum $p^2 > 0$,
amputated to remove the spinors,
\begin{align} \label{eq:qbeam_2}
 \tilde \Lambda^\lambda_\xi(b, \eps, p, L) = \Bigl< q_s(p) \Big| &\bar \psi(b^\mu) W_{\hat z}(b^\mu;L-b^z) \frac{\gamma^\lambda}{2}
 W_T(L \hat z; \bt, \vec{0}_T) W^\dagger_{\hat z}(0;L) \psi(0) \Big| q_s(p) \Bigr>_{\rm amp}
\,.\end{align}
The full set of possible projection operators is 
\begin{align} \label{eq:cP}
  \cP = \frac12 \Big\{ 1,{\gamma^5},  \gamma^\rho, \gamma^\rho\gamma^5,
    \sigma^{\rho\sigma} \Big\} \,.
\end{align}
Note that only $\cP_1 = \gamma^\rho$ with $\rho=\lambda$ yields a nonvanishing tree-level result and thus a valid renormalization. However, it is also interesting to study the mixing between different Dirac structures, and hence we also consider all other projectors \eq{cP} yielding nonvanishing one-loop results.
For our continuum analysis, this includes only the axial and vector projection operators,
so we consider two different projections of $\tilde \Lambda^\lambda_\xi$ to define the off-shell matrix element of the quasi-beam function:
\begin{alignat}{3} \label{eq:trace}
 \cP_1&=\frac12 \gamma^\rho\,:\qquad &&
 \tilde q^{\rho\lambda}_\xi(b,p,\eps,L) = \frac12 \tr\bigl[ \gamma^\rho \tilde \Lambda^\lambda_\xi(b,p,\eps,L) \bigr]
\,,\nn\\
 \cP_2&=\frac12 \gamma^\rho \gamma^5 \,:\qquad &&
 \tilde q^{\rho\lambda}_{a,\xi}(b,p,\eps,L) = \frac12 \tr\bigl[ \gamma^\rho \gamma^5 \tilde \Lambda^\lambda_\xi(b,p,\eps,L) \bigr]
\,.\end{alignat}
Here, the subscript ``$a$'' refers to axial.
We split all results into a piece corresponding to Feynman gauge $(\xi = 1$) plus a correction for $\xi\ne1$,
\begin{align}
 \tilde q^{\rho\lambda}_\xi(b,p,\eps,L) = \tilde q^{\rho\lambda}(b,p,\eps,L) + (1-\xi) \, \Delta \tilde q^{\rho\lambda}(b,p,\eps,L)
\,,\end{align}
and similarly for the axial projection.
Here, $\xi=0$ corresponds to the Landau gauge most relevant for lattice.

\begin{figure}[t]
 \centering
 \begin{subfigure}{0.3\textwidth}
  \includegraphics[width=\textwidth]{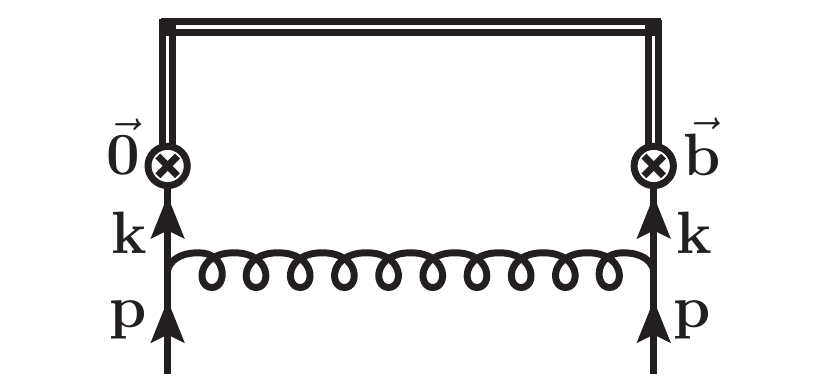}
  \caption{Vertex diagram}
  \label{fig:TMD a}
 \end{subfigure}
 \quad
 \begin{subfigure}{0.3\textwidth}
  \includegraphics[width=\textwidth]{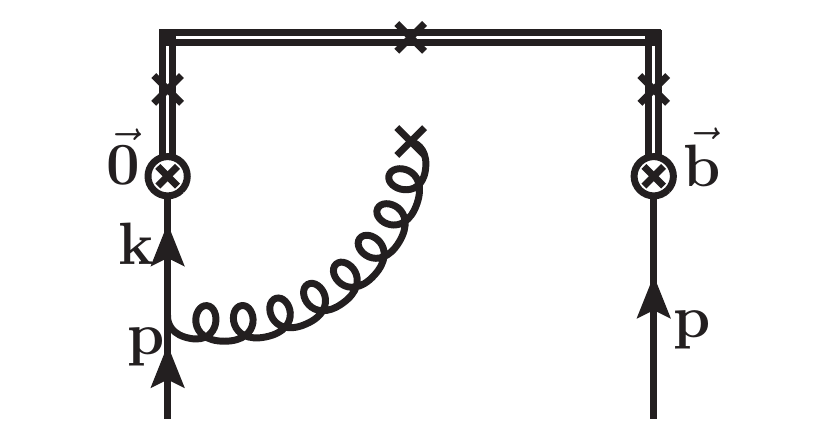}
  \caption{Sail topology}
  \label{fig:TMD b}
 \end{subfigure}
 \quad
 \begin{subfigure}{0.3\textwidth}
  \includegraphics[width=\textwidth]{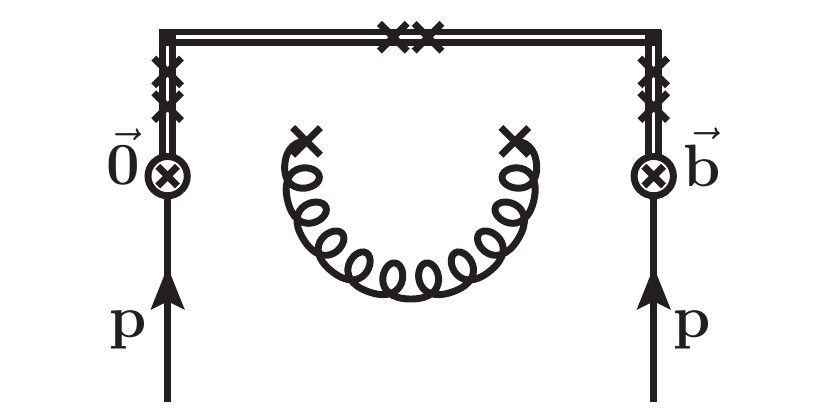}
  \caption{Wilson line self energy}
  \label{fig:TMD c}
 \end{subfigure}
 \caption{One-loop diagrams contributing to the quasi TMD beam function (with the mirrored sail diagram not shown).
          The double line represents straight Wilson line segments and the $\otimes$ are the quark fields.
          The crosses in (b) and (c) denote the set of locations for attaching the gluon that have to be added.}
 \label{fig:qbeamfunc_nlo}
\end{figure}

The tree level results are given by
\begin{align} \label{eq:q_lo}
 &\tilde q^{(0)\,\rho\lambda}(b,p,\eps,L)
 = \frac12 \tr\Bigl[ \gamma^\rho  \frac{\gamma^\lambda}{2} e^{\img \pb} \Bigr]
 \quad
 = \delta^{\rho\lambda} e^{\img \pb}
 \,,\qquad
 \Delta \tilde q^{(0)\,\rho\lambda}(b,p,\eps,L) = 0
\,,\nn\\
 &\tilde q^{(0)\,\rho\lambda}_a(b,p,\eps,L)
 = \frac12 \tr\Bigl[ \gamma^\rho \gamma^5 \frac{\gamma^\lambda}{2} e^{\img \pb} \Bigr]
 = 0
 \,,\qquad\qquad~\,
 \Delta \tilde q^{(0)\,\rho\lambda}_a(b,p,\eps,L) = 0
\,.\end{align}
At one loop, there are four topologies contributing to $\tilde q^{\rho\lambda}_\xi$ and $\tilde q^{\rho\lambda}_{a,\xi}$, as shown in \fig{qbeamfunc_nlo}. To evaluate these, we introduce two master integrals,
\begin{align} \label{eq:master_integrals}
 \MI_{i,j}^{\mu \nu \cdots}(b,p) &
 = - (4\pi\mu_0^{\eps})^2 \int\!\frac{\df^d k}{(2\pi)^d} \frac{k^\mu k^\nu \cdots}{(k^2)^i [(p-k)^2]^j} \, e^{\img k \cdot b}
\,,\nn\\
 \MI_{i,j}^{\mu \nu \cdots}(p) &
 = - (4\pi\mu_0^{\eps})^2 \int\!\frac{\df^d k}{(2\pi)^d} \frac{k^\mu k^\nu \cdots}{(k^2)^i [(p-k)^2]^j}
\,.\end{align}
Explicit results for these are collected in \app{master_integrals}.
In \eq{master_integrals}, $\mu_0$ is the MS renormalization scale, which is related to the $\MS$ scale by
\begin{align}
 \mu^2 \equiv \mu^2_\MS = \frac{4\pi}{e^{\gamma_E}} \mu_0^2
\,.\end{align}
In the following, we derive results for the different diagram topologies in terms of these master integrals,
keeping the Dirac indices $\rho$ and $\lambda$ as well as the gauge parameter $\xi$ generic.
Note that only the sail diagram is nonvanishing for the axial projection $\tilde q^{\rho\lambda}_{a,\xi}$,
and hence for the other diagrams we only discuss $\tilde q^{\rho\lambda}_\xi$.

\subsubsection{Vertex diagram}
\label{sec:vertex_diagram}

The vertex diagram shown in \fig{TMD a} is given by
\begin{align} \label{eq:vertex_1}
 \tilde q^{(1)\,\rho\lambda}_{{\rm v}\,,\,\xi}(b,p,\mu,\eps) &
 = \frac12 g^2 C_F \mu_0^{2\eps} \int\frac{\df^d k}{(2\pi)^d} e^{\img k \cdot b}
   \frac{\frac12 \tr[\gamma^\rho  \gamma^\mu \slashed{k} \gamma^\lambda \slashed{k} \gamma^\nu]}{k^4 (p-k)^2} \biggl[ \delta_{\mu\nu} - (1-\xi) \frac{(p-k)_\mu (p-k)_\nu}{(p-k)^2} \biggr]
\nn\\&
 \equiv \frac{\as C_F}{4\pi} \Bigl[ \tilde q^{(1)\,\rho\lambda}_{\rm v}(b,p,\mu,\eps) + (1-\xi) \Delta \tilde q^{(1)\,\rho\lambda}_{\rm v}(b,p,\mu,\eps) \Bigr]
\,.\end{align}
After evaluating the Dirac trace, the integrals can be expressed in terms of the master integrals
defined in \eq{master_integrals} as
\begin{align} \label{eq:vertex_final}
 \tilde q^{(1)\,\rho\lambda}_{\rm v}(b,p,\mu,\eps) &
 = (2-2\eps) (4\pi \mu_0^\eps)^2 \int\frac{\df^d k}{(2\pi)^d} e^{\img k \cdot b}
   \left[ \frac{\delta^{\rho\lambda}}{k^2 (p-k)^2} - \frac{2 k^\rho k^\lambda}{k^4 (p-k)^2} \right]
 \\\nn&
 = -2 \delta^{\rho\lambda} I_{1,1}(b,p) + 4 I_{2,1}^{\rho\lambda}(b,p) 
\,,\\ \label{eq:vertex_cov_final}
 \Delta \tilde q^{(1)\,\rho\lambda}_{\rm v}(b,p,\mu,\eps) &
 = - (4\pi \mu_0^\eps)^2 \int\frac{\df^d k}{(2\pi)^d} e^{\img k \cdot b}
   \biggl[
      \frac{\delta^{\rho\lambda}}{k^2 (p-k)^2}
    - \frac{2 k^\lambda p^\rho}{k^4 (p-k)^2}
    + \frac{2 p^\lambda (k^\rho-p^\rho)}{k^2 (p-k)^4}
    \nn \\\nn&\hspace{4.5cm}
    + \frac{2 p^2 k^\lambda (p^\rho-k^\rho)}{k^4 (p-k)^4}
   \biggr]
 \\\nn&
 = \delta^{\rho\lambda} I_{1,1}(b,p) - 2 p^\rho I_{2,1}^\lambda(b,p)
   + 2 p^\lambda \Bigl[ I_{1,2}^\rho(b,p) - p^\rho I_{1,2}(b,p) \Bigr]
   \nn\\&\quad
   + 2 p^2 \Bigl[ p^\rho I^\lambda_{2,2}(b,p) -I_{2,2}^{\rho\lambda}(b,p) \Bigr] 
\,.\end{align}
Note that all poles explicitly cancel between the different master integrals,
as infrared poles are regulated by the offshellness $p^2>0$ and UV poles are regulated by $b^2>0$.

\subsubsection{Sail diagram}
\label{sec:sail_diagram}

The sail topology of \fig{TMD b} and its mirror diagram are given by
\begin{align} \label{eq:sail_1}
 \tilde q^{(1)\,\rho\lambda}_{{\rm s}\,,\,\xi}(b,p,\mu,\eps,L) &
 = -\frac{\img}{2} g^2 C_F \mu_0^{2\eps} e^{\img \pb} \int_0^1 \df s \, \gamma^{\prime \mu}(s)
   \int\frac{\df^d k}{(2\pi)^d} \biggl[ \delta^{\mu\nu} - (1-\xi) \frac{k^\mu k^\nu}{k^2} \biggr]
   \nn\\&~\quad \times \biggl[
    e^{- \img k \cdot \gamma(s)} \frac{\frac12 \tr[\gamma^\rho \gamma^\lambda (\slashed{p}-\slashed{k}) \gamma_\nu]}{k^2 (p-k)^2}
  + e^{\img k \cdot [\gamma(s) - b]} \frac{\frac12 \tr[\gamma^\rho \gamma_\nu (\slashed{p}-\slashed{k}) \gamma^\lambda]}{k^2 (p-k)^2}\biggr]
\nn\\&
 \equiv \frac{\as C_F}{4\pi} \Bigl[ \tilde q^{(1)\,\rho\lambda}_{\rm s}(b,p,\mu,\eps,L) + (1-\xi) \Delta \tilde q^{(1)\,\rho\lambda}_{\rm s}(b,p,\mu,\eps,L) \Bigr]
\,.\end{align}
For compactness, we parameterize the Wilson lines by a path $\gamma(s)$, such that
\begin{align}
 W_\gamma &
 = P \exp\left[ - \img g \int_0^1 \df s \, \gamma^\prime(s) \cdot \cA[\gamma(s)] \right]
\,,\end{align}
where $\gamma^{\prime\mu}(s) = \df \gamma^\mu(s)/\df s$ and $\gamma(s)$ is composed of three straight segments given by
\begin{align} \label{eq:qtmd_path}
 \gamma_1(s) &= \left(\begin{matrix} 0 \\ \vec 0_T \\ L s \end{matrix}\right)
\,,\quad
 \gamma_2(s) = \left(\begin{matrix} 0 \\ s \bt  \\ L \end{matrix}\right)
\,,\quad
 \gamma_3(s) = \left(\begin{matrix} 0 \\ \bt \\ L  + s (b^z - L) \end{matrix}\right)
\,,\quad s \in [0,1]
\,.\end{align}
For brevity, here we suppress the explicit dependence of the $\gamma_i(s)$ on $b^\mu$ and $L$.
After evaluating the Dirac trace in \eq{sail_1}, the Feynman gauge piece can be expressed as
\begin{align} \label{eq:sail_Feynman_2}
 \tilde q^{(1)\,\rho\lambda}_{\rm s}(b,p,\mu,\eps,L) &
 = -\img e^{\img \pb} (4\pi\mu_0^\eps)^2 \int\!\frac{\df^d k}{(2\pi)^d}
   \frac{\delta^{\rho\lambda} (p^\mu-k^\mu) - \delta^{\mu\lambda} (p^\rho-k^\rho) + \delta^{\mu\rho} (p^\lambda-k^\lambda)}{k^2 (p-k)^2}
   \nn\\&\hspace{3.cm}\times
   \sum_{i=1}^3 \int_0^1 \df s \, \gamma_i^{\prime \mu}(s) \Bigl( e^{- \img k \cdot \gamma_i(s)} + e^{\img k \cdot [\gamma_i(s) - b]} \Bigr)
\,.\end{align}
Note that the terms proportional to $\delta^{\mu\lambda}$ and $\delta^{\mu\rho}$ cancel each other in the case $\rho=\lambda$,
and that only the term proportional to $k^\mu$ yields a 
$1/\epsilon$ pole, which can easily be extracted since $k \cdot \gamma^\prime(s)$ involves
a total derivative in $s$. We find
\begin{align} \label{eq:sail_final}
 \tilde q^{(1)\,\rho\lambda}_{\rm s}(b,p,\mu,\eps,L) &
 = \delta^{\rho\lambda} e^{\img \pb} \biggl[ \frac{2}{\eps} - \ln\frac{p^2}{\mu^2} + 4 + 2 I_{1,1}(-b,p) \biggr]
 \nn\\&\quad
 + \img e^{\img \pb} \sum_{i=1}^3 \int_0^1 \df s \, \gamma_i^{\prime \mu}(s)
   \Bigl[ \delta^{\rho\lambda} p^\mu \MI_{1,1}[-\gamma_i(s), p]
        \nn\\&\hspace{4.5cm}
        - \delta^{\mu\lambda} \bigl( p^\rho \MI_{1,1}[-\gamma_i(s), p] - \MI_{1,1}^\rho[-\gamma_i(s), p] \bigr)
        \nn\\&\hspace{4.5cm}
        + \delta^{\mu\rho} \bigl( p^\lambda \MI_{1,1}[-\gamma_i(s), p] - \MI_{1,1}^\lambda[-\gamma_i(s), p] \bigr)
        \nn\\&\hspace{4.5cm}
        + \gamma_i(s) \to b - \gamma_i(s) \Bigr]
\,,\end{align}
where we have made the UV pole in $1/\eps$ explicit.

In the covariant-gauge piece in \eq{sail_1}, the derivatives of the path always combine to $k \cdot \gamma_i(s)$,
such that the $\df s$ integration only involves a total derivative, {\it i.e.}\ one only encounters
\begin{align}
 \int_0^1 \df s \, [k \cdot \gamma^\prime(s)] e^{\pm \img k\cdot\gamma(s)}
 = \mp \img \int_0^1 \df s \, \frac{\df}{\df s} e^{\pm\img k\cdot\gamma(s)}
 = \pm \img \bigl( 1 - e^{\pm\img k \cdot b} \bigr)
\,.\end{align}
This gives a simple result in terms of master integrals,
\begin{align} \label{eq:sail_cov_final}
 &\Delta \tilde q^{(1)\,\rho\lambda}_{\rm s}(b,p,\mu,\eps)
 \nn\\&
 = -e^{\img \pb} (4\pi\mu_0^\eps)^2 \int\frac{\df^d k}{(2\pi)^d} \bigl(1 - e^{-\img b \cdot k}\bigr)
  \biggl[ \frac{\delta^{\rho\lambda}}{k^4} +  \frac{\delta^{\rho\lambda}}{k^2 (p-k)^2} -  \frac{\delta^{\rho\lambda} p^2}{k^4 (p-k)^2} + 2 \frac{k^\lambda p^\rho - k^\rho p^\lambda}{k^4 (p-k)^2} \biggr]
\nn\\&
 = \delta^{\rho\lambda} e^{\img \pb} \Bigl\{ \bigl[\MI_{2,0}(p) - \MI_{2,0}(-b,p)\bigr] + \bigl[\MI_{1,1}(p) - \MI_{1,1}(-b,p) \bigr] - p^2 \bigl[ \MI_{2,1}(p) - \MI_{2,1}(-b,p)\bigr] \Bigr\}
 \nn\\&\quad
 + 2 e^{\img \pb} \Bigl\{ p^\rho \bigl[ \MI_{2,1}^\lambda(p) - \MI_{2,1}^\lambda(-b,p) \bigr] - p^\lambda \bigl[  \MI_{2,1}^\rho(p) -  \MI_{2,1}^\rho(-b,p) \bigr] \Bigr\}
\,.\end{align}
This result contains a UV pole inducing a logarithmic contribution, given by
\begin{align}
 \Delta \tilde q^{(1)\,\rho\lambda}_{\rm s}(b,p,\mu,\eps) &
 = \delta^{\rho\lambda} e^{\img \pb} \left[ - \frac{2}{\eps} - \ln\frac{\mu^2 b^2}{b_0^2} - \ln\frac{\mu^2}{p^2} - 2  + \cO(\eps^0) \right]
\,.\end{align}

\paragraph{Axial projection.}
The sail diagram is the only diagram contributing for the axial projector $\cP = \frac12 \gamma^\rho \gamma^5$.
It is obtained similar to \eq{sail_1} as
\begin{align} \label{eq:sail_axial_1}
 \tilde q^{(1)\,\rho\lambda}_{a\,{\rm s}\,,\,\xi}(b,p,\mu,\eps,L) &
 = -\frac{\img}{2} g^2 C_F \mu_0^{2\eps} e^{\img \pb} \int_0^1 \df s \, \gamma^{\prime\mu}(s)
   \int\frac{\df^d k}{(2\pi)^d} \biggl[ \delta^{\mu\nu} - (1-\xi) \frac{k^\mu k^\nu}{k^2} \biggr]
   \nn\\&~\quad \times \biggl[
    e^{- \img k \cdot \gamma(s)} \frac{\frac12 \tr[\gamma^\rho \gamma^5 \gamma^\lambda (\slashed{p}-\slashed{k}) \gamma_\nu]}{k^2 (p-k)^2}
  + e^{\img k \cdot [\gamma(s) - b]} \frac{\frac12 \tr[\gamma^\rho \gamma^5 \gamma_\nu (\slashed{p}-\slashed{k}) \gamma^\lambda]}{k^2 (p-k)^2}\biggr]
\nn\\&
 \equiv \frac{\as C_F}{4\pi} \Bigl[ \tilde q^{(1)\,\rho\lambda}_{a\,\rm s}(b,p,\mu,\eps,L) + (1-\xi) \Delta \tilde q^{(1)\,\rho\lambda}_{a\,\rm s}(b,p,\mu,\eps,L) \Bigr]
\,.\end{align}
The gauge-dependent piece is easily seen to vanish,
\begin{align} \label{eq:sail_axial_cov_final}
 \Delta \tilde q^{(1)\,\rho\lambda}_{a\,\rm s}(b,p,\mu,\eps,L) = 0
\,,\end{align}
such that only the Feynman piece needs to be considered.
The relevant traces are given by
\begin{align}
 \frac12 \tr[\gamma^\rho \gamma^5 \gamma^\lambda (\slashed{p}-\slashed{k}) \gamma^\mu]
 = - \frac12 \tr[\gamma^\rho \gamma^5 \gamma^\mu (\slashed{p}-\slashed{k}) \gamma^\lambda]
 = 2 (k^\nu - p^\nu) \eps^{\rho\lambda\nu\mu}
\,,\end{align}
where the antisymmetric tensor is normalized such that $\eps^{0123} = 1$.
Inserting into \eq{sail_axial_1}, we obtain
\begin{align} \label{eq:sail_axial_2}
 \tilde q^{(1)\,\rho\lambda}_{a\,\rm s}(b,p,\mu,\eps,L) &
 = \img e^{\img \pb} \eps^{\rho\lambda\nu\mu} \int_0^1 \df s \, \gamma^{\prime\mu}(s) \,
   (4\pi\mu_0^\eps)^2 \!\! \int\!\!\frac{\df^d k}{(2\pi)^d} \frac{p^\nu - k^\nu}{k^2 (p-k)^2}
   \big( e^{- \img k \cdot \gamma(s)} - e^{\img k \cdot [\gamma(s) - b]} \bigr)
\nn\\&
 = \img e^{\img \pb} \eps^{\rho\lambda\nu\mu} \sum_{i=1}^3 \int_0^1 \!\!\df s \, \gamma_i^{\prime\mu}(s)
   \Bigl[ p^\nu \bigl(\MI_{1,1}[\gamma_i(s)-b,p] - \MI_{1,1}[-\gamma_i(s),p]\bigr)
   \nn\\&\hspace{4.8cm}
         - \MI^\nu_{1,1}[\gamma_i(s)-b,p] + \MI^\nu_{1,1}[-\gamma_i(s),p] \Bigr]
.\end{align}
Here, the $\MI_{1,1}$ are the usual master integrals defined in \eq{master_integrals}.
Note that both $\MI_{1,1}$ and $\MI_{1,1}^\nu$ are IR and UV finite, and hence \eq{sail_axial_2} does not contain any poles in $\eps$.
In particular, this implies that there is no ambiguity in defining $\gamma^5$ in $d=4-2\eps$ dimensions for this calculation.
The $\gamma_i(s)$ in \eq{sail_axial_2} are the three line segments defined in \eq{qtmd_path}.

\subsubsection{Tadpole diagram}
\label{sec:tadpole_diagram}

The Wilson line self energy, \fig{TMD c}, is given by
\begin{align} \label{eq:q_c}
 \tilde q^{(1)\,\rho\lambda}_{{\rm tp}\,,\,\xi}(b,p,\mu,\eps,L) &
 = - \frac12 \delta^{\rho\lambda} e^{\img b \cdot p}  g^2 C_F \mu_0^{2\eps}
   \int_0^1 \df s \, \df t \, \gamma^{\prime\mu}(s) \gamma^{\prime\nu}(t)
   \nn\\&\hspace{3.2cm}\times
   \int\!\!\frac{\df^d k}{(2\pi)^d} \frac{e^{\img k \cdot [\gamma(t)-\gamma(s)]}}{k^2}
   \biggl[ \delta^{\mu\nu} - (1-\xi) \frac{k^\mu k^\nu}{k^2}  \biggr]
\nn\\&
 \equiv \frac{\as C_F}{4\pi} \Bigl[ \tilde q^{(1)\,\rho\lambda}_{\rm tp}(b,p,\mu,\eps,L) + (1-\xi) \Delta\tilde q^{(1)\,\rho\lambda}_{\rm tp}(b,p,\mu,\eps,L) \Bigr]
\,,\end{align}
where as in \sec{sail_diagram} $\gamma(s)$ is the Wilson line path and we included a symmetry factor $1/2$.
The Feynman piece can be obtained from \mycite{Ebert:2019okf},
\begin{align} \label{eq:tadpole_final}
 \tilde q^{(1)\,\rho\lambda}_{\rm tp}(b,p,\mu,\eps,L)
 = 2 \delta^{\rho\lambda} e^{\img b \cdot p} \biggl[&
    \frac{3}{\eps} + \ln\frac{L^2 \mu^2}{b_0^2}  + \ln\frac{b_T^2 \mu^2}{b_0^2} + \ln\frac{(b^z-L)^2 \mu^2}{b_0^2} + 6
   \nn\\&
   - 2\frac{b^z}{b_T} \arctan\frac{b^z}{b_T} + 2\frac{L}{b_T} \arctan\frac{L}{b_T} + 2 \frac{L - b^z}{b_T} \arctan\frac{L-b^z}{b_T}
   \nn\\&
   - \ln\frac{[b_T^2 + L^2] [ b_T^2 + (L - b^z)^2]}{b_T^2 [b_T^2 + (b^z)^2]}
 \biggr]
\,.\end{align}
The covariant piece only involves an integral over a total derivative in $s$, and is given by
\begin{align} \label{eq:tadpole_cov_final}
 \Delta\tilde q^{(1)\,\rho\lambda}_{\rm tp}(b,p,\mu,\eps,L) &
 = \delta^{\rho\lambda} e^{\img b \cdot p} (4\pi \mu_0^{\eps})^2 \int\frac{\df^d k}{(2\pi)^d} \frac{1 - e^{\img k \cdot b}}{k^4}
 \nn\\&
 = \delta^{\rho\lambda} e^{\img b \cdot p} \left[  \frac{1}{\eps} + \ln\frac{b^2 \mu^2}{b_0^2} + \cO(\eps) \right]
\,.\end{align}

\subsubsection[\texorpdfstring{Full $\cO(\as)$ result}{Full O(alphaS) result}]{\boldmath Full $\cO(\as)$ result}
\label{sec:full_nlo}

The full one-loop result for the amputated Green's function defined in \eqs{qbeam_2}{trace} is given by%
\footnote{Note that the all-order bare result $\tilde q_\xi^{\rho\lambda}$ is formally independent of $\mu$,
while its perturbative expansion at each order in $\alpha_s(\mu)$ acquires an explicit scale dependence.}
\begin{align}
 \tilde q^{\rho\lambda}_\xi(b,p,\eps,L) &=
 \delta^{\rho\lambda} e^{\img \pb} + \frac{\as(\mu) C_F}{4\pi} \tilde q^{(1)\, \rho\lambda}_\xi(b,p,\mu,\eps,L) + \cO(\as^2)
\,,\nn\\
 \tilde q^{(1)\,\rho\lambda}_\xi(b,p,\mu,\eps,L) &= \tilde q^{\rho\lambda\,(1)}(b,p,\mu,\eps,L) + (1-\xi) \, \Delta \tilde q^{\rho\lambda\,(1)}(b,p,\mu,\eps,L)
\,,\end{align}
where the two pieces are given by
\begin{align}
 \tilde q^{\rho\lambda\,(1)}(b,p,\mu,\eps,L) &=
   \tilde q^{\rho\lambda\,(1)}_{\rm v}(b,p,\mu,\eps)
 + \tilde q^{\rho\lambda\,(1)}_{\rm s}(b,p,\mu,\eps,L)
 + \tilde q^{\rho\lambda\,(1)}_{\rm tp}(b,p,\mu,\eps,L)
\,,\\\nn
 \Delta\tilde q^{\rho\lambda\,(1)}(b,p,\mu,\eps) &=
   \Delta\tilde q^{\rho\lambda\,(1)}_{\rm v}(b,p,\mu,\eps)
 + \Delta\tilde q^{\rho\lambda\,(1)}_{\rm s}(b,p,\mu,\eps,L)
 + \Delta\tilde q^{\rho\lambda\,(1)}_{\rm tp}(b,p,\mu,\eps,L)
\,.\end{align}
The individual pieces can be found in \eqss{vertex_final}{sail_final}{tadpole_final} for $\tilde q^{\rho\lambda\,(1)}$,
and in \eqss{vertex_cov_final}{sail_cov_final}{tadpole_cov_final} for $\Delta\tilde q^{\rho\lambda\,(1)}$, respectively.

We have checked that the poles in $\eps$ agree with those reported in \mycite{Ebert:2019okf,Constantinou:2019vyb},
and verified numerically that after dropping these poles our result at $b^z=0$ agrees with \mycite{Constantinou:2019vyb}.
Note that our results are significantly more involved than those in \mycite{Constantinou:2019vyb} because we keep $b^z \ne 0$, which is necessary for the quasi-beam function that is needed as input for the calculation of $\gamma_\zeta^q(\mu,b_T)$.

For the axial projection, there is only one nonvanishing contribution, such that
\begin{align}
 \tilde q^{\rho\lambda}_{a\,\xi}(b,p,\eps,L) &= \frac{\as(\mu) C_F}{4\pi} \tilde q^{\rho\lambda\,(1)}_{a\,\rm s}(b,p,\mu,\eps,L)+\cO(\as^2)
\,,\end{align}
where $q^{\rho\lambda\,(1)}_{a\,\rm s}$ is given in \eq{sail_axial_2}.

\subsection[RI'/MOM renormalization factor and conversion to \texorpdfstring{$\MS$}{MSbar}]
           {\boldmath \RIpMOM~renormalization factor and conversion to $\MS$}
\label{sec:RIMOM_nlo}

Having calculated the full one-loop result for the off-shell amputated Green's function $\tilde q^{\rho\lambda}$,
we can now proceed to calculate the \RIpMOM~renormalization and the conversion to the $\MS$ scheme.
This also requires the one-loop wave function renormalization to account for the external state in the amputated Green's function.
In the \RIpMOM\ scheme it is given by~\cite{Martinelli:1994ty}
\begin{align} \label{eq:Zq}
 Z_{\rm wf}^{-1}(p,\eps) &
 = \frac{1}{4p^2} \Tr\bigl[ S^{-1}(p) \slashed{p}\bigr]
 = 1 - \frac{\as(\mu) C_F}{4\pi} Z_{\rm wf}^{(1)}(p,\mu,\eps) + \cO(\as^2)
  \,, \nn\\
 Z_{\rm wf}^{(1)}(p,\mu,\eps) &= -[1 - (1-\xi)] \Bigl(\frac{1}{\eps} + \ln\frac{\mu^2}{p^2} + 1 \Bigr)
\,.\end{align}
The \RIpMOM~renormalization of the quasi-TMDPDF also requires us to include the one-loop soft factor.
Using \eq{qDeltaS}, it can be written as
\begin{align} \label{eq:DeltaS_nlo}
 \tilde\Delta_S^q(b_T,\eps,L)  &
 = \frac{1}{\sqrt{\tilde S^q_{\rm bent}(b_T,\eps,L)}}
 = 1 - \frac12 \frac{\as(\mu) C_F}{4\pi} \tilde S^{q\,(1)}_{\rm bent}(b_T,\mu,\eps,L)
\,.\end{align}
The required one-loop result for the bent soft function can be obtained from~\mycite{Ebert:2019okf} as
\begin{align} \label{eq:qsoft_bent_nlo}
 \tilde S^{q\,(1)}_{\rm bent}(b_T,\mu,\eps,L) &
 = \frac{12}{\eps} + 12 \ln\frac{\mu^2 b_T^2}{b_0^2} + 16 \frac{L}{b_T} \arctan\frac{L}{b_T} + 24
   \nn\\&\quad
   + 4 \sqrt{2} \frac{b_T}{L} \arctan\frac{b_T}{\sqrt{2}L} - 4 \ln\frac{b_T^2+2L^2}{2 L^2} - 8 \ln\frac{b_T^2+L^2}{L^2}
\,.\end{align}
The \RIpMOM\ to $\MS$ conversion kernel follows from eqs.\ \eqref{eq:ZqRIMOM}, \eqref{eq:ZBRIMOM} and \eqref{eq:Zf_RIMOM_MS} as
\begin{align} \label{eq:Zq_nlo}
 \tilde Z_q^{\prime \rho\lambda}(b^z, \mu, b_T^R, p_R, L) &
 = \lim_{\eps\to0} \frac{\tilde Z_q^\MS(\mu,\eps)  Z_{\rm wf}(p_R,\eps) \tilde q^{\rho\lambda}_\xi(b^z,b_T^R,p_R,\eps,L) \tilde\Delta_S^q(b_T^R,\eps,L)}{\tilde q^{(0)\,\rho\lambda}_\xi(b^z, b_T^R, p_R)}
\nn\\&
= 1 + \frac{\as(\mu) C_F}{4\pi} \Bigl[ e^{-\img \vec b_T^R \cdot \vec p_T^R - \img b^z p^z} \, \tilde q^{\rho\lambda~(1)}_\xi(b^z, b_T^R,p_R,\mu,\eps,L) + Z_{\rm wf}^{(1)}(p_R,\mu,\eps)
   \nn\\&\hspace{3cm}
   - \frac12 \tilde S^{q\,(1)}_{\rm bent}(b_T^R,\mu,\eps,L) \Bigr]_{\cO(\eps^0)} + \cO(\as^2)
\,,\end{align}
where all the $1/\epsilon$ poles are canceled by $\tilde Z_q^\MS(\mu,\eps)$, so only the $\cO(\eps^0)$
terms are extracted from the terms in the square brackets.

Last but not least, one may note that $\tilde\Delta_S^q(b_T^R,\eps,L)$ also formally cancels out in the ratios in \eqs{gamma_zeta_2new}{gamma_zeta_4} due to its $b^z$-independence, and therefore equivalently we can drop it in \eq{Zq_nlo} and obtain the conversion factor that matches the \RIpMOM-renormalized quasi-beam function to the $\MS$ scheme,
\begin{align} \label{eq:ZB_nlo}
 \tilde Z_B^{\prime \rho\lambda}(b^z, \mu, b_T^R, p_R, L)
 = 1 + \frac{\as(\mu) C_F}{4\pi} \Bigl[& e^{-\img \vec b_T^R \cdot \vec p_T^R - \img b^z p^z} \, \tilde q^{\rho\lambda~(1)}_\xi(b^z, b_T^R,p_R,\mu,\eps,L)
 \nn\\&
 + Z_{\rm wf}^{(1)}(p_R,\mu,\eps) \Bigr]_{\cO(\eps^0)} + \cO(\as^2)
\,.\end{align}
However, this $\tilde Z_B^{\prime}$ will suffer from $L/b_T$ divergence that makes its numerical value much larger than one, indicating that the perturbation series does not converge. In constrast, $\tilde Z_q^{\prime}$ in \eq{Zq_nlo}, which includes the correction from $\tilde\Delta_S^q$, is free from such divergences and has good perturbative convergence, as we will demonstrate numerically in the following section.

\section{Numerical results}
\label{sec:numerics}

In this section, we numerically illustrate the importance of the perturbative matching from the \RIpMOM~to the $\MS$ scheme.
We assume a lattice with spacing $a = 0.06~\fm$ and size $L_{\rm lat} = 32\,a$, and set the length of the Wilson line to $L = 10\,a$.
The $\MS$ renormalization scale is chosen as $\mu = 3~\GeV$, with $\as(\mu) = 0.2492$ obtained using three-loop running from $\as(m_Z) = 0.118$. 
We always work in Landau gauge with $\xi=0$.
To show the effect of canceling linear divergences in $L/b_T$, we will consider both the conversion factor $\tilde Z_q^\prime$ for the quasi-TMDPDF and $\tilde Z_B^\prime$ for the quasi-beam function alone.

\begin{figure*}[t!]
 \includegraphics[width=0.48\textwidth]{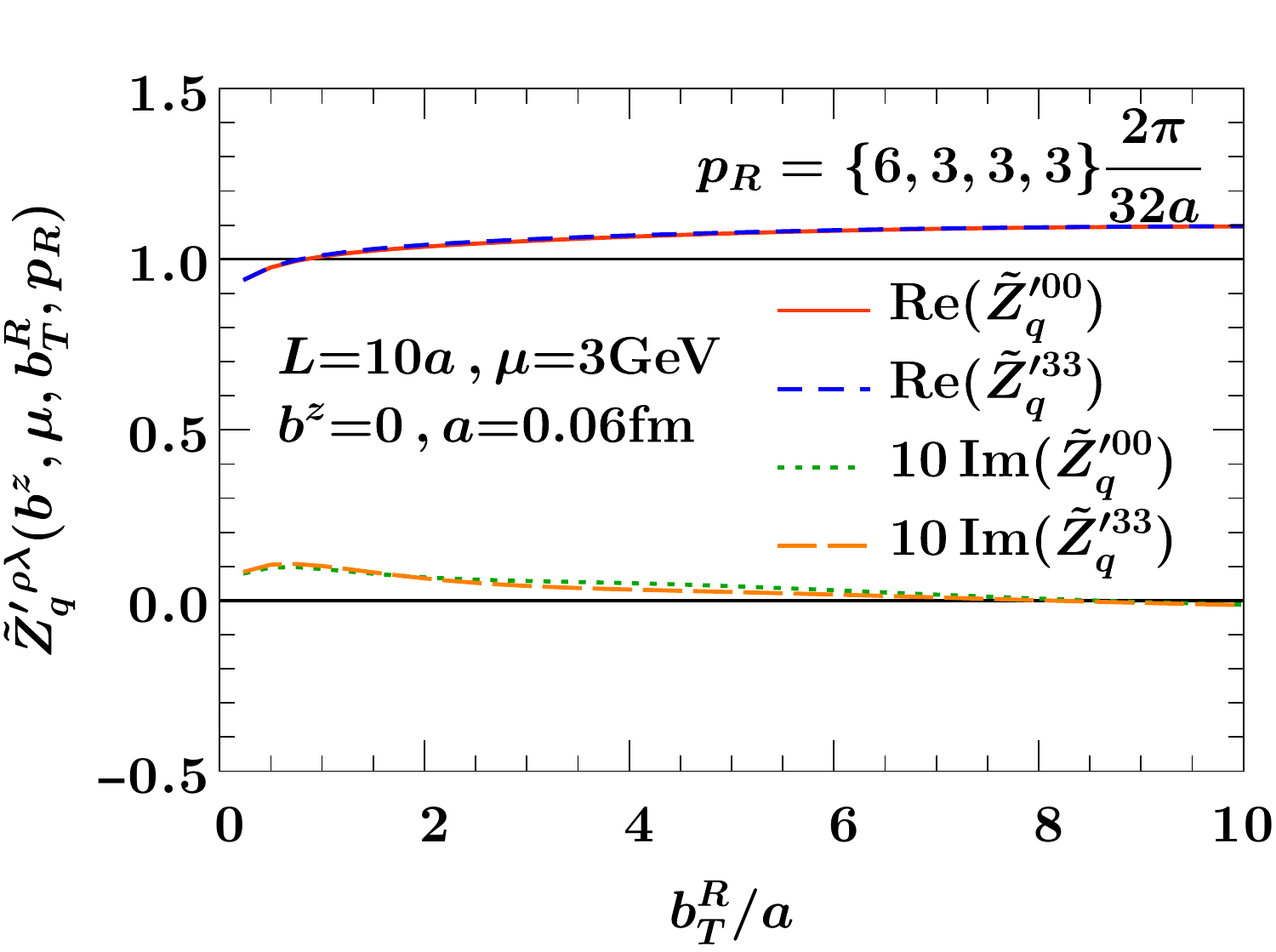}
\quad
 \includegraphics[width=0.48\textwidth]{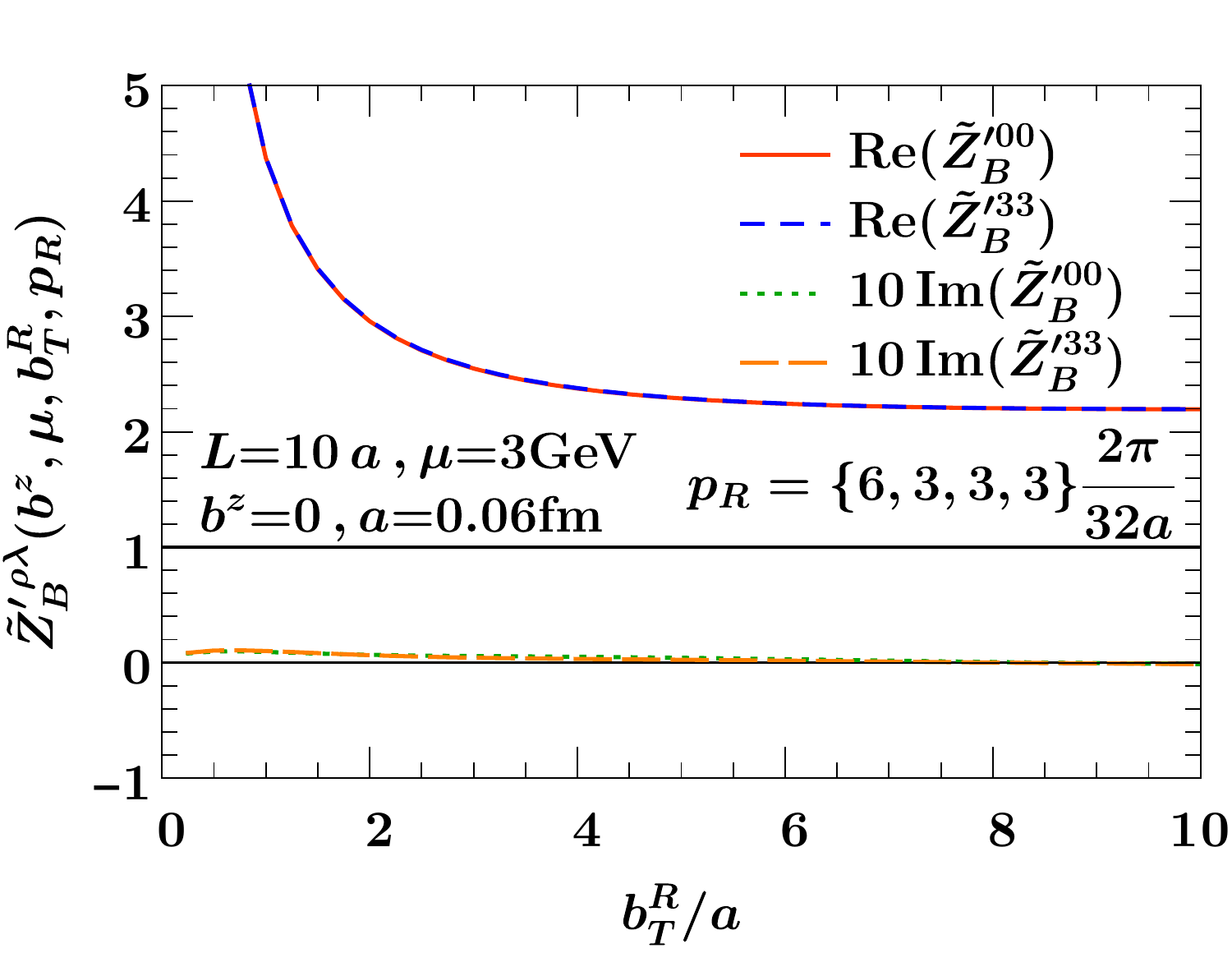}
\\
 \includegraphics[width=0.48\textwidth]{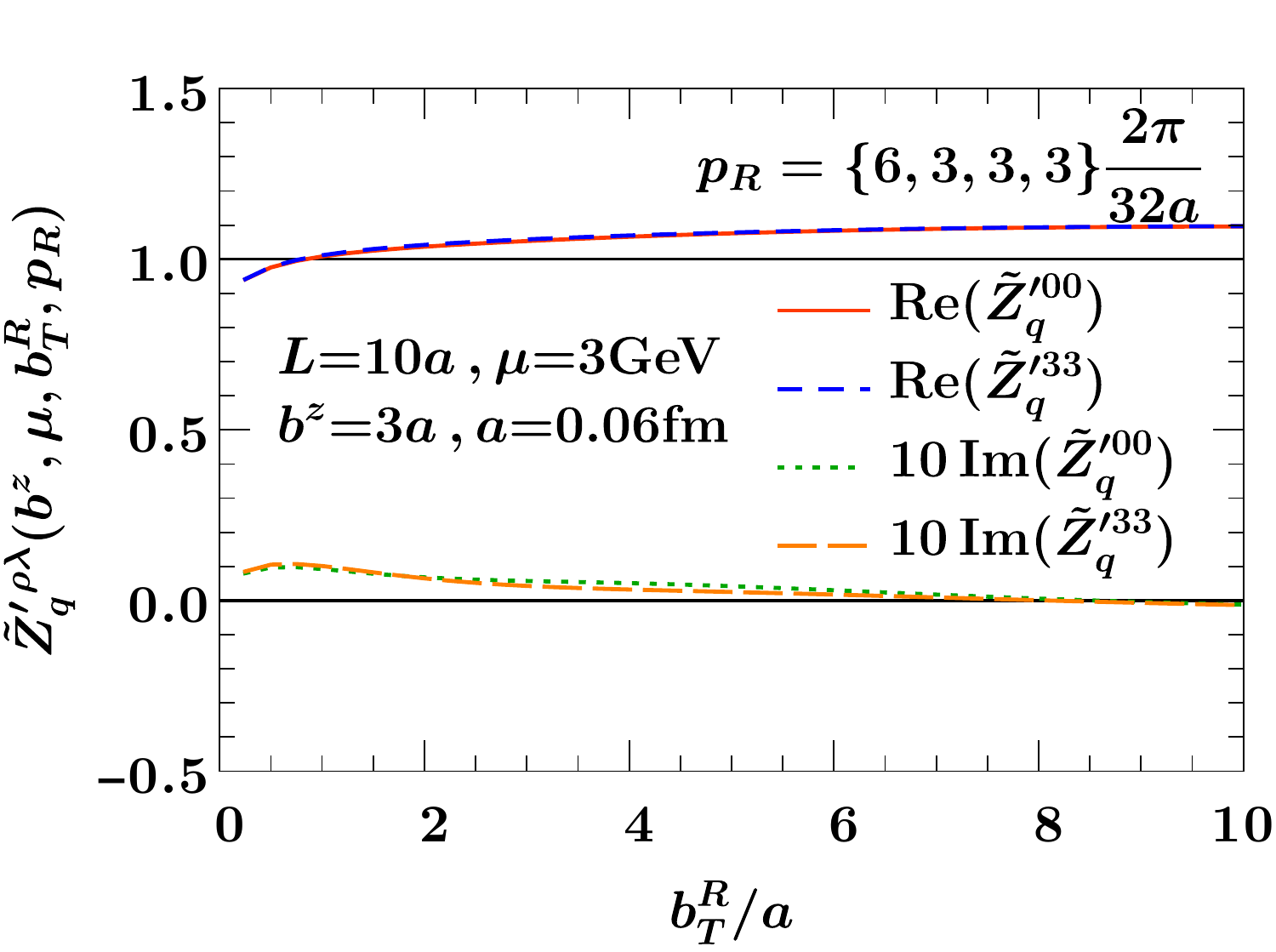}
\quad
 \includegraphics[width=0.48\textwidth]{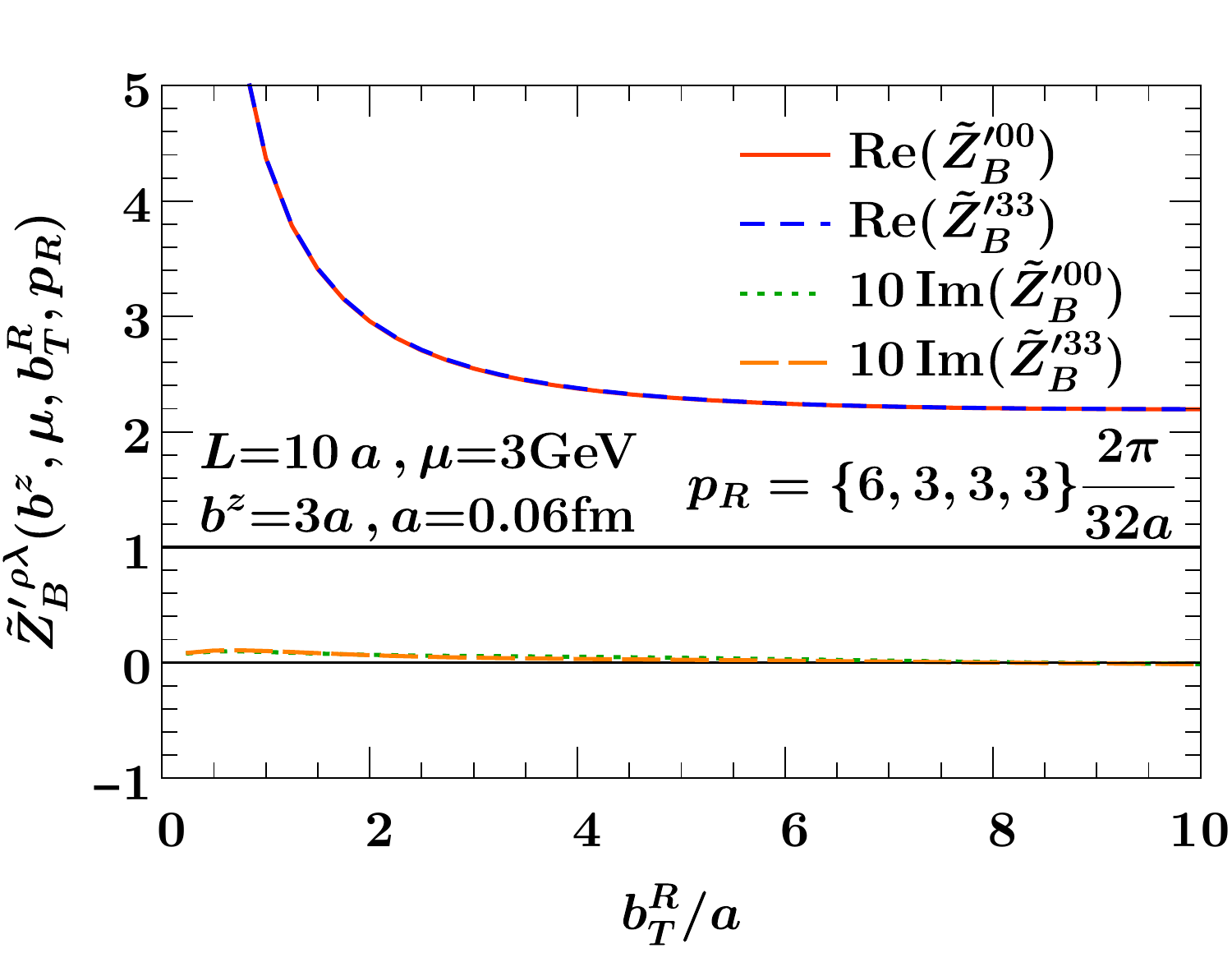}
\\
 \includegraphics[width=0.48\textwidth]{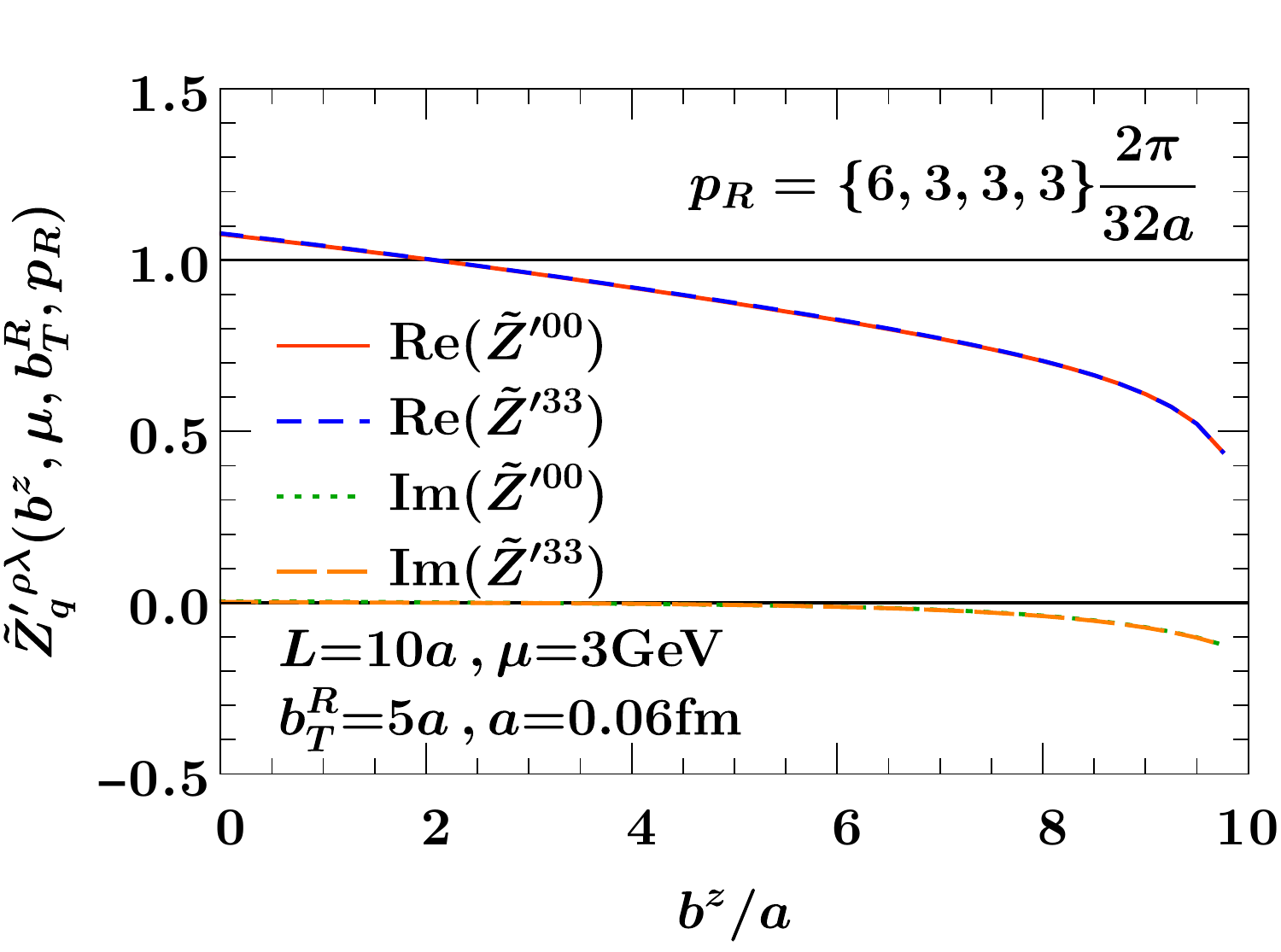}
\quad
 \includegraphics[width=0.48\textwidth]{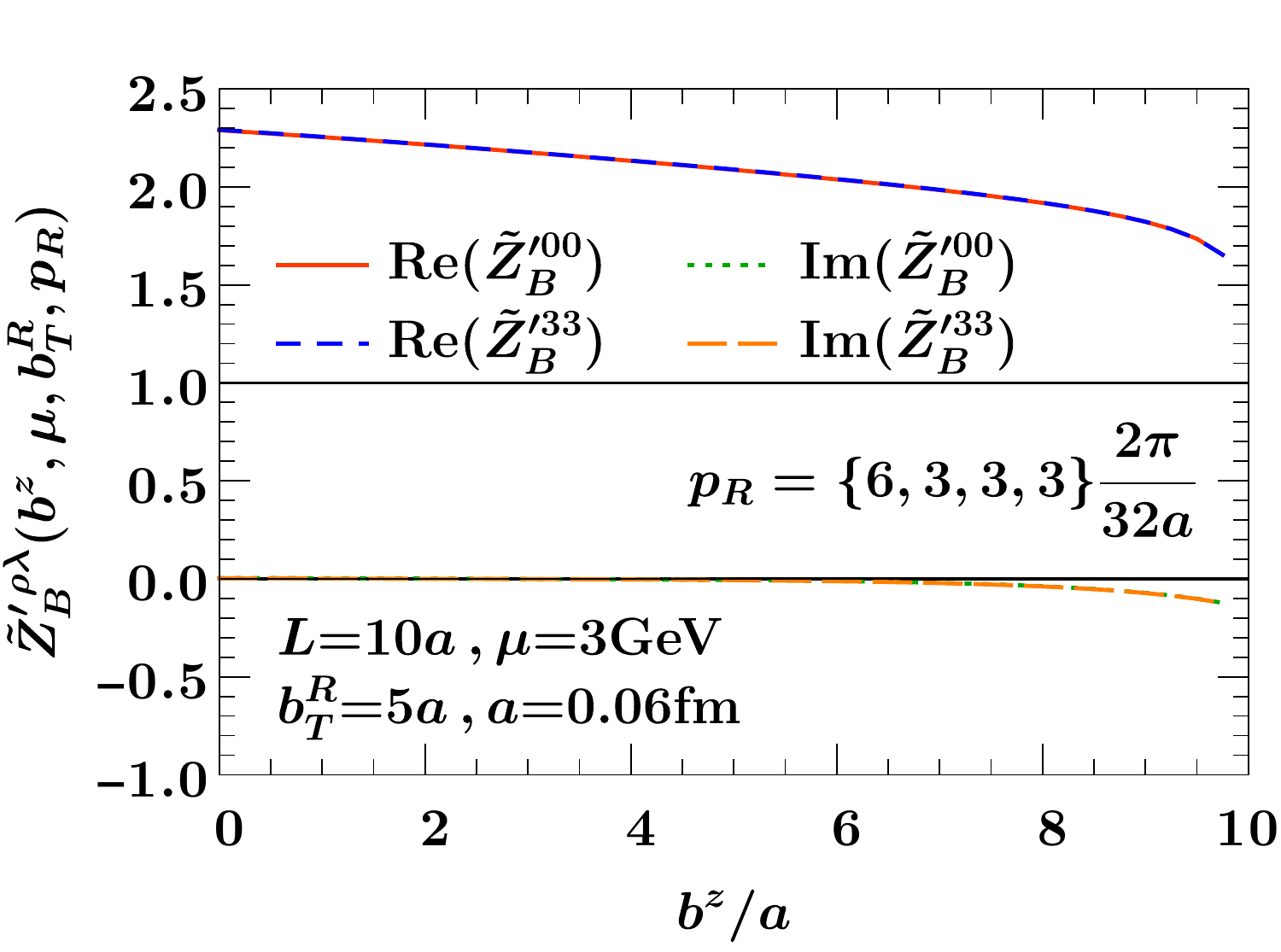}
 \caption{One-loop conversion factors $\tilde Z'$ from the \RIpMOM~scheme to the $\MS$ scheme for the quasi-TMDPDF (left) and the quasi-beam function (right), as a function of $b_T/a$ for $b^z=0$ (top), as a function of $b_T/a$ for $b^z = 3\,a$ (middle) and as a function of $b^z/a$ for $b_T = 5\,a$ (bottom). }
 \label{fig:C_fixed_p2}
\end{figure*}

We first consider the Euclidean momentum
\begin{align}
 p_R^\mu &= (p_R^0, \vec p_R\,) = (6, 3, 3, 3) \frac{2 \pi}{L_{\rm lat}} \,\approx (3.9, 1.9, 1.9, 1.0)~\GeV
\,,\nn\\
 p_R^2 &= (p_R^0)^2 + {\vec p}_R^{\,2} = 63 \Bigl(\frac{2 \pi}{L_{\rm lat}}\Bigr)^2 \approx (5.1~\GeV)^2
\,.\end{align}
In \fig{C_fixed_p2}, we show  $\tilde Z_q^\prime$ in the left panel and $\tilde Z_B^\prime$ in the right panel.
The $b_T$ dependence is shown for fixed $b^z=0$ (top row) and $b^z=3a$ (middle row),
while the bottom row shows the $b^z$ dependence for fixed $b_T=5a$.
In each plot, we show real and imaginary parts for the $\rho=\lambda=0$ Dirac structure in solid red and dotted green, respectively, as well as the real and imaginary parts for $\rho=\lambda=3$ in dashed blue and dashed orange, respectively. The imaginary parts in the first two rows are amplified by a factor of ten to increase their visibility.
The off-diagonal Dirac structures with $\rho\ne\lambda$ are very small and not shown here.
In all cases, we find a very small imaginary part of both $Z_B^\prime$ and $Z_q^\prime$,
and that the two choices $\lambda=\rho=0$ and $\lambda=\rho=3$ are very similar.
Hence in the following, we restrict our discussion to the real part and the choice $\rho=\lambda=0$ only.

For $\tilde Z_B^\prime$ (right panels of \fig{C_fixed_p2}), the presence of the $L/b_T^R$ divergence is clearly visible, and leads to large values for this factor. Since this coefficient is $1$ at lowest order, clearly perturbation theory is not converging for $\tilde Z_B^\prime$, as anticipated.

For $\tilde Z_q^\prime$ (left panels of \fig{C_fixed_p2}), we generically observe corrections close to $Z_q^\prime = 1$,
indicating that the $\cO(\as)$ corrections are rather moderate and of the expected size of a NLO correction.
However, there is a significant dependence on both $b_T^R$ and $b^z$.
In particular, one can observe a mild logarithmic dependence on $b_T^R$ as $b_T^R\to0$.
Since $b_T^R$ is a free parameter in the renormalization procedure, one can choose it freely to yield small matching corrections, as long as $b_T^R$ is perturbative.
The results in \fig{C_fixed_p2} indicate that $b_T^R \gtrsim a$ is a good choice.
In order to minimize lattice discretization effects, which are not captured in our analytic calculation,
one must choose $b_T^R \gg a$, so in practice we expect that $b_T^R \sim \cO(\mathrm{few}~a)$ is a reasonable choice.
There is also a significant $b^z$ dependence, arising from the fact that in the \RIpMOM\ scheme one fully absorbs the $b^z$-dependence at $p=p_R$ and $b_T=b_T^R$ into the UV renormalization, and this $b^z$ dependence must therefore be corrected perturbatively through the conversion to the $\MS$~scheme.

For larger $b^z$ the correction from $\tilde Z_q^\prime$ becomes numerically significant, as can seen from the bottom left panel of  \fig{C_fixed_p2}.
The impact of this large $b^z$ region is suppressed by the large parton momentum when the quasi-TMDPDF is Fourier transformed into the $x$-space, namely through the oscillation caused by the Fourier exponents involving $(xP^z b^z)$ in \eq{gamma_zeta_2new}. For the position space version in \eq{gamma_zeta_4} the analogous suppression of the large $b^z$ region occurs from the falling and oscillating  behavior of $\bar C_{\rm ns}$ with $(P^zb^z)$. The derivation of  \eq{relation_qTMD_1}, and thus both the momentum and position space formulae for $\gamma_\zeta^q$, assume the hierarchy $b^z \ll L$, and hence dominance of the integral away from the large $b^z\sim L$ region. Studying numerically the dependence of \eqs{gamma_zeta_2new}{gamma_zeta_4} on the upper limit of $b^z$ used in the integrations will help us understand how well this hierarchy is obeyed.

\begin{figure*}
 \includegraphics[width=0.48\textwidth]{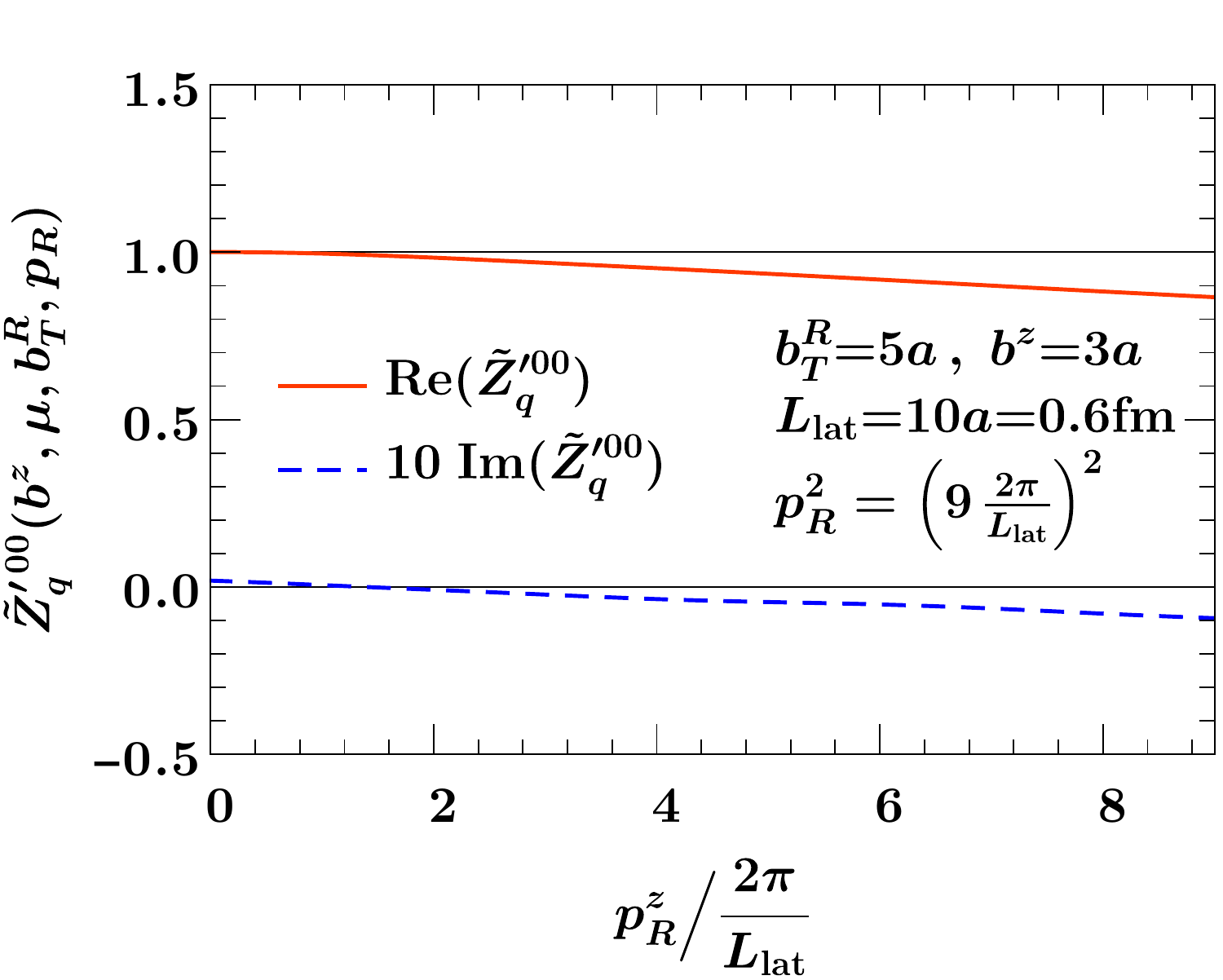}
\quad
 \includegraphics[width=0.48\textwidth]{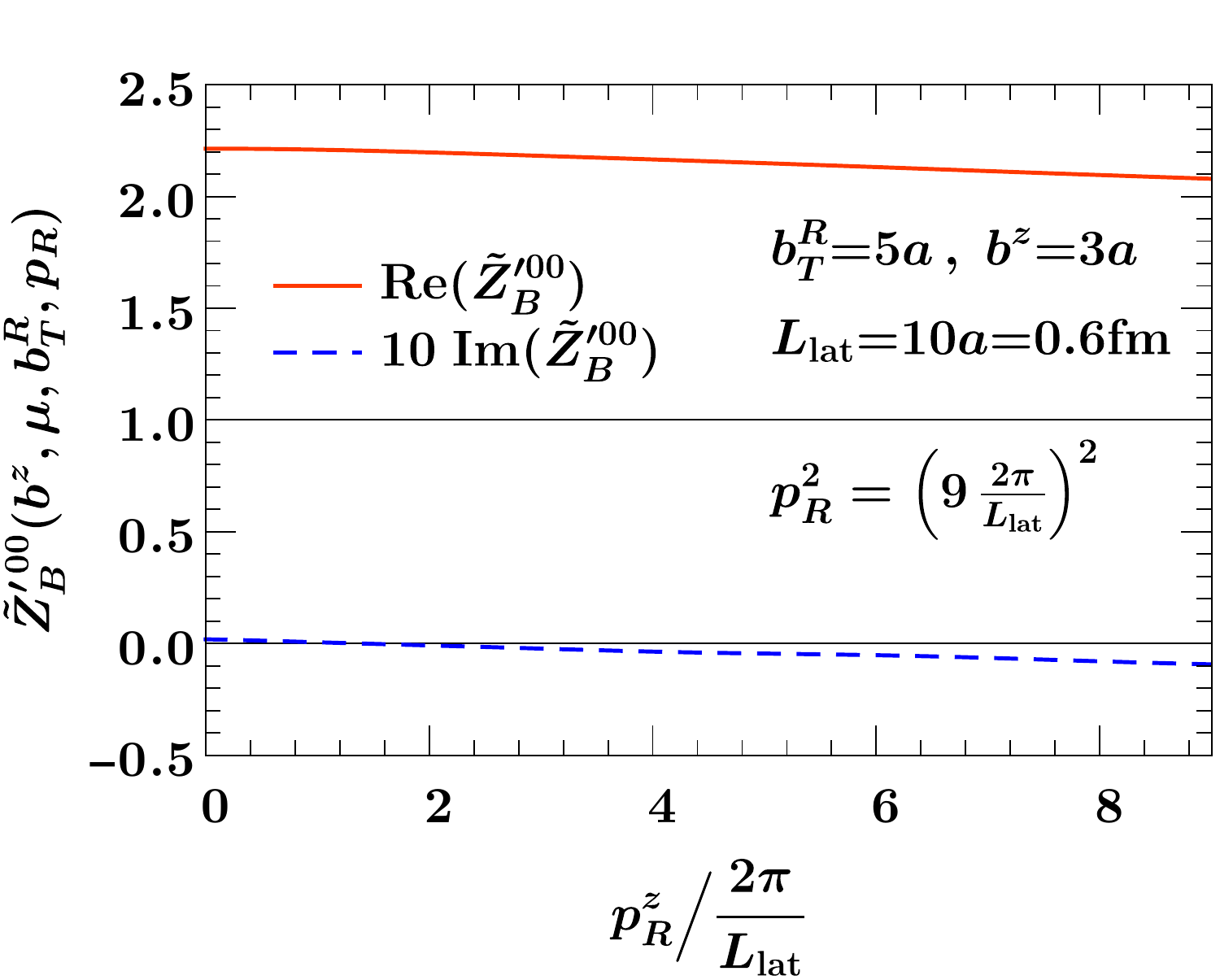}
 \caption{One-loop conversion factors $\tilde Z'$ from the \RIpMOM~scheme to the $\MS$ scheme for the quasi-TMDPDF (left) and the quasi-beam function (right),  as a function of the $z$ momentum, $p_R^z$, of the quark state. }
 \label{fig:ZS_pzSpectrum}
\end{figure*}

Finally, we study the $p_R^z$ dependence for fixed $b_T = 5\,a$ and $b^z = 3\,a$.
For a given $p_R^z$ there is still  considerable freedom for the other parameters of $p_R$, and we choose the Euclidean momentum
\begin{align}
 p_R^\mu = \bigl(\sqrt{79 - (p_R^z)^2}, 1, 1, p_R^z\bigr) \frac{2 \pi}{L_{\rm lat}}
 \,,\quad
 p_R^2 = \Bigl(9 \frac{2\pi}{L_{\rm lat}} \Bigr)^2 \approx (5.8~\GeV)^2
\,,\end{align}
where $p_R^0$ is a function of $p_R^z$ such that $p_R^2$ is fixed.
The largest value of $p_R^z$ yielding a real solution for $p_R^0$ is
then given by $p_R^z = \sqrt{79} (2\pi/L_{\rm lat}) \approx 5.7~\GeV$.
\fig{ZS_pzSpectrum} shows the resulting scheme conversion factors, with $\tilde Z_q^\prime$ on the left and $\tilde Z_B^\prime$ on the right.
As before, in both cases the imaginary part (blue dashed) is very small, and the real part (red) is close to unity for $\tilde Z_q^\prime$ indicating that perturbation theory is working as expected, while it has significant deviation from unity for $\tilde Z_B^\prime$ (where we see that perturbation theory is breaking down). 
For $\tilde Z_q^\prime$ it can also be observed that there is a relatively mild dependence on $p_R^z$.

\FloatBarrier

\section{Conclusion}
\label{sec:conclusion}

In this paper we have elaborated on the method to determine the Collins-Soper kernel using ratios of quasi-TMDPDFs.
Originally, in \mycite{Ebert:2018gzl} a method was proposed which used ratios of properly matched and renormalized quasi-TMDPDFs in momentum space. This
requires a Fourier transformation of spatial correlations obtained from the lattice to momentum space, which can be numerically challenging.
Here we have extended this proposal to demonstrate how to carry out the matching for renormalized ratios directly in position space. This trades the Fourier transformation for a convolution with a position space matching coefficient, which we expect will improve the numerical stability of the method.  The required position space matching coefficient $\bar C_{\rm ns}(\mu,y,P^z)$ was obtained here at ${\cal O}(\alpha_s)$.

In addition, we have calculated a renormalization scheme conversion factor that is needed for the lattice calculation. Renormalization on the lattice must necessarily be done nonperturbatively to properly handle power law divergences from spatial Wilson line self energies. Here
we calculated the one-loop renormalization factor for the transverse-momentum dependent quasi-TMDPDFs
in the regularization-independent momentum subtraction RI$^\prime$/MOM scheme with $b^z\ne 0$, and used this result to obtain
the one-loop conversion factor $\tilde Z_q^\prime(b^z,\mu,\tilde \mu)$ that converts from the \RIpMOM~scheme to the $\MS$~scheme.
This conversion factor is necessary to obtain results for the Collins-Soper kernel $\gamma_\zeta^q(\mu,b_T)$ in the desired $\MS$ scheme. Our results are thus key to determining the Collins-Soper kernel
from lattice QCD using ratios of quasi-TMDPDFs as proposed in \mycite{Ebert:2018gzl}, and elaborated on in \mycite{Ebert:2019okf}.
These results will also be used in the lattice study of nonperturbative renormalization of the quasi-beam functions~\cite{Shanahan:2019zcq}.

Together the results obtained here provide important ingredients to enable a first nonperturbative
determination of the Collins-Soper kernel from lattice QCD.

\begin{acknowledgments}
We thank Phiala Shanahan and Michael Wagman for useful discussions.
This work was supported by the U.S.\ Department of Energy, Office of Science,
Office of Nuclear Physics, from DE-SC0011090, DE-SC0012704
and within the framework of the TMD Topical Collaboration.
I.S.\ was also supported in part by the Simons Foundation through
the Investigator grant 327942.
M.E.\ was also supported by the Alexander von Humboldt Foundation
through a Feodor Lynen Research Fellowship.

\end{acknowledgments}

\appendix

\section{Master integrals}
\label{app:master_integrals}

The master integrals required in \sec{NLO} were defined in \eq{master_integrals} as
\begin{align}
 \MI_{i,j}^{\mu \nu \cdots}(b,p) &
 = - (4\pi\mu_0^{\eps})^2 \int\!\frac{\df^d k}{(2\pi)^d} \frac{k^\mu k^\nu \cdots}{(k^2)^i [(p-k)^2]^j} \, e^{\img k \cdot b}
\,,\nn\\
 \MI_{i,j}^{\mu \nu \cdots}(p) &
 = - (4\pi\mu_0^{\eps})^2 \int\!\frac{\df^d k}{(2\pi)^d} \frac{k^\mu k^\nu \cdots}{(k^2)^i [(p-k)^2]^j}
\,,\end{align}
where we employ a Euclidean metric.
All required tensor structures can be obtained from the scalar integrals through
\begin{align} \label{eq:master_integrals_tensor}
 \MI_{i,j}^\mu(p) &
 = \frac{p^\mu}{2} \biggl[ I_{i,j}(p) + \frac{1}{p^2} I_{i-1,j}(p) - \frac{1}{p^2} I_{i,j-1}(p) \biggr]
\,,\\
 \MI_{i,j}^\mu(b,p) &
 = -2 \img b^\mu \frac{\partial \MI_{i,j}(b,p)}{\partial (b^2)} -\img p^\mu \frac{\partial \MI_{i,j}(b,p)}{\partial (\pb)}
\,,\nn\\
 \MI_{i,j}^{\mu\nu}(b,p) &
 = -2 \delta^{\mu\nu} \frac{\partial \MI_{i,j}}{\partial (b^2)}
 - 2 (p^\mu b^\nu + b^\mu p^\nu) \frac{\partial^2 \MI_{i,j}}{\partial (b^2) \partial (\pb)}
 - 4 b^\mu b^\nu \frac{\partial^2 \MI_{i,j}}{\partial^2 (b^2)}
 - p^\mu p^\nu \frac{\partial^2 \MI_{i,j}}{\partial^2(\pb)}
\nn
\,.\end{align}
Here, we employed that $I_{i,j}(b,p)$ can only depend on the Lorentz scalars $b^2$, $\pb$ and $p^2$,
and for brevity suppressed the arguments in the last line of \eq{master_integrals_tensor}.

The scalar integrals $I_{i,j}(b,p)$ and $I_{i,j}(p)$ can be evaluated using Feynman parameters
and standard integral techniques. In the Euclidean regime, we have $b^2 > 0$ and $p^2 > 0$,
which yields the general results%
\footnote{Using Minkowski metric one obtains the same results in \eqs{master_integrals_scalar_1}{master_integrals_scalar_2},
up to a relative factor of $-\img$ and assuming $b^2 < 0$, $p^2 < 0$.}
\begin{align} \label{eq:master_integrals_scalar_1}
 \MI_{i,j}(p) &
 = -(p^2)^{2-i-j} \left(\frac{\mu^2 e^{\gamma_E}}{p^2}\right)^{\eps}
   \frac{\Gamma(i+j+\eps-2)}{\Gamma(4-i-j-2\eps)}
   \frac{\Gamma(2-i-\eps)}{\Gamma(i)} \frac{\Gamma(2-j-\eps)}{\Gamma(j)}
\,, \\ \label{eq:master_integrals_scalar_2}
 \MI_{i,j}(b,p) &
 = -\frac{2 (\mu^2 e^{\gamma_E})^\eps}{\Gamma(i)\Gamma(j)}
   \biggl(\frac{4 p^2}{b^2}\biggr)^{1 - (i+j+\eps)/2}
   \nn\\&\quad\times
   \int_0^1 \df x\, x^{(j-i-\eps)/2} (1-x)^{(i-j-\eps)/2} e^{\img x \pb}
   K_{i+j+\eps-2}\bigl(\sqrt{b^2 p^2 x (1-x)}\bigr)
\,.\end{align}
Here, $K_n(z)$ is the modified Bessel function of second kind.

The results in \eqs{master_integrals_scalar_1}{master_integrals_scalar_2} contain divergences as $\eps\to0$,
which typically cancel in the expressions for the individual diagrams given in \sec{NLO},
so that one can let $\eps\to0$ right away.
This cancellation can also be made manifest by extracting the explicit poles in $1/\eps$.
For the scalar integrals in \eq{master_integrals_scalar_1}, expanding in $\eps$ gives
\begin{align} \label{eq:MI_scalar_explicit}
 I_{0,j}(p) = I_{i,0}(p) &= 0
\,,\nn\\
 I_{1,1}(p) &= -\frac{1}{\eps} - \ln\frac{\mu^2}{p^2} - 2 + \cO(\eps)
\,,\nn\\
 I_{1,2}(p) = I_{2,1}(p) &= \frac{1}{p^2} \left[ \frac{1}{\eps} + \ln\frac{\mu^2}{p^2} + \cO(\eps) \right]
\,,\nn\\
 I_{2,2}(p) &= \frac{2}{(p^2)^2} \left[ \frac{1}{\eps} + \ln\frac{\mu^2}{p^2} + 1 + \cO(\eps) \right]
\,.\end{align}

The integral over the Feynman parameter $x$ in \eq{master_integrals_scalar_2}
is not known for arbitrary parameters $i$ and $j$, and has to be evaluated numerically.
Infrared divergences as $x \to 0$ to $x \to 1$ can be extracted using the asymptotic limit of $K_n(z\to0)$,
\begin{alignat}{3}
 \MI_{2,0}(b,p) &
 = \frac{1}{\eps} + \ln\frac{b^2 \mu^2}{b_0^2}
 \nn\\
 \MI_{1,1}(b,p) &
 = &&\int_0^1 \df x \, \cI_{1,1}(x, b^2, \pb, p^2)
 \nn\\
 \MI_{1,2}(b,p) &
 = \frac{e^{\img \pb}}{p^2} \biggl[ \frac{1}{\eps} + \ln\frac{\mu^2}{p^2} \biggr]
 &&+ \int_0^1 \df x \, \cI_{1,2}(x, b^2, \pb, p^2)
 \nn\\
 \MI_{2,1}(b,p) &
 = \frac{1}{p^2} \biggl[ \frac{1}{\eps} + \ln\frac{\mu^2}{p^2} \biggr]
 &&+ \int_0^1 \df x \, \cI_{2,1}(x, b^2, \pb, p^2)
 \nn\\
 \MI_{2,2}(b,p) &
 =  \frac{1 + e^{\img \pb}}{p^4}  \biggl[ \frac{1}{\eps} + \ln\frac{\mu^2}{p^2} + 1 \biggr]
 &&+ \int_0^1 \df x \, \cI_{2,2}(x, b^2, \pb, p^2)
\,,\end{alignat}
where the integral kernels are defined as
\begin{align}
 \cI_{1,1}(x, b^2, \pb, p^2) &
 = -2 e^{\img x \pb} K_0\Bigl(\sqrt{b^2 p^2 x (1-x)}\Bigr)
 \nn\\
 \tilde\cI_{1,2}(x, b^2, \pb, p^2) &
 = -\frac{e^{\img \pb}}{p^2} \frac{1}{x} \left[ e^{- \img x \pb}  \sqrt{b^2 p^2 x (1-x)} K_{1}\Bigl(\sqrt{b^2 p^2 x (1-x)}\Bigr) - 1 \right]
 \nn\\
 \cI_{2,1}(x, b^2, \pb, p^2) &
 = - \frac{1}{p^2} \frac{1}{x} \left[ e^{+\img x \pb} \sqrt{b^2 p^2 x (1-x)} K_{1}\Bigl(\sqrt{b^2 p^2 x (1-x)}\Bigr) - 1 \right]
 \nn\\
 \cI_{2,2}(x, b^2, \pb, p^2) &
 =  -\frac{1 + e^{\img \pb}}{4 p^4} \frac{1}{x (1-x)}
 \nn\\&\quad\times
 \biggl[ \frac{e^{\img x \pb} + e^{\img (1-x) \pb}}{1 + e^{\img \pb}} b^2 p^2 x (1-x) K_{2}\Bigl(\sqrt{b^2 p^2 x (1-x)}\Bigr) - 2  \biggr]
\,.\end{align}

\section{\boldmath Alternative determination of \texorpdfstring{$\gamma_\zeta$}{gamma\_zeta} in position space}
\label{app:CS_position_space}

Here, we present a slightly modified method compared to that presented in \sec{CS_position_space} for determining $\gamma_\zeta$ in position space.
There, the Fourier transform of the \emph{inverse} of the kernel $C_\ns$ was employed.
Here, we directly Fourier transform the kernel $C_\ns$,
\begin{align} \label{eq:FT_C_alt}
 C_\ns(\mu, x P^z_i) &= \int\frac{\df (b^z_i P^z_i)}{2\pi} \, e^{\img x (b^z_i P^z_i)} \, \bar C_\ns^\prime(\mu, b^z_i P^z_i, P^z_i)
\,,\nn\\
 \bar C_\ns^\prime(\mu, b^z_i P^z_i, P^z_i) &= \int\!\df x \, e^{-\img x (b^z_i P^z_i)} \, C_\ns(\mu, x P^z_i)
\,.\end{align}
As before, we Fourier transform with respect to $b^z_i P^z_i$ to absorb superfluous factors of $P^z_i$.

Plugging \eqs{FT_fTMD}{FT_C_alt} into \eq{relation_qTMD_2}, we get
\begin{align} \label{eq:app:relation_qTMD_2}
 &P^z_2 \int\!\df b^z_1 \, \df b^z_2 \, e^{\img x (b^z_1 P^z_1 + b^z_2 P^z_2)}
  \bar C_\ns^\prime(\mu, b^z_2 P^z_2, P^z_2) \tilde f_\ns(b^z_1, \bt, \mu, P^z_1)
 \\\nn=~&
  P^z_1 \int\!\df b^z_1 \, \df b^z_2 \, e^{\img x (b^z_1 P^z_1 + b^z_2 P^z_2)}
  \bar C_\ns^\prime(\mu, b^z_1 P^z_1, P^z_1) \tilde f_\ns(b^z_2, \bt, \mu, P^z_2)
  \ \exp\biggl[ \gamma_\zeta^q(\mu, b_T) \ln\frac{P^z_1}{P^z_2} \biggr]
\,.\end{align}
Next, we Fourier transform both sides from $x$ to $y$ by integrating over $x$ against $e^{-\img x y}$, obtaining
\begin{align} \label{eq:app:relation_qTMD_3}
 & \int\!\df b^z_1 \, \bar C_\ns^\prime(\mu, y - b^z_1 P^z_1, P^z_2) \tilde f_\ns(b^z_1, \bt, \mu, P^z_1)
 \\\nn=~&
  \int\! \df b^z_2 \, \bar C_\ns^\prime(\mu, y - b^z_2 P^z_2, P^z_1) \tilde f_\ns(b^z_2, \bt, \mu, P^z_2)
  \times \exp\biggl[ \gamma_\zeta^q(\mu, b_T) \ln\frac{P^z_1}{P^z_2} \biggr]
\,.\end{align}
This can trivially be solved for $\gamma_\zeta^q$ as
\begin{align} \label{eq:app:gamma_zeta_2}
 \gamma_\zeta^q(\mu, b_T) &=
 \frac{1}{\ln(P^z_1 / P^z_2)}
 \ln\frac{\int\!\df b^z \, \bar C_\ns^\prime(\mu, y - b^z P^z_1, P^z_2) \tilde f_\ns(b^z, \bt, \mu, P^z_1)}
         {\int\! \df b^z \, \bar C_\ns^\prime(\mu, y - b^z P^z_2, P^z_1) \tilde f_\ns(b^z, \bt, \mu, P^z_2)}
\,.\end{align}
Using the expression \eq{qtmdpdf} for $\tilde f_\ns$ and inserting a factor $\cR_B$
to separately cancel divergences in $b_T/a$, $L/a$ and $L/b_T$ in numerator and denominator, we obtain the final expression
\begin{align} \label{eq:app:gamma_zeta_3}
 &\gamma_\zeta^q(\mu, b_T) = \frac{1}{\ln(P^z_1/P^z_2)}
 \\\nn
&\times
 \ln\frac{\int\!\df b^z \, \bar C_\ns^\prime(\mu, y - b^z P^z_1, P^z_2) \tilde Z'_q(b^z,\mu,\tilde \mu)\,
 	\tilde Z_{\rm uv}^q(b^z,\tilde\mu,a) \, \cR_B(b_T,\tilde\mu,a,L) \, \tilde B_{\rm ns}(b^z, \bt, a, P^z_1, L)}
         {\int\! \df b^z \, \bar C_\ns^\prime(\mu, y - b^z P^z_2, P^z_1) \tilde Z'_q(b^z,\mu,\tilde \mu)\,
 	\tilde Z_{\rm uv}^q(b^z,\tilde\mu,a) \, \cR_B(b_T,\tilde\mu,a,L) \, \tilde B_{\rm ns}(b^z, \bt, a, P^z_2, L)}
\,.\end{align}
The key difference to \eq{gamma_zeta_4} is that in \eq{app:gamma_zeta_3}, both numerator and denominator depend on $P^z_1$ \emph{and} $P^z_2$, since $\bar C_\ns^\prime$ depends on both momenta.
In contrast, in \eq{gamma_zeta_4} the numerator only depends on $P^z_1$ and the denominator only depends on $P^z_2$, which makes the bookkeeping simpler for an analysis that separately determines the numerator and denominator before taking ratios.

\bibliographystyle{JHEP}
\bibliography{../literature}

\providecommand{\href}[2]{#2}\begingroup\raggedright\begin{thebibliography}{100}

\bibitem{Boer:2011fh}
D.~Boer et~al., \emph{{Gluons and the quark sea at high energies:
  Distributions, polarization, tomography}},
  \href{https://arxiv.org/abs/1108.1713}{{\ttfamily 1108.1713}}.

\bibitem{Accardi:2012qut}
A.~Accardi et~al., \emph{{Electron Ion Collider: The Next QCD Frontier}},
  {\emph{Eur. Phys. J.} {\bfseries A52} (2016) 268}
  [\href{https://arxiv.org/abs/1212.1701}{{\ttfamily 1212.1701}}].

\bibitem{Affolder:1999jh}
{\scshape CDF} collaboration, T.~Affolder et~al., \emph{{The transverse
  momentum and total cross section of $e^+e^-$ pairs in the $Z$ boson region
  from $p\bar{p}$ collisions at $\sqrt{s} = 1.8$ TeV}}, {\emph{Phys. Rev.
  Lett.} {\bfseries 84} (2000) 845}
  [\href{https://arxiv.org/abs/hep-ex/0001021}{{\ttfamily hep-ex/0001021}}].

\bibitem{Abbott:1999yd}
{\scshape D0} collaboration, B.~Abbott et~al., \emph{{Differential production
  cross section of $Z$ bosons as a function of transverse momentum at $\sqrt{s}
  = 1.8$ TeV}}, {\emph{Phys. Rev. Lett.} {\bfseries 84} (2000) 2792}
  [\href{https://arxiv.org/abs/hep-ex/9909020}{{\ttfamily hep-ex/9909020}}].

\bibitem{Abazov:2007ac}
{\scshape D0} collaboration, V.~M. Abazov et~al., \emph{{Measurement of the
  shape of the boson transverse momentum distribution in $p \bar{p} \to Z /
  \gamma^{*} \to e^+ e^- + X$ events produced at $\sqrt{s}=1.96$ TeV}},
  {\emph{Phys. Rev. Lett.} {\bfseries 100} (2008) 102002}
  [\href{https://arxiv.org/abs/0712.0803}{{\ttfamily 0712.0803}}].

\bibitem{Abazov:2010kn}
{\scshape D0} collaboration, V.~M. Abazov et~al., \emph{{Measurement of the
  normalized $Z/\gamma^* -> \mu^+\mu^-$ transverse momentum distribution in
  $p\bar{p}$ collisions at $\sqrt{s}=1.96$ TeV}}, {\emph{Phys. Lett.}
  {\bfseries B693} (2010) 522}
  [\href{https://arxiv.org/abs/1006.0618}{{\ttfamily 1006.0618}}].

\bibitem{Aad:2011gj}
{\scshape ATLAS} collaboration, G.~Aad et~al., \emph{{Measurement of the
  transverse momentum distribution of $Z/\gamma^*$ bosons in proton--proton
  collisions at $\sqrt{s}$=7 TeV with the ATLAS detector}}, {\emph{Phys. Lett.}
  {\bfseries B705} (2011) 415}
  [\href{https://arxiv.org/abs/1107.2381}{{\ttfamily 1107.2381}}].

\bibitem{Chatrchyan:2011wt}
{\scshape CMS} collaboration, S.~Chatrchyan et~al., \emph{{Measurement of the
  Rapidity and Transverse Momentum Distributions of $Z$ Bosons in $pp$
  Collisions at $\sqrt{s}=7$ TeV}}, {\emph{Phys. Rev.} {\bfseries D85} (2012)
  032002} [\href{https://arxiv.org/abs/1110.4973}{{\ttfamily 1110.4973}}].

\bibitem{Aad:2014xaa}
{\scshape ATLAS} collaboration, G.~Aad et~al., \emph{{Measurement of the
  $Z/\gamma^*$ boson transverse momentum distribution in $pp$ collisions at
  $\sqrt{s}$ = 7 TeV with the ATLAS detector}}, {\emph{JHEP} {\bfseries 09}
  (2014) 145} [\href{https://arxiv.org/abs/1406.3660}{{\ttfamily 1406.3660}}].

\bibitem{Khachatryan:2015oaa}
{\scshape CMS} collaboration, V.~Khachatryan et~al., \emph{{Measurement of the
  Z boson differential cross section in transverse momentum and rapidity in
  proton--proton collisions at 8 TeV}}, {\emph{Phys. Lett.} {\bfseries B749}
  (2015) 187} [\href{https://arxiv.org/abs/1504.03511}{{\ttfamily
  1504.03511}}].

\bibitem{Aad:2015auj}
{\scshape ATLAS} collaboration, G.~Aad et~al., \emph{{Measurement of the
  transverse momentum and $\phi ^*_{\eta }$ distributions of Drell--Yan lepton
  pairs in proton--proton collisions at $\sqrt{s}=8$ TeV with the ATLAS
  detector}}, {\emph{Eur. Phys. J.} {\bfseries C76} (2016) 291}
  [\href{https://arxiv.org/abs/1512.02192}{{\ttfamily 1512.02192}}].

\bibitem{Khachatryan:2016nbe}
{\scshape CMS} collaboration, V.~Khachatryan et~al., \emph{{Measurement of the
  transverse momentum spectra of weak vector bosons produced in proton-proton
  collisions at $ \sqrt{s}=8 $ TeV}}, {\emph{JHEP} {\bfseries 02} (2017) 096}
  [\href{https://arxiv.org/abs/1606.05864}{{\ttfamily 1606.05864}}].

\bibitem{Ashman:1991cj}
{\scshape European Muon} collaboration, J.~Ashman et~al., \emph{{Forward
  produced hadrons in $\mu p$ and $\mu d$ scattering and investigation of the
  charge structure of the nucleon}}, {\emph{Z. Phys.} {\bfseries C52} (1991)
  361}.

\bibitem{Derrick:1995xg}
{\scshape ZEUS} collaboration, M.~Derrick et~al., \emph{{Inclusive charged
  particle distributions in deep inelastic scattering events at HERA}},
  {\emph{Z. Phys.} {\bfseries C70} (1996) 1}
  [\href{https://arxiv.org/abs/hep-ex/9511010}{{\ttfamily hep-ex/9511010}}].

\bibitem{Adloff:1996dy}
{\scshape H1} collaboration, C.~Adloff et~al., \emph{{Measurement of charged
  particle transverse momentum spectra in deep inelastic scattering}},
  {\emph{Nucl. Phys.} {\bfseries B485} (1997) 3}
  [\href{https://arxiv.org/abs/hep-ex/9610006}{{\ttfamily hep-ex/9610006}}].

\bibitem{Aaron:2008ad}
{\scshape H1} collaboration, F.~D. Aaron et~al., \emph{{Measurement of the
  Proton Structure Function $F_L(x,Q^2)$ at Low $x$}}, {\emph{Phys. Lett.}
  {\bfseries B665} (2008) 139}
  [\href{https://arxiv.org/abs/0805.2809}{{\ttfamily 0805.2809}}].

\bibitem{Airapetian:2012ki}
{\scshape HERMES} collaboration, A.~Airapetian et~al., \emph{{Multiplicities of
  charged pions and kaons from semi-inclusive deep-inelastic scattering by the
  proton and the deuteron}}, {\emph{Phys. Rev.} {\bfseries D87} (2013) 074029}
  [\href{https://arxiv.org/abs/1212.5407}{{\ttfamily 1212.5407}}].

\bibitem{Adolph:2013stb}
{\scshape COMPASS} collaboration, C.~Adolph et~al., \emph{{Hadron Transverse
  Momentum Distributions in Muon Deep Inelastic Scattering at 160 GeV/$c$}},
  {\emph{Eur. Phys. J.} {\bfseries C73} (2013) 2531}
  [\href{https://arxiv.org/abs/1305.7317}{{\ttfamily 1305.7317}}].

\bibitem{Aghasyan:2017ctw}
{\scshape COMPASS} collaboration, M.~Aghasyan et~al.,
  \emph{{Transverse-momentum-dependent Multiplicities of Charged Hadrons in
  Muon-Deuteron Deep Inelastic Scattering}}, {\emph{Phys. Rev.} {\bfseries D97}
  (2018) 032006} [\href{https://arxiv.org/abs/1709.07374}{{\ttfamily
  1709.07374}}].

\bibitem{Catani:2011kr}
S.~Catani and M.~Grazzini, \emph{{Higgs Boson Production at Hadron Colliders:
  Hard-Collinear Coefficients at the NNLO}}, {\emph{Eur. Phys. J.} {\bfseries
  C72} (2012) 2013} [\href{https://arxiv.org/abs/1106.4652}{{\ttfamily
  1106.4652}}].

\bibitem{Catani:2012qa}
S.~Catani, L.~Cieri, D.~de~Florian, G.~Ferrera and M.~Grazzini, \emph{{Vector
  boson production at hadron colliders: hard-collinear coefficients at the
  NNLO}}, {\emph{Eur. Phys. J.} {\bfseries C72} (2012) 2195}
  [\href{https://arxiv.org/abs/1209.0158}{{\ttfamily 1209.0158}}].

\bibitem{Gehrmann:2014yya}
T.~Gehrmann, T.~Luebbert and L.~L. Yang, \emph{{Calculation of the transverse
  parton distribution functions at next-to-next-to-leading order}},
  {\emph{JHEP} {\bfseries 06} (2014) 155}
  [\href{https://arxiv.org/abs/1403.6451}{{\ttfamily 1403.6451}}].

\bibitem{Luebbert:2016itl}
T.~L{\"u}bbert, J.~Oredsson and M.~Stahlhofen, \emph{{Rapidity renormalized TMD
  soft and beam functions at two loops}}, {\emph{JHEP} {\bfseries 03} (2016)
  168} [\href{https://arxiv.org/abs/1602.01829}{{\ttfamily 1602.01829}}].

\bibitem{Echevarria:2015byo}
M.~G. Echevarria, I.~Scimemi and A.~Vladimirov, \emph{{Universal transverse
  momentum dependent soft function at NNLO}}, {\emph{Phys. Rev.} {\bfseries
  D93} (2016) 054004} [\href{https://arxiv.org/abs/1511.05590}{{\ttfamily
  1511.05590}}].

\bibitem{Echevarria:2016scs}
M.~G. Echevarria, I.~Scimemi and A.~Vladimirov, \emph{{Unpolarized Transverse
  Momentum Dependent Parton Distribution and Fragmentation Functions at
  next-to-next-to-leading order}}, {\emph{JHEP} {\bfseries 09} (2016) 004}
  [\href{https://arxiv.org/abs/1604.07869}{{\ttfamily 1604.07869}}].

\bibitem{Luo:2019hmp}
M.-X. Luo, X.~Wang, X.~Xu, L.~L. Yang, T.-Z. Yang and H.~X. Zhu,
  \emph{{Transverse Parton Distribution and Fragmentation Functions at NNLO:
  the Quark Case}}, \href{https://doi.org/10.1007/JHEP10(2019)083}{\emph{JHEP}
  {\bfseries 10} (2019) 083}
  [\href{https://arxiv.org/abs/1908.03831}{{\ttfamily 1908.03831}}].

\bibitem{Luo:2019bmw}
M.-X. Luo, T.-Z. Yang, H.~X. Zhu and Y.~J. Zhu, \emph{{Transverse Parton
  Distribution and Fragmentation Functions at NNLO: the Gluon Case}},
  \href{https://arxiv.org/abs/1909.13820}{{\ttfamily 1909.13820}}.

\bibitem{Landry:1999an}
F.~Landry, R.~Brock, G.~Ladinsky and C.~P. Yuan, \emph{{New fits for the
  nonperturbative parameters in the CSS resummation formalism}},
  \href{https://doi.org/10.1103/PhysRevD.63.013004}{\emph{Phys. Rev.}
  {\bfseries D63} (2001) 013004}
  [\href{https://arxiv.org/abs/hep-ph/9905391}{{\ttfamily hep-ph/9905391}}].

\bibitem{Landry:2002ix}
F.~Landry, R.~Brock, P.~M. Nadolsky and C.~P. Yuan, \emph{{Tevatron Run-1 $Z$
  boson data and Collins-Soper-Sterman resummation formalism}},
  \href{https://doi.org/10.1103/PhysRevD.67.073016}{\emph{Phys. Rev.}
  {\bfseries D67} (2003) 073016}
  [\href{https://arxiv.org/abs/hep-ph/0212159}{{\ttfamily hep-ph/0212159}}].

\bibitem{Konychev:2005iy}
A.~V. Konychev and P.~M. Nadolsky, \emph{{Universality of the
  Collins-Soper-Sterman nonperturbative function in gauge boson production}},
  \href{https://doi.org/10.1016/j.physletb.2005.12.063}{\emph{Phys. Lett.}
  {\bfseries B633} (2006) 710}
  [\href{https://arxiv.org/abs/hep-ph/0506225}{{\ttfamily hep-ph/0506225}}].

\bibitem{DAlesio:2014mrz}
U.~D'Alesio, M.~G. Echevarria, S.~Melis and I.~Scimemi, \emph{{Non-perturbative
  QCD effects in $q_{T}$ spectra of Drell-Yan and Z-boson production}},
  \href{https://doi.org/10.1007/JHEP11(2014)098}{\emph{JHEP} {\bfseries 11}
  (2014) 098} [\href{https://arxiv.org/abs/1407.3311}{{\ttfamily 1407.3311}}].

\bibitem{Bacchetta:2017gcc}
A.~Bacchetta, F.~Delcarro, C.~Pisano, M.~Radici and A.~Signori,
  \emph{{Extraction of partonic transverse momentum distributions from
  semi-inclusive deep-inelastic scattering, Drell-Yan and Z-boson production}},
  {\emph{JHEP} {\bfseries 06} (2017) 081}
  [\href{https://arxiv.org/abs/1703.10157}{{\ttfamily 1703.10157}}].

\bibitem{Scimemi:2017etj}
I.~Scimemi and A.~Vladimirov, \emph{{Analysis of vector boson production within
  TMD factorization}}, {\emph{Eur. Phys. J.} {\bfseries C78} (2018) 89}
  [\href{https://arxiv.org/abs/1706.01473}{{\ttfamily 1706.01473}}].

\bibitem{Musch:2010ka}
B.~U. Musch, P.~H{\"a}gler, J.~W. Negele and A.~Sch{\"a}fer, \emph{{Exploring
  quark transverse momentum distributions with lattice QCD}}, {\emph{Phys.
  Rev.} {\bfseries D83} (2011) 094507}
  [\href{https://arxiv.org/abs/1011.1213}{{\ttfamily 1011.1213}}].

\bibitem{Musch:2011er}
B.~U. Musch, P.~H{\"a}gler, M.~Engelhardt, J.~W. Negele and A.~Sch{\"a}fer,
  \emph{{Sivers and Boer-Mulders observables from lattice QCD}}, {\emph{Phys.
  Rev.} {\bfseries D85} (2012) 094510}
  [\href{https://arxiv.org/abs/1111.4249}{{\ttfamily 1111.4249}}].

\bibitem{Engelhardt:2015xja}
M.~Engelhardt, P.~H{\"a}gler, B.~Musch, J.~Negele and A.~Sch{\"a}fer,
  \emph{{Lattice QCD study of the Boer-Mulders effect in a pion}}, {\emph{Phys.
  Rev.} {\bfseries D93} (2016) 054501}
  [\href{https://arxiv.org/abs/1506.07826}{{\ttfamily 1506.07826}}].

\bibitem{Yoon:2016dyh}
B.~Yoon, T.~Bhattacharya, M.~Engelhardt, J.~Green, R.~Gupta, P.~H{\"a}gler
  et~al., \emph{{Lattice QCD calculations of nucleon transverse
  momentum-dependent parton distributions using clover and domain wall
  fermions}},  in \emph{{Proceedings, 33rd International Symposium on Lattice
  Field Theory (Lattice 2015): Kobe, Japan, July 14-18, 2015}}, SISSA, SISSA,
  2015, \href{https://arxiv.org/abs/1601.05717}{{\ttfamily 1601.05717}}.

\bibitem{Yoon:2017qzo}
B.~Yoon, M.~Engelhardt, R.~Gupta, T.~Bhattacharya, J.~R. Green, B.~U. Musch
  et~al., \emph{{Nucleon Transverse Momentum-dependent Parton Distributions in
  Lattice QCD: Renormalization Patterns and Discretization Effects}},
  {\emph{Phys. Rev.} {\bfseries D96} (2017) 094508}
  [\href{https://arxiv.org/abs/1706.03406}{{\ttfamily 1706.03406}}].

\bibitem{Collins:1981va}
J.~C. Collins and D.~E. Soper, \emph{{Back-To-Back Jets: Fourier Transform from
  B to K-Transverse}}, {\emph{Nucl. Phys.} {\bfseries B197} (1982) 446}.

\bibitem{Collins:1981uk}
J.~C. Collins and D.~E. Soper, \emph{{Back-To-Back Jets in QCD}}, {\emph{Nucl.
  Phys.} {\bfseries B193} (1981) 381}.

\bibitem{Korchemsky:1987wg}
G.~P. Korchemsky and A.~V. Radyushkin, \emph{{Renormalization of the Wilson
  Loops Beyond the Leading Order}},
  \href{https://doi.org/10.1016/0550-3213(87)90277-X}{\emph{Nucl. Phys.}
  {\bfseries B283} (1987) 342}.

\bibitem{Moch:2004pa}
S.~Moch, J.~A.~M. Vermaseren and A.~Vogt, \emph{{The Three loop splitting
  functions in QCD: The Nonsinglet case}},
  \href{https://doi.org/10.1016/j.nuclphysb.2004.03.030}{\emph{Nucl. Phys.}
  {\bfseries B688} (2004) 101}
  [\href{https://arxiv.org/abs/hep-ph/0403192}{{\ttfamily hep-ph/0403192}}].

\bibitem{Vogt:2004mw}
A.~Vogt, S.~Moch and J.~A.~M. Vermaseren, \emph{{The Three-loop splitting
  functions in QCD: The Singlet case}},
  \href{https://doi.org/10.1016/j.nuclphysb.2004.04.024}{\emph{Nucl. Phys.}
  {\bfseries B691} (2004) 129}
  [\href{https://arxiv.org/abs/hep-ph/0404111}{{\ttfamily hep-ph/0404111}}].

\bibitem{Davies:1984hs}
C.~T.~H. Davies and W.~J. Stirling, \emph{{Nonleading Corrections to the
  Drell-Yan Cross-Section at Small Transverse Momentum}}, {\emph{Nucl. Phys.}
  {\bfseries B244} (1984) 337}.

\bibitem{Davies:1984sp}
C.~T.~H. Davies, B.~R. Webber and W.~J. Stirling, \emph{{Drell-Yan
  Cross-Sections at Small Transverse Momentum}}, {\emph{Nucl. Phys.} {\bfseries
  B256} (1985) 413}.

\bibitem{deFlorian:2000pr}
D.~de~Florian and M.~Grazzini, \emph{{Next-to-next-to-leading logarithmic
  corrections at small transverse momentum in hadronic collisions}},
  {\emph{Phys. Rev. Lett.} {\bfseries 85} (2000) 4678}
  [\href{https://arxiv.org/abs/hep-ph/0008152}{{\ttfamily hep-ph/0008152}}].

\bibitem{Becher:2010tm}
T.~Becher and M.~Neubert, \emph{{{Drell-Yan} Production at Small $q_T$,
  Transverse Parton Distributions and the Collinear Anomaly}}, {\emph{Eur.
  Phys. J.} {\bfseries C71} (2011) 1665}
  [\href{https://arxiv.org/abs/1007.4005}{{\ttfamily 1007.4005}}].

\bibitem{Li:2016axz}
Y.~Li, D.~Neill and H.~X. Zhu, \emph{{An Exponential Regulator for Rapidity
  Divergences}},  \href{https://arxiv.org/abs/1604.00392}{{\ttfamily
  1604.00392}}.

\bibitem{Li:2016ctv}
Y.~Li and H.~X. Zhu, \emph{{Bootstrapping Rapidity Anomalous Dimensions for
  Transverse-Momentum Resummation}}, {\emph{Phys. Rev. Lett.} {\bfseries 118}
  (2017) 022004} [\href{https://arxiv.org/abs/1604.01404}{{\ttfamily
  1604.01404}}].

\bibitem{Vladimirov:2016dll}
A.~A. Vladimirov, \emph{{Correspondence between Soft and Rapidity Anomalous
  Dimensions}}, {\emph{Phys. Rev. Lett.} {\bfseries 118} (2017) 062001}
  [\href{https://arxiv.org/abs/1610.05791}{{\ttfamily 1610.05791}}].

\bibitem{Moch:2017uml}
S.~Moch, B.~Ruijl, T.~Ueda, J.~A.~M. Vermaseren and A.~Vogt, \emph{{Four-Loop
  Non-Singlet Splitting Functions in the Planar Limit and Beyond}},
  \href{https://doi.org/10.1007/JHEP10(2017)041}{\emph{JHEP} {\bfseries 10}
  (2017) 041} [\href{https://arxiv.org/abs/1707.08315}{{\ttfamily
  1707.08315}}].

\bibitem{Moch:2018wjh}
S.~Moch, B.~Ruijl, T.~Ueda, J.~A.~M. Vermaseren and A.~Vogt, \emph{{On quartic
  colour factors in splitting functions and the gluon cusp anomalous
  dimension}},
  \href{https://doi.org/10.1016/j.physletb.2018.06.017}{\emph{Phys. Lett.}
  {\bfseries B782} (2018) 627}
  [\href{https://arxiv.org/abs/1805.09638}{{\ttfamily 1805.09638}}].

\bibitem{Grozin:2016ydd}
A.~Grozin, \emph{{Leading and next-to-leading large-$n_f$ terms in the cusp
  anomalous dimension and quark-antiquark potential}},
  \href{https://doi.org/10.22323/1.260.0053}{\emph{PoS} {\bfseries LL2016}
  (2016) 053} [\href{https://arxiv.org/abs/1605.03886}{{\ttfamily
  1605.03886}}].

\bibitem{Henn:2016men}
J.~M. Henn, A.~V. Smirnov, V.~A. Smirnov and M.~Steinhauser, \emph{{A planar
  four-loop form factor and cusp anomalous dimension in QCD}},
  \href{https://doi.org/10.1007/JHEP05(2016)066}{\emph{JHEP} {\bfseries 05}
  (2016) 066} [\href{https://arxiv.org/abs/1604.03126}{{\ttfamily
  1604.03126}}].

\bibitem{Davies:2016jie}
J.~Davies, A.~Vogt, B.~Ruijl, T.~Ueda and J.~A.~M. Vermaseren,
  \emph{{Large-$n_f$ contributions to the four-loop splitting functions in
  QCD}}, \href{https://doi.org/10.1016/j.nuclphysb.2016.12.012}{\emph{Nucl.
  Phys.} {\bfseries B915} (2017) 335}
  [\href{https://arxiv.org/abs/1610.07477}{{\ttfamily 1610.07477}}].

\bibitem{Lee:2016ixa}
J.~Henn, A.~V. Smirnov, V.~A. Smirnov, M.~Steinhauser and R.~N. Lee,
  \emph{{Four-loop photon quark form factor and cusp anomalous dimension in the
  large-$N_c$ limit of QCD}},
  \href{https://doi.org/10.1007/JHEP03(2017)139}{\emph{JHEP} {\bfseries 03}
  (2017) 139} [\href{https://arxiv.org/abs/1612.04389}{{\ttfamily
  1612.04389}}].

\bibitem{Grozin:2018vdn}
A.~Grozin, \emph{{Four-loop cusp anomalous dimension in QED}},
  \href{https://doi.org/10.1007/JHEP06(2018)073,
  10.1007/JHEP01(2019)134}{\emph{JHEP} {\bfseries 06} (2018) 073}
  [\href{https://arxiv.org/abs/1805.05050}{{\ttfamily 1805.05050}}].

\bibitem{Lee:2019zop}
R.~N. Lee, A.~V. Smirnov, V.~A. Smirnov and M.~Steinhauser, \emph{{Four-loop
  quark form factor with quartic fundamental colour factor}},
  \href{https://doi.org/10.1007/JHEP02(2019)172}{\emph{JHEP} {\bfseries 02}
  (2019) 172} [\href{https://arxiv.org/abs/1901.02898}{{\ttfamily
  1901.02898}}].

\bibitem{Henn:2019rmi}
J.~M. Henn, T.~Peraro, M.~Stahlhofen and P.~Wasser, \emph{{Matter dependence of
  the four-loop cusp anomalous dimension}},
  \href{https://doi.org/10.1103/PhysRevLett.122.201602}{\emph{Phys. Rev. Lett.}
  {\bfseries 122} (2019) 201602}
  [\href{https://arxiv.org/abs/1901.03693}{{\ttfamily 1901.03693}}].

\bibitem{Bruser:2019auj}
R.~Br{\"u}ser, A.~Grozin, J.~M. Henn and M.~Stahlhofen, \emph{{Matter
  dependence of the four-loop QCD cusp anomalous dimension: from small angles
  to all angles}}, \href{https://doi.org/10.1007/JHEP05(2019)186}{\emph{JHEP}
  {\bfseries 05} (2019) 186}
  [\href{https://arxiv.org/abs/1902.05076}{{\ttfamily 1902.05076}}].

\bibitem{Scimemi:2016ffw}
I.~Scimemi and A.~Vladimirov, \emph{{Power corrections and renormalons in
  Transverse Momentum Distributions}}, {\emph{JHEP} {\bfseries 03} (2017) 002}
  [\href{https://arxiv.org/abs/1609.06047}{{\ttfamily 1609.06047}}].

\bibitem{Bertone:2019nxa}
V.~Bertone, I.~Scimemi and A.~Vladimirov, \emph{{Extraction of unpolarized
  quark transverse momentum dependent parton distributions from
  Drell-Yan/Z-boson production}},
  \href{https://doi.org/10.1007/JHEP06(2019)028}{\emph{JHEP} {\bfseries 06}
  (2019) 028} [\href{https://arxiv.org/abs/1902.08474}{{\ttfamily
  1902.08474}}].

\bibitem{Vladimirov:2019bfa}
A.~Vladimirov, \emph{{Pion-induced Drell-Yan processes within TMD
  factorization}}, \href{https://doi.org/10.1007/JHEP10(2019)090}{\emph{JHEP}
  {\bfseries 10} (2019) 090}
  [\href{https://arxiv.org/abs/1907.10356}{{\ttfamily 1907.10356}}].

\bibitem{Ji:2013dva}
X.~Ji, \emph{{Parton Physics on a Euclidean Lattice}}, {\emph{Phys. Rev. Lett.}
  {\bfseries 110} (2013) 262002}
  [\href{https://arxiv.org/abs/1305.1539}{{\ttfamily 1305.1539}}].

\bibitem{Ji:2014gla}
X.~Ji, \emph{{Parton Physics from Large-Momentum Effective Field Theory}},
  {\emph{Sci. China Phys. Mech. Astron.} {\bfseries 57} (2014) 1407}
  [\href{https://arxiv.org/abs/1404.6680}{{\ttfamily 1404.6680}}].

\bibitem{Xiong:2013bka}
X.~Xiong, X.~Ji, J.-H. Zhang and Y.~Zhao, \emph{{One-loop matching for parton
  distributions: Nonsinglet case}}, {\emph{Phys. Rev.} {\bfseries D90} (2014)
  014051} [\href{https://arxiv.org/abs/1310.7471}{{\ttfamily 1310.7471}}].

\bibitem{Ma:2014jla}
Y.-Q. Ma and J.-W. Qiu, \emph{{Extracting Parton Distribution Functions from
  Lattice QCD Calculations}}, {\emph{Phys. Rev.} {\bfseries D98} (2018) 074021}
  [\href{https://arxiv.org/abs/1404.6860}{{\ttfamily 1404.6860}}].

\bibitem{Ma:2014jga}
Y.-Q. Ma and J.-W. Qiu, \emph{{QCD Factorization and PDFs from Lattice QCD
  Calculation}}, {\emph{Int. J. Mod. Phys. Conf. Ser.} {\bfseries 37} (2015)
  1560041} [\href{https://arxiv.org/abs/1412.2688}{{\ttfamily 1412.2688}}].

\bibitem{Ji:2015jwa}
X.~Ji and J.-H. Zhang, \emph{{Renormalization of quasiparton distribution}},
  {\emph{Phys. Rev.} {\bfseries D92} (2015) 034006}
  [\href{https://arxiv.org/abs/1505.07699}{{\ttfamily 1505.07699}}].

\bibitem{Ji:2015qla}
X.~Ji, A.~Sch{\"a}fer, X.~Xiong and J.-H. Zhang, \emph{{One-Loop Matching for
  Generalized Parton Distributions}}, {\emph{Phys. Rev.} {\bfseries D92} (2015)
  014039} [\href{https://arxiv.org/abs/1506.00248}{{\ttfamily 1506.00248}}].

\bibitem{Xiong:2015nua}
X.~Xiong and J.-H. Zhang, \emph{{One-loop matching for transversity generalized
  parton distribution}}, {\emph{Phys. Rev.} {\bfseries D92} (2015) 054037}
  [\href{https://arxiv.org/abs/1509.08016}{{\ttfamily 1509.08016}}].

\bibitem{Li:2016amo}
H.-n. Li, \emph{{Nondipolar Wilson links for quasiparton distribution
  functions}}, {\emph{Phys. Rev.} {\bfseries D94} (2016) 074036}
  [\href{https://arxiv.org/abs/1602.07575}{{\ttfamily 1602.07575}}].

\bibitem{Ishikawa:2016znu}
T.~Ishikawa, Y.-Q. Ma, J.-W. Qiu and S.~Yoshida, \emph{{Practical quasi parton
  distribution functions}},  \href{https://arxiv.org/abs/1609.02018}{{\ttfamily
  1609.02018}}.

\bibitem{Chen:2016fxx}
J.-W. Chen, X.~Ji and J.-H. Zhang, \emph{{Improved quasi parton distribution
  through Wilson line renormalization}}, {\emph{Nucl. Phys.} {\bfseries B915}
  (2017) 1} [\href{https://arxiv.org/abs/1609.08102}{{\ttfamily 1609.08102}}].

\bibitem{Carlson:2017gpk}
C.~E. Carlson and M.~Freid, \emph{{Lattice corrections to the quark
  quasidistribution at one-loop}}, {\emph{Phys. Rev.} {\bfseries D95} (2017)
  094504} [\href{https://arxiv.org/abs/1702.05775}{{\ttfamily 1702.05775}}].

\bibitem{Briceno:2017cpo}
R.~A. Brice{\~n}o, M.~T. Hansen and C.~J. Monahan, \emph{{Role of the Euclidean
  signature in lattice calculations of quasidistributions and other nonlocal
  matrix elements}}, {\emph{Phys. Rev.} {\bfseries D96} (2017) 014502}
  [\href{https://arxiv.org/abs/1703.06072}{{\ttfamily 1703.06072}}].

\bibitem{Xiong:2017jtn}
X.~Xiong, T.~Luu and U.-G. Mei{\ss}ner, \emph{{Quasi-Parton Distribution
  Function in Lattice Perturbation Theory}},
  \href{https://arxiv.org/abs/1705.00246}{{\ttfamily 1705.00246}}.

\bibitem{Constantinou:2017sej}
M.~Constantinou and H.~Panagopoulos, \emph{{Perturbative renormalization of
  quasi-parton distribution functions}}, {\emph{Phys. Rev.} {\bfseries D96}
  (2017) 054506} [\href{https://arxiv.org/abs/1705.11193}{{\ttfamily
  1705.11193}}].

\bibitem{Rossi:2017muf}
G.~C. Rossi and M.~Testa, \emph{{Note on lattice regularization and equal-time
  correlators for parton distribution functions}}, {\emph{Phys. Rev.}
  {\bfseries D96} (2017) 014507}
  [\href{https://arxiv.org/abs/1706.04428}{{\ttfamily 1706.04428}}].

\bibitem{Ji:2017rah}
X.~Ji, J.-H. Zhang and Y.~Zhao, \emph{{More On Large-Momentum Effective Theory
  Approach to Parton Physics}}, {\emph{Nucl. Phys.} {\bfseries B924} (2017)
  366} [\href{https://arxiv.org/abs/1706.07416}{{\ttfamily 1706.07416}}].

\bibitem{Ji:2017oey}
X.~Ji, J.-H. Zhang and Y.~Zhao, \emph{{Renormalization in Large Momentum
  Effective Theory of Parton Physics}}, {\emph{Phys. Rev. Lett.} {\bfseries
  120} (2018) 112001} [\href{https://arxiv.org/abs/1706.08962}{{\ttfamily
  1706.08962}}].

\bibitem{Ishikawa:2017faj}
T.~Ishikawa, Y.-Q. Ma, J.-W. Qiu and S.~Yoshida, \emph{{Renormalizability of
  quasiparton distribution functions}}, {\emph{Phys. Rev.} {\bfseries D96}
  (2017) 094019} [\href{https://arxiv.org/abs/1707.03107}{{\ttfamily
  1707.03107}}].

\bibitem{Green:2017xeu}
J.~Green, K.~Jansen and F.~Steffens, \emph{{Nonperturbative Renormalization of
  Nonlocal Quark Bilinears for Parton Quasidistribution Functions on the
  Lattice Using an Auxiliary Field}}, {\emph{Phys. Rev. Lett.} {\bfseries 121}
  (2018) 022004} [\href{https://arxiv.org/abs/1707.07152}{{\ttfamily
  1707.07152}}].

\bibitem{Wang:2017qyg}
W.~Wang, S.~Zhao and R.~Zhu, \emph{{Gluon quasidistribution function at one
  loop}}, {\emph{Eur. Phys. J.} {\bfseries C78} (2018) 147}
  [\href{https://arxiv.org/abs/1708.02458}{{\ttfamily 1708.02458}}].

\bibitem{Chen:2017mie}
J.-W. Chen, T.~Ishikawa, L.~Jin, H.-W. Lin, J.-H. Zhang and Y.~Zhao,
  \emph{{Symmetry Properties of Nonlocal Quark Bilinear Operators on a
  Lattice}}, \href{https://doi.org/10.1088/1674-1137/43/10/103101}{\emph{Chin.
  Phys.} {\bfseries C43} (2019) 103101}
  [\href{https://arxiv.org/abs/1710.01089}{{\ttfamily 1710.01089}}].

\bibitem{Stewart:2017tvs}
I.~W. Stewart and Y.~Zhao, \emph{{Matching the quasiparton distribution in a
  momentum subtraction scheme}}, {\emph{Phys. Rev.} {\bfseries D97} (2018)
  054512} [\href{https://arxiv.org/abs/1709.04933}{{\ttfamily 1709.04933}}].

\bibitem{Wang:2017eel}
W.~Wang and S.~Zhao, \emph{{On the power divergence in quasi gluon distribution
  function}}, {\emph{JHEP} {\bfseries 05} (2018) 142}
  [\href{https://arxiv.org/abs/1712.09247}{{\ttfamily 1712.09247}}].

\bibitem{Spanoudes:2018zya}
G.~Spanoudes and H.~Panagopoulos, \emph{{Renormalization of Wilson-line
  operators in the presence of nonzero quark masses}}, {\emph{Phys. Rev.}
  {\bfseries D98} (2018) 014509}
  [\href{https://arxiv.org/abs/1805.01164}{{\ttfamily 1805.01164}}].

\bibitem{Izubuchi:2018srq}
T.~Izubuchi, X.~Ji, L.~Jin, I.~W. Stewart and Y.~Zhao, \emph{{Factorization
  Theorem Relating Euclidean and Light-Cone Parton Distributions}},
  {\emph{Phys. Rev.} {\bfseries D98} (2018) 056004}
  [\href{https://arxiv.org/abs/1801.03917}{{\ttfamily 1801.03917}}].

\bibitem{Xu:2018mpf}
J.~Xu, Q.-A. Zhang and S.~Zhao, \emph{{Light-cone distribution amplitudes of
  vector meson in a large momentum effective theory}},
  \href{https://doi.org/10.1103/PhysRevD.97.114026}{\emph{Phys. Rev.}
  {\bfseries D97} (2018) 114026}
  [\href{https://arxiv.org/abs/1804.01042}{{\ttfamily 1804.01042}}].

\bibitem{Rossi:2018zkn}
G.~Rossi and M.~Testa, \emph{{Euclidean versus Minkowski short distance}},
  \href{https://doi.org/10.1103/PhysRevD.98.054028}{\emph{Phys. Rev.}
  {\bfseries D98} (2018) 054028}
  [\href{https://arxiv.org/abs/1806.00808}{{\ttfamily 1806.00808}}].

\bibitem{Zhang:2018diq}
J.-H. Zhang, X.~Ji, A.~Sch{\"a}fer, W.~Wang and S.~Zhao, \emph{{Accessing Gluon
  Parton Distributions in Large Momentum Effective Theory}}, {\emph{Phys. Rev.
  Lett.} {\bfseries 122} (2019) 142001}
  [\href{https://arxiv.org/abs/1808.10824}{{\ttfamily 1808.10824}}].

\bibitem{Li:2018tpe}
Z.-Y. Li, Y.-Q. Ma and J.-W. Qiu, \emph{{Multiplicative Renormalizability of
  Operators defining Quasiparton Distributions}}, {\emph{Phys. Rev. Lett.}
  {\bfseries 122} (2019) 062002}
  [\href{https://arxiv.org/abs/1809.01836}{{\ttfamily 1809.01836}}].

\bibitem{Liu:2018tox}
Y.-S. Liu, W.~Wang, J.~Xu, Q.-A. Zhang, S.~Zhao and Y.~Zhao, \emph{{Matching
  the meson quasidistribution amplitude in the RI/MOM scheme}}, {\emph{Phys.
  Rev.} {\bfseries D99} (2019) 094036}
  [\href{https://arxiv.org/abs/1810.10879}{{\ttfamily 1810.10879}}].

\bibitem{Chen:2016utp}
J.-W. Chen, S.~D. Cohen, X.~Ji, H.-W. Lin and J.-H. Zhang, \emph{{Nucleon
  Helicity and Transversity Parton Distributions from Lattice QCD}},
  {\emph{Nucl. Phys.} {\bfseries B911} (2016) 246}
  [\href{https://arxiv.org/abs/1603.06664}{{\ttfamily 1603.06664}}].

\bibitem{Radyushkin:2017ffo}
A.~Radyushkin, \emph{{Target Mass Effects in Parton Quasi-Distributions}},
  {\emph{Phys. Lett.} {\bfseries B770} (2017) 514}
  [\href{https://arxiv.org/abs/1702.01726}{{\ttfamily 1702.01726}}].

\bibitem{Braun:2018brg}
V.~M. Braun, A.~Vladimirov and J.-H. Zhang, \emph{{Power corrections and
  renormalons in parton quasidistributions}}, {\emph{Phys. Rev.} {\bfseries
  D99} (2019) 014013} [\href{https://arxiv.org/abs/1810.00048}{{\ttfamily
  1810.00048}}].

\bibitem{Lin:2014zya}
H.-W. Lin, J.-W. Chen, S.~D. Cohen and X.~Ji, \emph{{Flavor Structure of the
  Nucleon Sea from Lattice QCD}}, {\emph{Phys. Rev.} {\bfseries D91} (2015)
  054510} [\href{https://arxiv.org/abs/1402.1462}{{\ttfamily 1402.1462}}].

\bibitem{Alexandrou:2015rja}
C.~Alexandrou, K.~Cichy, V.~Drach, E.~Garcia-Ramos, K.~Hadjiyiannakou,
  K.~Jansen et~al., \emph{{Lattice calculation of parton distributions}},
  {\emph{Phys. Rev.} {\bfseries D92} (2015) 014502}
  [\href{https://arxiv.org/abs/1504.07455}{{\ttfamily 1504.07455}}].

\bibitem{Alexandrou:2016jqi}
C.~Alexandrou, K.~Cichy, M.~Constantinou, K.~Hadjiyiannakou, K.~Jansen,
  F.~Steffens et~al., \emph{{Updated Lattice Results for Parton
  Distributions}}, {\emph{Phys. Rev.} {\bfseries D96} (2017) 014513}
  [\href{https://arxiv.org/abs/1610.03689}{{\ttfamily 1610.03689}}].

\bibitem{Zhang:2017bzy}
J.-H. Zhang, J.-W. Chen, X.~Ji, L.~Jin and H.-W. Lin, \emph{{Pion Distribution
  Amplitude from Lattice QCD}}, {\emph{Phys. Rev.} {\bfseries D95} (2017)
  094514} [\href{https://arxiv.org/abs/1702.00008}{{\ttfamily 1702.00008}}].

\bibitem{Alexandrou:2017huk}
C.~Alexandrou, K.~Cichy, M.~Constantinou, K.~Hadjiyiannakou, K.~Jansen,
  H.~Panagopoulos et~al., \emph{{A complete non-perturbative renormalization
  prescription for quasi-PDFs}}, {\emph{Nucl. Phys.} {\bfseries B923} (2017)
  394} [\href{https://arxiv.org/abs/1706.00265}{{\ttfamily 1706.00265}}].

\bibitem{Chen:2017mzz}
J.-W. Chen, T.~Ishikawa, L.~Jin, H.-W. Lin, Y.-B. Yang, J.-H. Zhang et~al.,
  \emph{{Parton distribution function with nonperturbative renormalization from
  lattice QCD}}, {\emph{Phys. Rev.} {\bfseries D97} (2018) 014505}
  [\href{https://arxiv.org/abs/1706.01295}{{\ttfamily 1706.01295}}].

\bibitem{Chen:2017lnm}
T.~Ishikawa, L.~Jin, H.-W. Lin, A.~Sch{\"a}fer, Y.-B. Yang, J.-H. Zhang et~al.,
  \emph{{Gaussian-weighted parton quasi-distribution (Lattice Parton Physics
  Project (LP$^{3}$))}}, {\emph{Sci. China Phys. Mech. Astron.} {\bfseries 62}
  (2019) 991021} [\href{https://arxiv.org/abs/1711.07858}{{\ttfamily
  1711.07858}}].

\bibitem{Chen:2017gck}
{\scshape LP3} collaboration, J.-H. Zhang, L.~Jin, H.-W. Lin, A.~Sch{\"a}fer,
  P.~Sun, Y.-B. Yang et~al., \emph{{Kaon Distribution Amplitude from Lattice
  QCD and the Flavor SU(3) Symmetry}}, {\emph{Nucl. Phys.} {\bfseries B939}
  (2019) 429} [\href{https://arxiv.org/abs/1712.10025}{{\ttfamily
  1712.10025}}].

\bibitem{Alexandrou:2018pbm}
C.~Alexandrou, K.~Cichy, M.~Constantinou, K.~Jansen, A.~Scapellato and
  F.~Steffens, \emph{{Light-Cone Parton Distribution Functions from Lattice
  QCD}}, {\emph{Phys. Rev. Lett.} {\bfseries 121} (2018) 112001}
  [\href{https://arxiv.org/abs/1803.02685}{{\ttfamily 1803.02685}}].

\bibitem{Chen:2018xof}
J.-W. Chen, L.~Jin, H.-W. Lin, Y.-S. Liu, Y.-B. Yang, J.-H. Zhang et~al.,
  \emph{{Lattice Calculation of Parton Distribution Function from LaMET at
  Physical Pion Mass with Large Nucleon Momentum}},
  \href{https://arxiv.org/abs/1803.04393}{{\ttfamily 1803.04393}}.

\bibitem{Chen:2018fwa}
J.-H. Zhang, J.-W. Chen, L.~Jin, H.-W. Lin, A.~Sch{\"a}fer and Y.~Zhao,
  \emph{{First direct lattice-QCD calculation of the $x$-dependence of the pion
  parton distribution function}},
  \href{https://doi.org/10.1103/PhysRevD.100.034505}{\emph{Phys. Rev.}
  {\bfseries D100} (2019) 034505}
  [\href{https://arxiv.org/abs/1804.01483}{{\ttfamily 1804.01483}}].

\bibitem{Alexandrou:2018eet}
C.~Alexandrou, K.~Cichy, M.~Constantinou, K.~Jansen, A.~Scapellato and
  F.~Steffens, \emph{{Transversity parton distribution functions from lattice
  QCD}}, \href{https://doi.org/10.1103/PhysRevD.98.091503}{\emph{Phys. Rev.}
  {\bfseries D98} (2018) 091503}
  [\href{https://arxiv.org/abs/1807.00232}{{\ttfamily 1807.00232}}].

\bibitem{Liu:2018uuj}
{\scshape Lattice Parton} collaboration, Y.-S. Liu et~al., \emph{{Unpolarized
  isovector quark distribution function from lattice QCD: A systematic analysis
  of renormalization and matching}},
  \href{https://doi.org/10.1103/PhysRevD.101.034020}{\emph{Phys. Rev.}
  {\bfseries D101} (2020) 034020}
  [\href{https://arxiv.org/abs/1807.06566}{{\ttfamily 1807.06566}}].

\bibitem{Lin:2018qky}
H.-W. Lin, J.-W. Chen, X.~Ji, L.~Jin, R.~Li, Y.-S. Liu et~al., \emph{{Proton
  Isovector Helicity Distribution on the Lattice at Physical Pion Mass}},
  {\emph{Phys. Rev. Lett.} {\bfseries 121} (2018) 242003}
  [\href{https://arxiv.org/abs/1807.07431}{{\ttfamily 1807.07431}}].

\bibitem{Fan:2018dxu}
Z.-Y. Fan, Y.-B. Yang, A.~Anthony, H.-W. Lin and K.-F. Liu, \emph{{Gluon
  Quasi-PDF From Lattice QCD}},
  \href{https://doi.org/10.1103/PhysRevLett.121.242001}{\emph{Phys. Rev. Lett.}
  {\bfseries 121} (2018) 242001}
  [\href{https://arxiv.org/abs/1808.02077}{{\ttfamily 1808.02077}}].

\bibitem{Liu:2018hxv}
Y.-S. Liu, J.-W. Chen, L.~Jin, R.~Li, H.-W. Lin, Y.-B. Yang et~al.,
  \emph{{Nucleon Transversity Distribution at the Physical Pion Mass from
  Lattice QCD}},  \href{https://arxiv.org/abs/1810.05043}{{\ttfamily
  1810.05043}}.

\bibitem{Cichy:2019ebf}
K.~Cichy, L.~Del~Debbio and T.~Giani, \emph{{Parton distributions from lattice
  data: the nonsinglet case}},
  \href{https://doi.org/10.1007/JHEP10(2019)137}{\emph{JHEP} {\bfseries 10}
  (2019) 137} [\href{https://arxiv.org/abs/1907.06037}{{\ttfamily
  1907.06037}}].

\bibitem{Chai:2019rer}
Y.~Chai et~al., \emph{{Parton distribution functions of $\Delta^+$ on the
  lattice}},  2019, \href{https://arxiv.org/abs/1907.09827}{{\ttfamily
  1907.09827}}.

\bibitem{Ji:2014hxa}
X.~Ji, P.~Sun, X.~Xiong and F.~Yuan, \emph{{Soft factor subtraction and
  transverse momentum dependent parton distributions on the lattice}},
  {\emph{Phys. Rev.} {\bfseries D91} (2015) 074009}
  [\href{https://arxiv.org/abs/1405.7640}{{\ttfamily 1405.7640}}].

\bibitem{Ji:2018hvs}
X.~Ji, L.-C. Jin, F.~Yuan, J.-H. Zhang and Y.~Zhao, \emph{{Transverse Momentum
  Dependent Quasi-Parton-Distributions}}, {\emph{Phys. Rev.} {\bfseries D99}
  (2019) 114006} [\href{https://arxiv.org/abs/1801.05930}{{\ttfamily
  1801.05930}}].

\bibitem{Ebert:2018gzl}
M.~A. Ebert, I.~W. Stewart and Y.~Zhao, \emph{{Determining the Nonperturbative
  Collins-Soper Kernel From Lattice QCD}}, {\emph{Phys. Rev.} {\bfseries D99}
  (2019) 034505} [\href{https://arxiv.org/abs/1811.00026}{{\ttfamily
  1811.00026}}].

\bibitem{Ebert:2019okf}
M.~A. Ebert, I.~W. Stewart and Y.~Zhao, \emph{{Towards Quasi-Transverse
  Momentum Dependent PDFs Computable on the Lattice}},
  \href{https://doi.org/10.1007/JHEP09(2019)037}{\emph{JHEP} {\bfseries 09}
  (2019) 037} [\href{https://arxiv.org/abs/1901.03685}{{\ttfamily
  1901.03685}}].

\bibitem{Martinelli:1994ty}
G.~Martinelli, C.~Pittori, C.~T. Sachrajda, M.~Testa and A.~Vladikas, \emph{{A
  General method for nonperturbative renormalization of lattice operators}},
  {\emph{Nucl. Phys.} {\bfseries B445} (1995) 81}
  [\href{https://arxiv.org/abs/hep-lat/9411010}{{\ttfamily hep-lat/9411010}}].

\bibitem{Constantinou:2019vyb}
M.~Constantinou, H.~Panagopoulos and G.~Spanoudes, \emph{{One-loop
  renormalization of staple-shaped operators in continuum and lattice
  regularizations}}, {\emph{Phys. Rev.} {\bfseries D99} (2019) 074508}
  [\href{https://arxiv.org/abs/1901.03862}{{\ttfamily 1901.03862}}].

\bibitem{Soper:1979fq}
D.~E. Soper, \emph{{Partons and Their Transverse Momenta in {QCD}}},
  {\emph{Phys. Rev. Lett.} {\bfseries 43} (1979) 1847}.

\bibitem{Collins:1992tv}
J.~C. Collins and F.~V. Tkachov, \emph{{Breakdown of dimensional regularization
  in the Sudakov problem}}, {\emph{Phys. Lett.} {\bfseries B294} (1992) 403}
  [\href{https://arxiv.org/abs/hep-ph/9208209}{{\ttfamily hep-ph/9208209}}].

\bibitem{Collins:2008ht}
J.~Collins, \emph{{Rapidity divergences and valid definitions of parton
  densities}}, {\emph{PoS} {\bfseries LC2008} (2008) 028}
  [\href{https://arxiv.org/abs/0808.2665}{{\ttfamily 0808.2665}}].

\bibitem{GarciaEchevarria:2011rb}
M.~G. Echevarria, A.~Idilbi and I.~Scimemi, \emph{{Factorization Theorem For
  Drell-Yan At Low $q_T$ And Transverse Momentum Distributions
  On-The-Light-Cone}}, {\emph{JHEP} {\bfseries 07} (2012) 002}
  [\href{https://arxiv.org/abs/1111.4996}{{\ttfamily 1111.4996}}].

\bibitem{Chiu:2011qc}
J.-y. Chiu, A.~Jain, D.~Neill and I.~Z. Rothstein, \emph{{The Rapidity
  Renormalization Group}}, {\emph{Phys. Rev. Lett.} {\bfseries 108} (2012)
  151601} [\href{https://arxiv.org/abs/1104.0881}{{\ttfamily 1104.0881}}].

\bibitem{Chiu:2012ir}
J.-Y. Chiu, A.~Jain, D.~Neill and I.~Z. Rothstein, \emph{{A Formalism for the
  Systematic Treatment of Rapidity Logarithms in Quantum Field Theory}},
  {\emph{JHEP} {\bfseries 05} (2012) 084}
  [\href{https://arxiv.org/abs/1202.0814}{{\ttfamily 1202.0814}}].

\bibitem{Capitani:2002mp}
S.~Capitani, \emph{{Lattice perturbation theory}},
  \href{https://doi.org/10.1016/S0370-1573(03)00211-4}{\emph{Phys. Rept.}
  {\bfseries 382} (2003) 113}
  [\href{https://arxiv.org/abs/hep-lat/0211036}{{\ttfamily hep-lat/0211036}}].

\bibitem{Reisz:1988kk}
T.~Reisz, \emph{{Lattice Gauge Theory: Renormalization to All Orders in the
  Loop Expansion}},
  \href{https://doi.org/10.1016/0550-3213(89)90613-5}{\emph{Nucl. Phys.}
  {\bfseries B318} (1989) 417}.

\bibitem{Luscher:1995vs}
M.~Luscher and P.~Weisz, \emph{{Background field technique and renormalization
  in lattice gauge theory}},
  \href{https://doi.org/10.1016/0550-3213(95)00346-T}{\emph{Nucl. Phys.}
  {\bfseries B452} (1995) 213}
  [\href{https://arxiv.org/abs/hep-lat/9504006}{{\ttfamily hep-lat/9504006}}].

\bibitem{Dorn:1986dt}
H.~Dorn, \emph{{Renormalization of Path Ordered Phase Factors and Related
  Hadron Operators in Gauge Field Theories}}, {\emph{Fortsch. Phys.} {\bfseries
  34} (1986) 11}.

\bibitem{Bauer:2001yt}
C.~W. Bauer, D.~Pirjol and I.~W. Stewart, \emph{{Soft collinear factorization
  in effective field theory}}, {\emph{Phys. Rev.} {\bfseries D65} (2002)
  054022} [\href{https://arxiv.org/abs/hep-ph/0109045}{{\ttfamily
  hep-ph/0109045}}].

\bibitem{Bauer:2002nz}
C.~W. Bauer, S.~Fleming, D.~Pirjol, I.~Z. Rothstein and I.~W. Stewart,
  \emph{{Hard scattering factorization from effective field theory}},
  \href{https://doi.org/10.1103/PhysRevD.66.014017}{\emph{Phys. Rev.}
  {\bfseries D66} (2002) 014017}
  [\href{https://arxiv.org/abs/hep-ph/0202088}{{\ttfamily hep-ph/0202088}}].

\bibitem{Shanahan:2019zcq}
P.~Shanahan, M.~Wagman and Y.~Zhao, \emph{{Nonperturbative renormalization of
  staple-shaped Wilson line operators in lattice QCD}},
  \href{https://arxiv.org/abs/1911.00800}{{\ttfamily 1911.00800}}.

\end{thebibliography}\endgroup

\end{document}